\documentclass[useAMS,fleqn, usenatbib, final]{mnras}

\usepackage{amssymb}
\usepackage{amsmath}
\usepackage{array}
\usepackage{graphicx}
\usepackage{color,units}
\usepackage[dvipsnames]{xcolor} 
\usepackage{aas_macros}
\usepackage{lineno}
\usepackage{xspace}
\usepackage{dcolumn}
\usepackage{longtable}
\usepackage[normalem]{ulem} 
\usepackage{subfigure}
\usepackage[T1]{fontenc}
\usepackage{hyperref}

\newcommand{\bilby}{{\sc Bilby}\xspace}
\newcommand{\bilbypipe}{{\sc bilby\_pipe}\xspace}
\newcommand{\bilbypipegdb}{\texttt{bilby\_pipe\_gracedb}\xspace}
\newcommand{\pbilby}{{\sc pBilby}\xspace}
\newcommand{\dynesty}{{\sc dynesty}\xspace}
\newcommand{\cpnest}{{\sc cpnest}\xspace}
\newcommand{\ptemcee}{{\sc ptemcee}\xspace}
\newcommand{\gwpy}{{\sc GWpy}\xspace}
\newcommand{\lvalert}{{\sc lvalert}\xspace}
\newcommand{\gwcelery}{{\sc GWCelery}\xspace}
\newcommand{\pesummary}{{\sc PESummary}\xspace}
\newcommand{\scipy}{{\sc scipy}\xspace}
\newcommand{\lal}{{\sc LAL}\xspace}
\newcommand{\lalinference}{{\sc LALInference}\xspace}
\newcommand{\lalsuite}{{\sc LALSuite}\xspace}
\newcommand{\lalsimulation}{{\sc LALSimulation}\xspace}
\newcommand{\rapidpe}{{\sc RapidPE}\xspace}
\newcommand{\rift}{{\sc RIFT}\xspace}
\newcommand{\bayestar}{{\sc bayestar}\xspace}
\newcommand{\bayeswave}{{\sc BayesWave}\xspace}
\newcommand{\bayesline}{{\sc BayesLine}\xspace}
\newcommand{\astropy}{{\tt Astropy}\xspace}
\newcommand{\pycbcinference}{{\sc PyCBCInference}\xspace}

\newcommand{\skymap}{{\tt ligo.skymap}\xspace}
\newcommand{\python}{{\sc Python}}

\newcommand{\imrphenomp}{{\sc IMRPhenomPv2}\xspace}
\newcommand{\imrphenompvtwonrtidal}{{\sc IMRPhenomPv2\_NRTidalv2}\xspace}

\newcommand{\pptestpvalue}{0.7206\xspace}

\LTcapwidth=\textwidth

\title[\bilby Gravitational-Wave catalogue]{
Bayesian inference for compact binary coalescences with \bilby:
Validation and application to the first LIGO--Virgo
gravitational-wave transient catalogue}

\newcommand{\SPAno}{1}
\newcommand{\OzGravMonashno}{2}
\newcommand{\CITno}{3}
\newcommand{\MITno}{4}
\newcommand{\Kavlino}{5}
\newcommand{\Cardiffno}{6}
\newcommand{\CIERAno}{7}
\newcommand{\SUPAno}{8}
\newcommand{\NIKHEFno}{9}
\newcommand{\UVAno}{10}
\newcommand{\IITKno}{11}
\newcommand{\IUCAAno}{12}
\newcommand{\Lancasterno}{13}
\newcommand{\Swinno}{14}
\newcommand{\OzGravSwinno}{15}
\newcommand{\apcno}{16}
\newcommand{\Minnesotano}{17}
\newcommand{\Bombayno}{18}
\newcommand{\CCAno}{19}
\newcommand{\GeorgiaTechno}{20}
\newcommand{\Mallorcano}{21}
\newcommand{\MaxPlanckno}{22}
\newcommand{\Oregonno}{23}
\newcommand{\Milwaukeeno}{24}
\newcommand{\Barcelonano}{25}
\newcommand{\CSIROno}{26}
\newcommand{\Floridano}{27}
\newcommand{\LaSapienzano}{28}
\newcommand{\INFNno}{29}
\newcommand{\UniMelbno}{30}
\newcommand{\OzGravUniMelbno}{31}
\newcommand{\ICRRno}{32}
\newcommand{\NIMSno}{33}
\newcommand{\UoBno}{34}
\newcommand{\SPA}{School of Physics and Astronomy, Monash University, Clayton VIC 3800, Australia}
\newcommand{\OzGravMonash}{OzGrav: The ARC Centre of Excellence for Gravitational Wave Discovery, Clayton VIC 3800, Australia}
\newcommand{\OzGravSwin}{OzGrav: The ARC Centre of Excellence for Gravitational Wave Discovery, Hawthorn VIC 3122, Australia}
\newcommand{\Swin}{Centre for Astrophysics and Supercomputing, Swinburne University of Technology, Hawthorn VIC 3122, Australia}
\newcommand{\UniMelb}{School of Physics, University of Melbourne, Parkville, Victoria 3010, Australia}
\newcommand{\OzGravUniMelb}{OzGrav: The ARC Centre of Excellence for Gravitational Wave Discovery, University of Melbourne, Parkville,\\ Victoria 3010, Australia}

\newcommand{\CIERA}{Center for Interdisciplinary Exploration and Research in Astrophysics (CIERA), \\Northwestern University, 1800 Sherman Ave, Evanston, IL 60201, USA}
\newcommand{\MIT}{LIGO Laboratory, Massachusetts Institute of Technology, Cambridge, MA 02139, USA}
\newcommand{\Kavli}{Department of Physics and Kavli Institute for Astrophysics and Space Research, Massachusetts Institute of Technology, \\ 77 Massachusetts Ave, Cambridge, MA 02139, USA}
\newcommand{\Cardiff}{Cardiff University, Cardiff CF24 3AA, UK}
\newcommand{\IUCAA}{Inter-University Centre for Astronomy and Astrophysics, Pune 411007, India}
\newcommand{\Lancaster}{Department of Physics, Lancaster University, Lancaster, LA1 4YB, UK}
\newcommand{\Minnesota}{University of Minnesota, Minneapolis, MN 55455, USA}
\newcommand{\CCA}{Center for Computational Astrophysics, Flatiron Institute, 162 5th Ave, New York, NY 10010}

\newcommand{\GeorgiaTech}{School of Physics, Georgia Institute of Technology, Atlanta, GA 30332, USA}
\newcommand{\MaxPlanck}{Max Planck Institute for Gravitational Physics (Albert Einstein Institute), D-14476 Potsdam-Golm, Germany}

\newcommand{\Oregon}{University of Oregon, Eugene, OR 97403, USA}
\newcommand{\Milwaukee}{University of Wisconsin-Milwaukee, Milwaukee, WI 53201, USA}
\newcommand{\SUPA}{SUPA, University of Glasgow, Glasgow G12 8QQ, UK}

\newcommand{\ICRR}{Institute for Cosmic Ray Research, The University of Tokyo, 5-1-5 Kashiwanoha, Kashiwa, Chiba 277-8582, Japan}

\newcommand{\NIMS}{National Institute for Mathematical Sciences, Daejeon 34047, South Korea}
\newcommand{\CIT}{LIGO Laboratory, California Institute of Technology, Pasadena, CA 91125, USA}
\newcommand{\NIKHEF}{Nikhef, Science Park 105, 1098 XG Amsterdam, The Netherlands}
\newcommand{\UVA}{Institute for High-Energy Physics, University of Amsterdam, Science Park 904, 1098 XH Amsterdam, The Netherlands}
\newcommand{\IITK}{Department of Physics, Indian Institute of Technology, Kanpur 208016, India}
\newcommand{\Bombay}{Department of Physics, Indian Institute of Technology Bombay, Mumbai, Maharashtra 400076, India}
\newcommand{\apc}{Université de Paris, CNRS, Astroparticule et Cosmologie, F-75006 Paris, France}
\newcommand{\Mallorca}{Universitat de les Illes Balears, IAC3---IEEC, E-07122 Palma de Mallorca, Spain}
\newcommand{\Barcelona}{Institut de F\'isica d'Altes Energies (IFAE), Barcelona Institute of Science and Technology, and ICREA, \\ E-08193 Barcelona, Spain}
\newcommand{\CSIRO}{CSIRO Astronomy and Space Science, Australia Telescope National Facility, Epping, NSW 1710, Australia}
\newcommand{\Florida}{University of Florida, Gainesville, FL 32611, USA}
\newcommand{\LaSapienza}{Universit\`a di Roma ``La Sapienza,'' I-00185 Roma, Italy}
\newcommand{\INFN}{INFN, Sezione di Roma, I-00185 Roma, Italy}
\newcommand{\UoB}{School of Physics and Astronomy and Institute for Gravitational Wave Astronomy, University of Birmingham, \\ Edgbaston, Birmingham, B15 9TT, United Kingdom}

\author[Romero-Shaw et al.]{
\parbox{\textwidth}{
I.~M.~Romero-Shaw$^{\SPAno,\OzGravMonashno}$\thanks{isobel.romero-shaw@monash.edu},
C.~Talbot$^{\CITno,\SPAno,\OzGravMonashno}$,
S.~Biscoveanu$^{\MITno,\Kavlino}$, 
V.~D'Emilio$^\Cardiffno$, 
G.~Ashton$^{\SPAno,\OzGravMonashno}$, 
C.~P.~L.~Berry$^{\CIERAno,\SUPAno}$, 
S.~Coughlin$^{\CIERAno}$, 
S.~Galaudage$^{\SPAno,\OzGravMonashno}$, 
C.~Hoy$^\Cardiffno$, 
M.~H\"ubner$^{\SPAno,\OzGravMonashno}$, 
K.~S.~Phukon$^{\NIKHEFno,\UVAno, \IITKno,\IUCAAno}$, 
M.~Pitkin$^{\Lancasterno}$, 
M.~Rizzo$^{\CIERAno}$, 
N.~Sarin$^{\SPAno,\OzGravMonashno}$, 
R.~Smith$^{\SPAno,\OzGravMonashno}$, 
S.~Stevenson$^{\Swinno,\OzGravSwinno}$, 
A.~Vajpeyi$^{\SPAno,\OzGravMonashno}$, 
M.~Ar\`ene$^{\apcno}$, 
K.~Athar$^{\SPAno}$,
S.~Banagiri$^{\Minnesotano}$, 
N.~Bose$^{\Bombayno}$, 
M.~Carney$^\CIERAno$, 
K.~Chatziioannou$^{\CCAno}$, 
J.~A.~Clark$^{\GeorgiaTechno}$, 
M.~Colleoni$^{\Mallorcano}$,
R.~Cotesta$^{\MaxPlanckno}$, 
B.~Edelman$^{\Oregonno}$, 
H.~Estell\'es$^{\Mallorcano}$
C.~Garc{\'i}a-Quir{\'o}s$^{\Mallorcano}$, 
Abhirup Ghosh$^{\MaxPlanckno}$, 
R.~Green$^{\Cardiffno}$, 
C.-J.~Haster$^{\MITno,\Kavlino}$, 
S.~Husa$^{\Mallorcano}$,
D.~Keitel$^{\Mallorcano}$,
A.~X.~Kim$^\CIERAno$, 
F.~Hernandez-Vivanco$^{\SPAno,\OzGravMonashno}$, 
I.~Maga\~na~Hernandez$^{\Milwaukeeno}$, 
C.~Karathanasis$^{\Barcelonano}$, 
P.~D.~Lasky$^{\SPAno,\OzGravMonashno}$, 
N.~De Lillo$^\SUPAno$, 
M.~E.~Lower$^{\Swinno,\OzGravSwinno,\CSIROno}$, 
D.~Macleod$^\Cardiffno$, 
M.~Mateu-Lucena$^{\Mallorcano}$, 
A.~Miller$^{\Floridano,\LaSapienzano,\INFNno}$, 
M.~Millhouse$^{\UniMelbno, \OzGravUniMelbno}$, 
S.~Morisaki$^{\ICRRno}$, 
S.~H.~Oh$^{\NIMSno}$, 
S.~Ossokine$^{\MaxPlanckno}$, 
E.~Payne$^{\SPAno,\OzGravMonashno}$, 
J.~Powell$^{\Swinno,\OzGravSwinno}$, 
G.~Pratten$^{\UoBno}$,
M.~P\"urrer$^{\MaxPlanckno}$, 
A.~Ramos-Buades$^{\Mallorcano}$, 
V.~Raymond$^\Cardiffno$, 
E.~Thrane$^{\SPAno,\OzGravMonashno}$, 
J.~Veitch$^\SUPAno$ 
D.~Williams$^\SUPAno$, 
M.~J.~Williams$^\SUPAno$, 
L.~Xiao$^\CITno$, 
}\vspace{0.2cm}\\
$^1$\SPA\\
$^2$\OzGravMonash\\
$^3$\CIT\\
$^4$\MIT\\
$^5$\Kavli\\
$^6$\Cardiff\\
$^7$\CIERA\\
$^8$\SUPA\\
$^9$\NIKHEF\\
$^{10}$\UVA\\
$^{11}$\IITK\\
$^{12}$\IUCAA\\
$^{13}$\Lancaster\\
$^{14}$\Swin\\
$^{15}$\OzGravSwin\\
$^{16}$\apc\\
$^{17}$\Minnesota\\
$^{18}$\Bombay\\
$^{19}$\CCA\\
$^{20}$\GeorgiaTech\\
$^{21}$\Mallorca\\
$^{22}$\MaxPlanck\\
$^{23}$\Oregon\\
$^{24}$\Milwaukee\\
$^{25}$\Barcelona\\
$^{26}$\CSIRO\\
$^{27}$\Florida\\
$^{28}$\LaSapienza\\
$^{29}$\INFN\\
$^{30}$\UniMelb\\
$^{31}$\OzGravUniMelb\\
$^{32}$\ICRR\\
$^{33}$\NIMS\\
$^{34}$\UoB
}

\date{Accepted XXX. Received YYY; in original form ZZZ}
\pubyear{2020}

\begin{document}
\label{firstpage}
\pagerange{\pageref{firstpage}--\pageref{lastpage}}
\maketitle
\clearpage

\begin{abstract}
Gravitational waves provide a unique tool for observational astronomy. 
While the first LIGO--Virgo catalogue of gravitational-wave transients (GWTC-1) contains eleven signals from black hole and neutron star binaries, the number of observations is increasing rapidly as detector sensitivity improves. 
To extract information from the observed signals, it is imperative to have fast, flexible, and scalable inference techniques. 
In a previous paper, we introduced \bilby: a modular and user-friendly Bayesian inference library adapted to address the needs of gravitational-wave inference. 
In this work, we demonstrate that \bilby produces reliable results for simulated gravitational-wave signals from compact binary mergers, and verify that it accurately reproduces results reported for the eleven GWTC-1 signals. 
Additionally, we provide configuration and output files for all analyses to allow for easy reproduction, modification, and future use. 
This work establishes that \bilby is primed and ready to analyse the rapidly growing population of compact binary coalescence gravitational-wave signals.
\end{abstract}


\section{Introduction}

Gravitational-wave astronomy presents a revolutionary opportunity to probe fundamental physics and astrophysics, ranging from the neutron star equation of state and stellar evolution to the expansion of the Universe.
The first direct observations of gravitational-wave signals have been made by Advanced LIGO~\citep{ligo} and Advanced Virgo~\citep{virgo}; their first gravitational-wave catalogue of transients~\citep[GWTC-1;][]{abbott2019_GWTC1} contains ten binary black hole coalescences and one binary neutron star coalescence.
The third observing run may yield $\mathcal{O}(10^2)$ additional observations~\citep{abbott_19_observing_scenarios}, with signals from a second binary neutron star merger \citep{abbott20_GW190425}, one merger of a black hole with a $\unit[2.6]{M_\odot}$ compact object, and an additional two binary black hole mergers~\citep{abbott20_GW190412, abbott20_GW190521} already confirmed.

Gravitational-wave signals encode information about their sources which can be difficult, if not impossible, to otherwise obtain. 
To extract information from the observed signals requires careful statistical inference. 
The inferred source parameters can inform our understanding of binary stellar evolution~\citep{stevenson15,abbott16_gw150914_astro,Zevin2017,abbott17_gw170817_progenitor,Barrett2018,belczynski18,bavera19}, the equation of state of neutron-star matter~\citep{abbott18_GW170817_NS_parameters,most18,essick19,abbott19_EOS_model-select}, and the nature of gravity~\citep{yunes13,abbott16_gw150914_testingGR,yunes16,abbott19_TGR,isi19}. 
Multimessenger observations of gravitational and electromagnetic radiation~\citep{abbott17_gw170817_multimessenger} can give an even richer understanding, enabling measurements of cosmological parameters~\citep{abbott17_gw170817_Hubble,abbott19_O2_cosmo,cantiello18,hotokezaka19,dhawan19,chen18}, insights into the structures of gamma-ray bursts~\citep{abbott17_gw170817_gwgrb,mooley18,margutti18,fong19,biscoveanu2019a}, and identifying the origins of heavy elements~\citep{abbott17_gw170817_ejecta,chornock17,tanvir17,kasliwal19,watson19}.
However, electromagnetic emission can fade rapidly, necessitating rapid localization of the gravitational-wave source~\citep{abbott_19_observing_scenarios}.
To maximize the scientific return of gravitational-wave observations, it is therefore of paramount importance to make use of and continue to develop efficient, reliable, and accurate computational inference.

\bilby is a user-friendly Bayesian inference library that can be used to analyse gravitational-wave signals to infer their source properties~\citep{ashton19}.
\bilby is modular and can be easily adapted to handle a range of inference problems in gravitational-wave astronomy and beyond~\citep[e.g.,][]{powell19,farah19,goncharov19, sarin20}.
In the context of gravitational-wave astrophysics and compact binary mergers, it has been used to extract information about short gamma-ray burst properties~\citep{biscoveanu2019a}, neutron star parameters~\citep{coughlin19,hernandezvivanco2019, hernandezvivanco2019b, biscoveanu2019b}, the formation history of binary compact objects~\citep{lower18,romero-shaw19,ramos-buades2019, romero-shaw20,zevin2020}, population properties using hierarchical inference~\citep{abbott19_O2_pops,talbot19,galaudage19,Kimball20}, and test general relativity~\citep[][]{keitel19, ashton2019b, payne2019, zhao19, huebner19, Wang2020}.
This paper concentrates on using \bilby to infer the properties of individual signals from compact binary coalescences---the inspiral, merger and ringdown of binaries composed of neutron stars and black holes. 

We outline the developments included in the \bilby software to accurately and efficiently infer the properties of compact binary coalescence (CBC) signals, and demonstrate their validity both through tests using simulated signals and via comparisons to existing observational results.
In Section~\ref{sec:bayesian-inference}, we describe the applications of Bayesian inference to compact binary coalescence events detected in gravitational waves.
In Section~\ref{sec:bilby-package}, we focus on the \bilby package, with particular emphasis on improvements made since the publication of \citet{ashton19} in Section~\ref{sec:changes}.
We outline our code validation tests in Section~\ref{subsec:validation}, and describe the automation of \bilby---allowing for efficient and immediate analysis of gravitational-wave event candidates---in Section~\ref{sec:automation}. 
In Section~\ref{sec:event-catalog}, we reanalyse the eleven signals from GWTC-1, ensuring that we use both identical data and identical data processing techniques as used to produce the public GWTC-1 results obtained using the Bayesian parameter estimation package \lalinference~\citep{veitch15}. 
We cross-validate our results for GWTC-1 against these previous results.
We defer analysis of detections from the third observing run in anticipation of a future \bilby catalogue.
Results of the analyses presented here, in a format matching recent releases of LIGO--Virgo posterior samples, are provided as accompaniments to this paper.
Our investigations confirm the effectiveness of \bilby as it begins to be used for LIGO--Virgo parameter estimation~\cite{abbott20_GW190425,abbott20_GW190412}.
Throughout this paper, we use notations for CBC source parameters that are defined in Appendix~\ref{sec:definitions}.

\section{Bayesian Inference for Compact Binaries}\label{sec:bayesian-inference}
In this section, we outline the fundamental procedures carried out by \bilby and provide a summary of new features implemented since the first \bilby paper~\citep{ashton19}. For a thorough and up-to-date description of \bilby, the reader is directed to the \bilby documentation.\footnote{\href{https://lscsoft.docs.ligo.org/bilby}{lscsoft.docs.ligo.org/bilby/}}

\subsection{Applications of Bayesian Inference to Compact Binary Coalescences}\label{subsec:bayesian_inference}
The primary objective of gravitational-wave inference for compact binary merger signals is to recover posterior probability densities for the source parameters $\boldsymbol{\theta}$ (defined in Appendix~\ref{sec:definitions}), like the masses and spins of the binary components, given the data and a model hypothesis. 
The posterior can be computed using Bayes' theorem~\citep{Bayes},
\begin{align}
    p(\boldsymbol{\theta}|d,\mathcal{H}) &= \frac{\mathcal{L}(d|\boldsymbol{\theta}, \mathcal{H})\pi(\boldsymbol{\theta}|\mathcal{H})}{\mathcal{Z}(d|\mathcal{H})},
    \label{eq:posterior}
\end{align}
where  $\mathcal{L}(d|\boldsymbol{\theta}, \mathcal{H})$ is the likelihood, $\pi(\boldsymbol{\theta}|\mathcal{H})$ is the prior, $\mathcal{Z}(d|\mathcal{H})$ is the evidence, and $\mathcal{H}$ is the model. 
The prior is chosen to incorporate any \textit{a priori} knowledge about the parameters.
The likelihood represents the probability of the detectors measuring data $d$, assuming a signal (described by the model hypothesis $\mathcal{H}$) with source properties $\boldsymbol{\theta}$. 
The evidence, or marginalized likelihood,
\begin{align}
\mathcal{Z}(d|\mathcal{H})=\int p(d|\boldsymbol{\theta},\mathcal{H})\pi(\boldsymbol{\theta}|\mathcal{H})\,\mathrm{d}\boldsymbol{\theta},
\end{align} 
serves as a measure of how well the data is modeled by the hypothesis; it acts as a normalization constant in parameter estimation, but is important in model selection.

The standard likelihood function used to analyse gravitational-wave transients is defined in, e.g.,~\citet{finn92, romano17}, where both the data and the model are expressed in the frequency domain.
This likelihood has stationary Gaussian noise, which is a good approximation in most cases~\citep[e.g.,][]{berry15,abbott17_gw150914_systematics, ligo2019guide} unless one of the instruments is affected by a glitch~\citep{pankow18,powell18_glitch}. 
We assume the noise power spectral density (PSD) is independent of the model parameters and therefore ignore the normalization term, yielding
\begin{align}
\ln\mathcal{L}(d|\boldsymbol{\theta}) &\propto -\sum_k \frac{2|d_k - h_k (\boldsymbol{\theta})|^2}{T S_k},
\label{eq:likelihood2}
\end{align}
where $k$ is the frequency bin index, $S$ is the PSD of the noise, $T$ is the duration of the analysis segment.
The data $d$ and waveform model $h(\boldsymbol{\theta})$ are the Fourier transforms of their time-domain counterparts.
Given the likelihood and the prior, we can calculate the posterior probability distribution for the source parameters.

There are multiple approaches to calculating the posterior probability distribution. 
For example, \linebreak{}
\rapidpe~\citep{pankow15} and its iterative spin-off \rift~\citep{lange2018} use highly-parallelized grid-based methods to compute the posterior probability distribution, while 
\bayestar \citep{singer16a,singer16b} rapidly localizes gravitational-wave sources, calculating probabilities on a multiresolution grid of the sky.
Bayesian inference schemes using various machine-learning algorithms are also being developed~\citep{george18,gabbard19}.
However, the majority of Bayesian inference analysis is done by stochastically sampling the posterior probability distribution. 

Over many years, Markov-chain Monte Carlo \citep[MCMC;][]{christensen98, christensen01, rover06, rover07, vandersluys08b, vandersluys08a} and nested sampling~\citep{veitch08,veitch10} algorithms for gravitational-wave inference have been developed. 
This work culminated in the development of \lalinference, a Bayesian inference library using custom-built Markov-chain Monte Carlo and nested sampling algorithms~\citep{veitch15}.\footnote{In this work, we focus on Bayesian inference for ground-based gravitational-wave detection. Similar techniques have been developed for studying the gravitational-wave observations of other instruments, such as pulsar timing arrays \citep{temponest, VigelandVallisneri} and future space-based detectors \citep{Babak08, Babak10, Marsat20}.} 
\lalinference has been the workhorse of gravitational-wave inference since the initial LIGO--Virgo era~\citep{aasi13_big_dog}, through the first observation~\citep{abbott16_gw150914_pe} to the production of GWTC-1~\citep{abbott2019_GWTC1}. 
Other stochastic sampling packages used for gravitational-wave inference include \pycbcinference~\citep{biwer18} and \citet{zackay18}, which uses relative-binning \citep{Cornish10, Cornish20} to reduce the computational cost of the likelihood.
In addition to these sampling packages which fit CBC waveform templates to the data, \bayeswave~\citep{Cornish14} uses a trans-dimensional MCMC to fit an \textit{a priori} unknown number of sine-Gaussian wavelets to the data.
\bayeswave also implements the \bayesline algorithm~\citep{Littenberg14} to generate a parameterised fit for the interferometer noise PSD.
Power spectral densities produced by \bayesline are widely used in gravitational-wave parameter estimation and are used in this work.
\bilby has been designed to adapt to the changing needs of the gravitational-wave inference community, emphasizing modularity and ease of accessibility.

While \lalinference implements customized stochastic samplers, \bilby employs external, off-the-shelf samplers, with some adaption.
This allows the user to easily switch between samplers with minimal disruption: a useful feature for cross validating results using different samplers.
Typically, external samplers need to be tuned and adapted for use in gravitational-wave inference.
In some cases, this is a simple case of choosing sensible settings; we provide details of the settings that have been verified for gravitational-wave analysis in Section~\ref{sec:event-catalog} and Appendix~\ref{appendix:run-settings}.
However, we also find cases where the off-the-shelf samplers themselves need to be adjusted.
Where possible, we propagate those proposed changes to the original sampling packages. 
Alternatively (e.g., when the change is perhaps gravitational-wave specific), we adjust the sampler from within \bilby.

\subsection{Stochastic Sampling}
Various Monte Carlo sampling schemes have been developed to solve the Bayesian inference problem and estimate the posterior distribution described by Eq.~(\ref{eq:posterior}).
For low-dimensional problems, a solution might be to estimate the best-fit parameters by computing the posterior probability for every point on a grid over the parameter space.
However, as the number of dimensions increases, this becomes exponentially inefficient.\footnote{Quasi-circular binary black hole coalescence waveform models typically have $n_{\mathrm{dim}}=15$, depending on the number of spin orientations included in the waveform model. Binary neutron star coalescence models include an additional two parameters that describe their tides. We provide definitions of all parameters describing binary compact objects in Appendix~\ref{sec:definitions}. There are a further $\approx20$ parameters per interferometer that describe uncertainties in detector calibration.} 
The common alternative to solve this problem has been to use stochastic samplers, which fall broadly into two (not mutually exclusive) categories:
MCMC \citep{metropolis1953equation, hastings1970monte} and nested sampling \citep{Skilling06}. In general terms, independent samples are drawn \textit{stochastically} from the posterior, such that the number of samples in the range $(\boldsymbol{\theta}, \boldsymbol{\theta} + \boldsymbol{\Delta \theta})$ is proportional to $p(\boldsymbol{\theta} |d, \mathcal{H})\Delta\theta$.

MCMC methods generate posterior samples by noting the positions of particles undergoing a biased random walk through the parameter space, with the probability of moving to a new point in the space given by the transition probability of the Markov chain.
Sampling is completed once some user-specified termination condition is reached, usually a threshold for the number of posterior samples that should be accumulated to provide an accurate representation of the posterior.

Nested sampling methods generate posterior samples as a byproduct of calculating the evidence integral $\mathcal{Z}(d|\mathcal{H})$.
A set of live points is drawn from the prior distribution, and at each iteration, the live point with the lowest likelihood is replaced by a new nested sample that lies in a part of the parameter space with a higher likelihood.
The evidence is approximated by summing the products of the likelihood at the discarded point and the difference in the prior volume between successive iterations.
The nested samples are converted to posterior samples by weighting by the posterior probability at that point in the parameter space.
The nested sampling algorithm stops once a predefined termination condition has been reached.
The most commonly used termination condition is when the fraction of the evidence in the remaining prior volume is smaller than a predefined amount.

For more details on both MCMC and nested sampling methods, we refer the reader to \cite{hogg18} and \cite{dynesty}, respectively. 

\section{The \bilby Package}\label{sec:bilby-package}

\bilby has a modular structure, allowing users to extend and develop it to suit their needs; examples include online \bilby (Section~\ref{sec:online_bilby}), \bilbypipe (Section~\ref{sec:bilbypipe}) and parallel \bilby \citep[\pbilby; Section~\ref{sec:parallel_bilby};][]{Smith:2019ucc}, amongst others \citep[e.g.,][]{talbot19}. 
\bilby comprises three main subpackages.
The \texttt{core} subpackage contains the basic implementation of likelihoods, priors, sampler interfaces, the result container class and a host of utilities.
The \texttt{gw} subpackage builds on \texttt{core} and contains gravitational-wave specific implementations of priors and likelihoods.
These implementations include a detailed detector and calibration model, an interface to waveform models, and a number of utilities.
Finally, the \texttt{hyper} subpackage implements hyper-parameter estimation in \bilby, which in the gravitational-wave context is used for population inference.

\subsection{Changes within \bilby}
\label{sec:changes}
Since the original \bilby paper~\citep{ashton19}, there have been a number of significant changes and added features to the code package. 
We describe these in the following subsections.
We discuss prior constraints in Section~\ref{sec:constrained_priors}, conditional priors in Section~\ref{sec:conditional_priors}, and the implementation of cosmological priors in Section~\ref{sec:cosmological_priors}.
We detail the custom jump proposals implemented for the \cpnest~\citep{Veitch17-cpnest} and {\sc ptmcmc}\xspace~\citep{ptmcmc} samplers in Section~\ref{sec:custom_jump_proposals}, and the various available prior boundary conditions in Section~\ref{sec:boundary_conditions}.
Sampling processes can be accelerated using likelihood marginalizations and reduced-order quadratures; we explore how these methods can be applied to \bilby analyses in Sections \ref{sec:marginalisation} and \ref{sec:roq}, respectively.
In Section~\ref{sec:calibration}, we explain how uncertainties in detector calibration are folded into \bilby parameter estimation.
Finally, in Section~\ref{sec:gwt_specific_plots} we present some of the gravitational-wave transient-specific plots that \bilby can create.
In addition to the changes described below, \bilby now also supports the {\sc kombine}\xspace~\citep{kombine}, {\sc ptmcmc}\xspace~\citep{ptmcmc}, {\sc PolyChord}\xspace~\citep{polychord1, polychord2}, and {\sc UltraNest}\xspace~\citep{ultranest1, ultranest2} samplers.
A full and up-to-date list of changes can be found in the \bilby changelog.\footnote{\href{https://git.ligo.org/lscsoft/bilby/blob/master/CHANGELOG.md}{git.ligo.org/lscsoft/bilby/blob/master/CHANGELOG.md}}

\subsubsection{Constrained priors}\label{sec:constrained_priors}
Each time the sampler chooses a new point to test from the multi-dimensional parameter space, it selects this point from within the region specified by the multi-dimensional prior.
It is often advantageous to be able to cut out parts of the prior space by placing restrictions on relationships between parameters.
For example, in gravitational-wave inference we frequently wish to specify a prior on the binary component masses, $m_1$ and $m_2$, while enforcing that $m_1 \geq m_2$, which is equivalent to the constraint that the mass ratio $q = m_2 / m_1 \leq 1$.

In \bilby, the collection of priors on all parameters is stored as a \texttt{PriorDict} object. 
In order to enforce a constraint, a \bilby user can add a \texttt{Constraint} prior object to the \texttt{PriorDict}.
It is necessary to tell the {\tt PriorDict} how to convert between its sampled parameters and its constrained parameters; this is done by passing a {\tt conversion\_function} at instantiation of the \texttt{PriorDict}.
The \bilby default binary black hole and binary neutron star prior set classes ({\tt BBHPriorDict} and {\tt BNSPriorDict}, respectively) can impose constraints on any of the known binary parameters.
This ensures that users can sample in the set of parameters that best suits their problem, while ensuring that the relevant indirectly-sampled quantities are constrained.
Without applying any prior constraints, all \bilby prior distributions are correctly normalised.
When constraints are imposed on the prior distribution, the updated normalisation is approximated using a Monte Carlo integral.

\subsubsection{Conditional priors}\label{sec:conditional_priors}

One may choose to make the prior for one parameter conditional on the value of another.
This can increase efficiency, particularly if large parts of the prior space would be forbidden by an equivalent constraint prior.
A commonly used parameterisation of the population distribution of binary black hole masses is
\begin{equation}
    \begin{split}
        p(m_1 | m_{\min}, m_{\max}, \alpha) &= (1 - \alpha) \frac{m_{1}^{-\alpha}}{m_{\max}^{1-\alpha} - m_{\min}^{1-\alpha}}, \\
        p(q | m_{1}, m_{\min}, \beta) &= (1 + \beta) \frac{m_{1}^{1 + \beta} q^{\beta}}{m_{1}^{1 + \beta} - m_{\min}^{1 + \beta}},
    \end{split}
\end{equation}
where $m_{\min}$ and $m_{\max}$ are the maximum and minimum allowed masses for the primary component, and $\alpha$ and $\beta$ are power-law indices~\citep{Fishbach17b,abbott19_O2_pops}. 
If we wish to use a similar prior to analyse individual binary black hole coalescences, we require a prior for mass ratio which is conditioned on the primary mass.
We provide a \texttt{ConditionalPriorDict} and conditional versions of all implemented priors within \bilby to facilitate analyses of this kind.
Further, \bilby is able to handle nested and multiple dependencies, and automatically resolves the order in which conditional priors need to be called. 
The conditional relationship between different priors can have any functional form specified by the user.

\subsubsection{Cosmological priors}\label{sec:cosmological_priors}

\begin{figure}
    \centering
    \includegraphics[width=\linewidth]{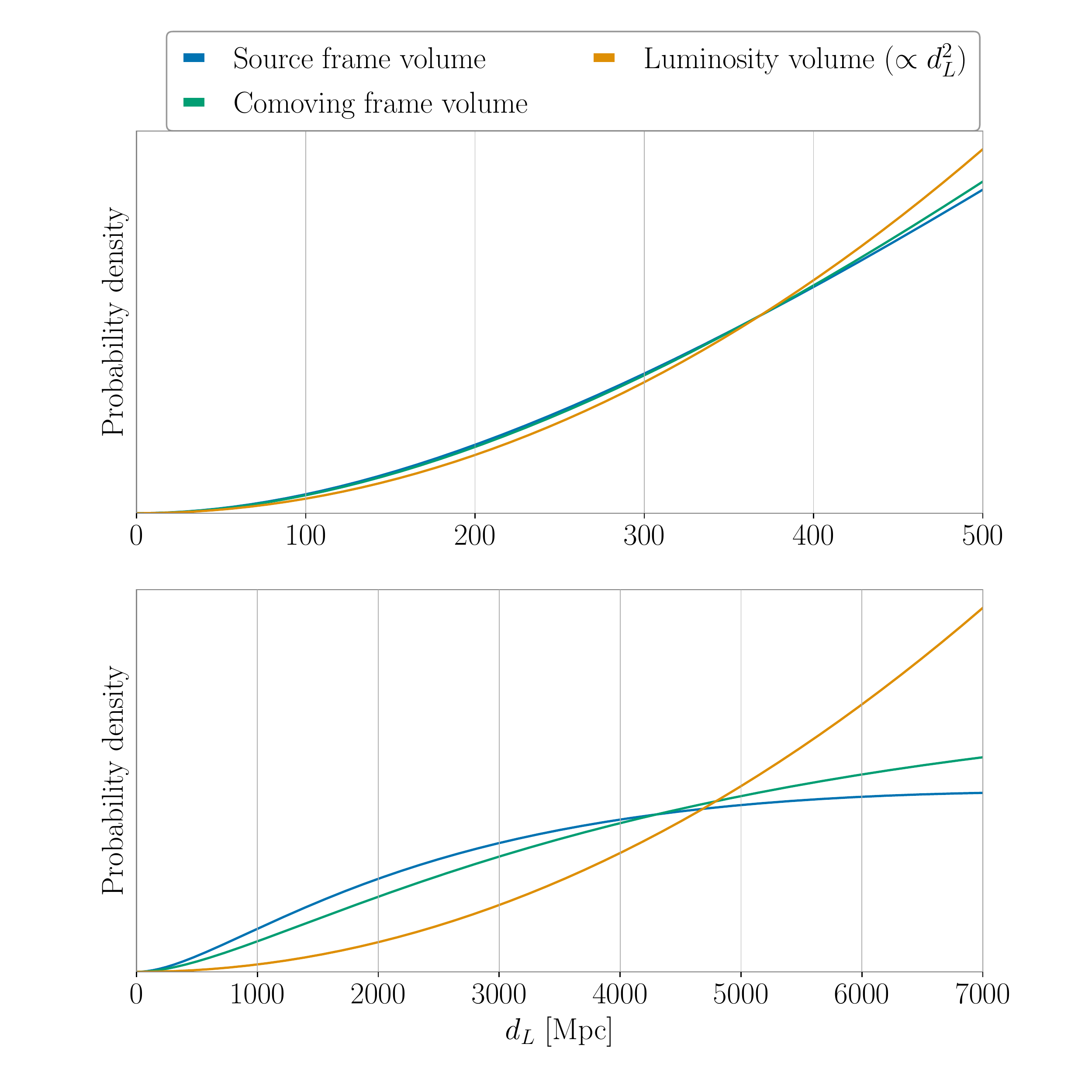}
    \caption{Comparison of distance priors out to redshift $z=0.10$ (top panel) and $z=1.02$ (bottom panel), respectively corresponding to $d_\mathrm{L}=500~\mathrm{Mpc}$ and $d_\mathrm{L}=7000~\mathrm{Mpc}$, according to \citet{ade2016} cosmology.
    The upper and lower panels show the range of the luminosity distance priors for the default \unit[128]{s} and high-mass prior sets, respectively. We display priors that are uniform in luminosity volume, comoving volume, and the (comoving) source frame.
    The probability density of each curve is normalized with respect to the upper limit cut-off displayed in that panel.}
    \label{fig:distance-prior-comparison}
\end{figure}

Most previous parameter estimation analyses of CBCs have assumed a prior on luminosity distance $d_\mathrm{L}$ which is $\pi(d_\mathrm{L}) \propto d_\mathrm{L}^2$ \citep[e.g.,][]{abbott16_gw150914_pe,abbott2019_GWTC1}. 
A $\pi(d_\mathrm{L}) \propto d_\mathrm{L}^2$ prior would distribute mergers uniformly throughout a Euclidean universe.
This is an adequate approximation at small redshifts, as illustrated in Figure~\ref{fig:distance-prior-comparison}; however, beyond a redshift of $\sim{}1$, the difference between a prior which is uniform in the comoving (source) frame volume and uniform in luminosity volume is large.
We therefore implement a range of cosmologically-informed prior classes.

The {\tt Cosmological} base class allows the user to specify a prior in either luminosity distance, comoving distance, or redshift using any cosmology supported in \astropy~\citep{astropy1, astropy2}.\footnote{By default, \bilby uses the \citet{ade2016} cosmology.}
Additionally, users can specify the prior in terms of redshift and then convert to an equivalent prior on luminosity distance if desired.
We implement two new source distance priors: a \texttt{UniformComovingVolume} prior, defined as
\begin{equation}
    \pi(z) \propto \frac{\mathrm{d}V_\mathrm{c}}{\mathrm{d}z},
\end{equation}
where $V_\mathrm{c}$ is the comoving volume, and a \linebreak{}\texttt{UniformSourceFrame} prior, defined as
\begin{equation}
    \pi(z) \propto \frac{1}{1 + z}\frac{\mathrm{d}V_{c}}{\mathrm{d}z}.
\end{equation}
The additional factor of $(1 + z)^{-1}$ accounts for time dilation.

Additional \texttt{Cosmological} prior classes of the form 
\begin{equation}
    \pi(z) \propto \frac{\mathrm{d}V_{c}}{\mathrm{d}z} f(z)
\end{equation}
can be defined by providing $f(z)$.

\subsubsection{Joint priors}

In cases where one requires more complex priors that depend on multiple parameters we implemented the \texttt{JointPrior} class in which the user can define a distribution that describes the prior on multiple parameters.
This is implemented in \bilby in the \texttt{MultivariateGaussian} prior that lets the user define multi-modal and multivariate Gaussian priors.
It is also used in the \texttt{HEALPixMap} prior in which a user can implement a prior on the sky position and optionally distance according to a given \texttt{HEALPix}~\citep{healpix, healpy} map.

\subsubsection{Custom jump proposals}\label{sec:custom_jump_proposals}
Users of \bilby can define custom jump proposals through its interface to the \cpnest and {\sc ptmcmc} samplers.
Jump proposals describe how the sampler finds new points in the parameter space.
\cpnest has a defined cycle of proposals that can be changed by the user.
These proposals can be useful when there are known degeneracies in the parameter space, e.g., phase $\phi$ and polarization angle $\psi$ under a shift by $\pi/2$ in either parameter~\citep{veitch15}.
Sampling in right ascension $\alpha$ and declination $\delta$ can also be improved using custom jump proposals; degeneracy typically leads to a ring-shaped two-dimensional posterior in these parameters for signals detected by two detectors \citep{Singer:2014qca,berry15}.
We provide proposals for the above two cases in the \bilby implementation of \cpnest, while additional proposals can be defined by the user to suit their needs.

\subsubsection{Boundary conditions}\label{sec:boundary_conditions}

For many parameters, such as the mass ratio $q$ and spin magnitudes $a_1$, $a_2$, posterior distributions have significant support close to the prior boundaries.
This is expected behaviour and a direct result of the choice of prior (e.g., the choice to fix $m_1 \geq m_2$ ensures $q \leq 1$).
In \bilby, \texttt{Prior} objects have boundaries that can be specified by the user as \texttt{None}, \texttt{reflective}, or \texttt{periodic}.
For samplers which support these settings, these options specify the behaviour of the sampler when it proposes a point that is outside of the prior volume.
For a \texttt{None} boundary, such a point is rejected.
Priors that have \texttt{reflective} boundaries are reflected about the boundary (a proposed mass ratio of $1 + \epsilon$ is reflected to $1 - \epsilon$) while \texttt{periodic} boundaries wrap around (a proposed phase of $\pi + \epsilon$ is wrapped to $\epsilon$).

The \dynesty sampler~\citep{dynesty} supports all available parameters boundary settings.
The {\sc pymultinest}\xspace sampler~\citep{multinest1, multinest2, multinest3, pymultinest} can implement \texttt{periodic} boundary conditions, but not \texttt{reflective}, which are treated as \texttt{None}.
All other samplers implemented in \bilby treat all prior boundaries as \texttt{None}.

While {\tt reflective} boundaries are implemented, their usage is not recommended due to concerns that they break detailed balance~\citep[e.g.,][]{Suwa2010}. When using the \dynesty sampler, we recommend using {\tt periodic} boundaries for relevant parameters (e.g., the right ascension and phase). These recommendations are mirrored in our choices of default priors, discussed in Section~\ref{subsec:default_priors}.

\subsubsection{Alternative sky and time parameterisations}
\label{sec:skyframe}

The most common way to describe the location of the source on the sky and its time of arrival is with the equatorial coordinates right ascension $\alpha$ and declination $\delta$, and the coalescence time at the center of the Earth $t_\mathrm{c}$.
However, particularly when the signal is only observed in two detectors, the likelihood is determined primarily by the time delay between the arrival of the signal at each detector.
The posterior distribution on these parameters often assumes a broken ring shape misaligned with the equatorial coordinate system~\citep{Singer:2014qca,berry15}, making sampling difficult.
A more natural parameterisation of the problem is given by sampling in the time of arrival at one of the detectors (ideally the one with the largest SNR), and rotating the sky coordinates such that the ring structure is uncorrelated in the sampling parameters.

We allow the user to specify a {\tt reference\_frame} and {\tt time\_reference}.
The argument {\tt reference\_frame} can either be an {\tt InterferometerList}, a string with the names of two known detectors, e.g., {\tt H1L1}, or {\tt sky} to sample in $\alpha$ and $\delta$.
Cases where sampling in $\alpha$ and $\delta$ is preferred include when the astrophysical location of the source is exactly known, e.g., by using the location of the host galaxy of a binary neutron star merger, the user can sample in $\alpha$ and $\delta$ by specifying {\tt reference\_frame=sky}.
In this parameterisation the zenith angle $\kappa$ is related to the time delay of the merger between the two detectors and is therefore well measured.
The azimuthal angle $\epsilon$ is only weakly constrained for a two-detector network.
The argument {\tt time\_reference} can be the name of any known interferometer, e.g., {\tt H1}, or {\tt geocent} to sample in the time at the geocenter.

The detector-based sampling frame is defined in terms of the zenith $\kappa$ and azimuthal $\epsilon$ angles relative to the vector connecting the vertices of the two interferometers specified $\delta r$.
We perform the transformation from ($\kappa$, $\epsilon$) to ($\delta$, $\alpha$) by constructing the rotation matrix $R$ which maps $\hat{z}$ to the unit vector $\delta \hat{r}$.
The rotation matrix $R$ can be described by three Euler angles ($\alpha$, $\beta$, $\gamma$)
\begin{align}\label{eq:euler}
    R &= R_{3}(\gamma)R_{2}(\beta)R_{3}(\alpha), 
\end{align}
\begin{align}
    \tan \alpha = \frac{-\delta r_{y} \delta r_{z}}{\delta r_{x}}, \quad \nonumber
    \cos \beta = \delta r_{y}, \quad \nonumber
    \tan \gamma = \frac{\delta r_{y}}{\delta r_{x}}. 
\end{align}
Here $\delta r_{\{x,y,z\}}$ are the Cartesian components of $\delta r$ and $R_{2, 3}$ are rotation matrices about the $y$- and $z$-axes respectively.

\subsubsection{Analytic likelihood marginalizations}
\label{sec:marginalisation}

The likelihood in Eq.~\eqref{eq:likelihood2} can be costly to evaluate for some signal models, and the size of the coalescence-time posterior relative to its much wider prior can make sampling the entire space difficult.
Therefore, we reduce the dimensionality of the CBC problem by analytically marginalizing over certain parameters, speeding up computation and improving the sampler convergence.
The parameters we commonly marginalise over are the coalescence time, binary orbital phase, and luminosity distance. 
In the frequency domain, a waveform of total duration $T$ can be written in terms of a reference time $t_0$, phase $\phi_{0}$, and luminosity distance $d_{0}$ as
\begin{align}
    \label{eq:analytic-marg}
    h_{k}(\boldsymbol{\lambda}, t, \phi, d_\mathrm{L}) &= h(\boldsymbol{\lambda}, t_{0}, \phi_{0} = 0, d_{0})\times\\ &\exp{\left[-2\pi ik \frac{(t-t_{0})}{T}\right]}\exp{(2i\phi)} \frac{d_{0}}{d_\mathrm{L}} \nonumber,
\end{align}
where $k$ indicates the frequency bin and $\boldsymbol{\lambda}$ represents the set of the other binary parameters, including the masses and spins, whose contributions to the waveform cannot be separated and thus cannot be analytically marginalized. 
The phase dependence can only be factored out for waveforms that include just the dominant $\ell = 2,\ m = |2|$ mode;
however, this factorization has been shown to be a reasonable approximation in some cases when precession is not measurable~\citep{abbott17_gw170104_detection}. 
The marginalized likelihood is obtained by integrating the likelihood in Eq.~\eqref{eq:likelihood2} over phase, distance, and coalescence time after using the factorisation in Eq.~\eqref{eq:analytic-marg}. 
The phase integral simplifies to a modified Bessel function of the first kind, evaluated at the magnitude of the complex inner product of the waveform and the data~\citep{veitch13, veitch15}. 

The distance marginalization is performed numerically, using a Riemann sum in matched filter and optimal signal-to-noise ratio (SNR) over the range $\rho \in [10^{-5}, 10^{10}]$, spaced uniformly in log-space~\citep{singer16a, singer16b, thrane18}.
To improve efficiency at run-time, we build a lookup table which is interpolated and then evaluated.
The lookup table is computed before the sampling phase begins, and can be cached and reloaded from previous analyses that used the same distance prior.

The marginalization over time involves performing a quadrature integral over an evenly spaced array of times separated by the sampling frequency. This marginalization is enabled by the fact that the inner product of the time-domain waveform and data can be rewritten as a fast Fourier transform~\citep{farr14}.
The sky location inferred when sampling in the {\tt sky} frame and using the time-marginalised likelihood is not generally correct and we do not recommend combining these two features.

If the signal is loud and the sampling frequency is too low, the reconstructed coalescence-time posterior appears discrete, since each of the generated parameters lies on one of the nodes of the array.
One solution to this is to increase the resolution of the array times by increasing the sampling frequency. 
However, this increases the computational cost of the marginalized likelihood evaluation.
Additionally, gravitational-wave detector data is natively sampled at $16~\mathrm{kHz}$~\citep{abbott_19_gwosc}, so increasing the time resolution beyond this level would require a different technique, e.g., zero-padding. 
In order to avoid increasing the sampling frequency, we maintain a continuous coalescence-time posterior by introducing a \texttt{time\_jitter} $\delta t$.
This parameter varies the position of the time array over which the numerical integral is performed.
We apply a uniform prior with bounds such that 
\begin{equation}
\frac{-T}{2} \leq \delta t < \frac{T}{2},  
\end{equation}
thus reducing the prior space to be searched.

When using the analytically-marginalized likelihood, the sampler does not produce posterior samples for the marginalized parameters.
However, \bilby is able to generate samples for these parameters in post-processing.
Using \bilby, we recalculate the likelihood by recomputing the optimal matched filter signal-to-noise ratio and the inner product of the waveform and data.
We then obtain a posterior array for the marginalized parameter in question, evaluated at discrete points in the parameter's prior space.
We generate posterior samples by sampling from this interpolated posterior array.
By drawing a single sample for each of the marginalized parameters for each posterior sample we maintain the degeneracies between, e.g., distance and binary orbital inclination.
For detailed derivations of the analytically marginalized likelihood and the posterior sample reconstruction process, see \citet{thrane18}.

\subsubsection{Reduced-order quadrature}\label{sec:roq}

In order to reduce the number of frequencies at which the likelihood in Eq.~\eqref{eq:likelihood2} must be evaluated, we implement the reduced-order quadrature (ROQ) likelihood~\citep{smith16}.
This method works by identifying a reduced basis that can describe the signal model well over a certain range of the parameter space.
Application of reduced-order methods have been crucial for expediting inference for long duration signals, such as the binary neutron star merger GW170817 \citep{abbott2019_GWTC1}.
Evaluating the ROQ likelihood requires access to the appropriate basis.
A set of bases for the most commonly used waveform, \imrphenomp, are publicly available online.\footnote{\label{footnote:ROQ}\href{https://git.ligo.org/lscsoft/ROQ_data}{git.ligo.org/lscsoft/ROQ\_data}}

The {\tt ROQGravitationalWaveTransient} likelihood class in \bilby is able to analyse arbitrary reduced-order bases.
This likelihood can also be marginalized over phase and/or distance. A time-marginalized ROQ likelihood has not yet been implemented.

\subsubsection{Calibration}\label{sec:calibration}

The imperfect nature of the detector calibration introduces a systematic error in the measured astrophysical strain~\citep{abbott16_gw150914_astro}.
Following~\citet{farr14b}, we split this error into frequency-dependent amplitude and phase offsets, $\delta A(f)$ and $\delta \phi(f)$ respectively.
The observed strain can then be related to the true strain as
\begin{align}
    h_{\mathrm{obs}}(f) = h(f)\left[1+\delta A(f)\right]\exp\left[i\delta \phi(f)\right].
\end{align}
Since the calibration error is small, we perform a small angle expansion in the phase correction,
\begin{align}
    \exp\left[{i\delta \phi(f)}\right] = \frac{2+i \delta \phi(f)}{2 - i\delta \phi(f)} + \mathcal{O}\left(\delta \phi^{3}\right).
\end{align}
Substituting this, we obtain
\begin{align}
    h_{\mathrm{obs}}(f) = h(f)\left[1+\delta A(f)\right]\frac{2+i \delta \phi(f)}{2 - i\delta \phi(f)}.
\end{align}

The amplitude and phase uncertainty are modeled as cubic splines in \bilby,
\begin{align}
    \label{eq:calibration_factor_A}
    \delta A(f) &= s(f; \{f_{j}, \delta A_{j}\}), \\
    \label{eq:calibration_factor_phi}
    \delta \phi(f) &= s(f; \{f_{j}, \delta \phi_{j}\}),
\end{align}
where the spline nodes $f_{j}$ are fixed and distributed uniformly in log-space between the minimum and maximum frequencies included in the likelihood, and the values of the splines at the nodes, $\delta A_{j}$ and $\delta \phi_{j}$, are sampled parameters~\citep{vitale12}.

The priors on the spline values are taken to be normal distributions, with means and widths that can either be constant or loaded from a frequency-dependent calibration envelope file~\citep{cahillane17, viets18}.
The calibration factor defined in Eq.~\eqref{eq:calibration_factor_A} and Eq.~\eqref{eq:calibration_factor_phi} are applied to the waveform calculated for each prior sample before the likelihood is computed.
Figure~\ref{fig:calibration-posterior} shows an example plot of the calibration spline posterior for both the amplitude and phase uncertainties.

\subsubsection{Gravitational-wave transient-specific plots}\label{sec:gwt_specific_plots}

\begin{figure}
    \centering
    \includegraphics[width=\linewidth]{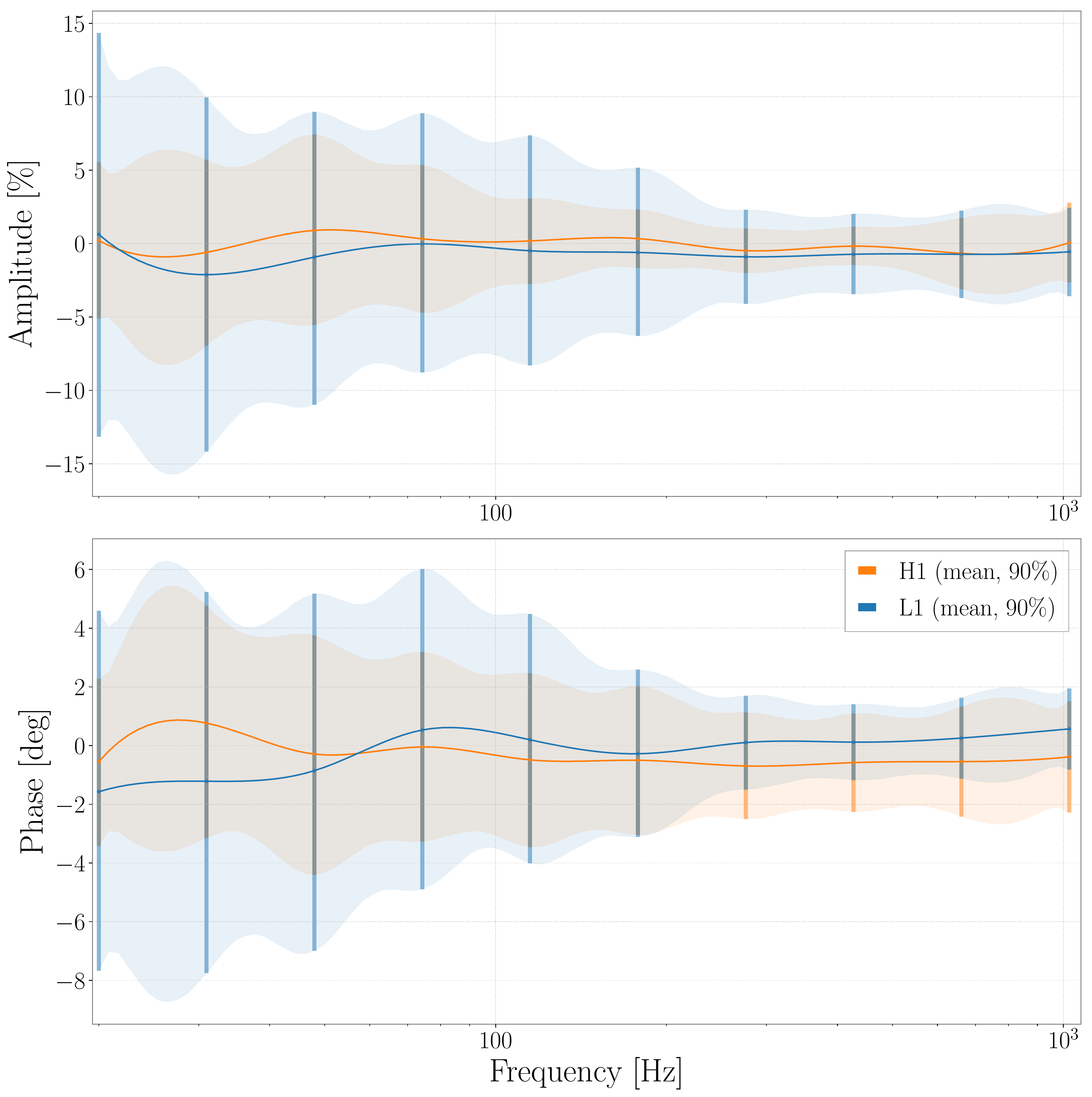}
    \caption{Calibration posteriors for the amplitude (top) and the phase uncertainty (bottom) for both LIGO Hanford (orange) and Livingston (blue) detectors for GW150914. The solid curves shows the mean, while the shaded region represents the $90\%$ confidence intervals. The vertical lines show the locations of the spline points.
    }
    \label{fig:calibration-posterior}
\end{figure}

\begin{figure}
    \centering
    \includegraphics[width=\linewidth]{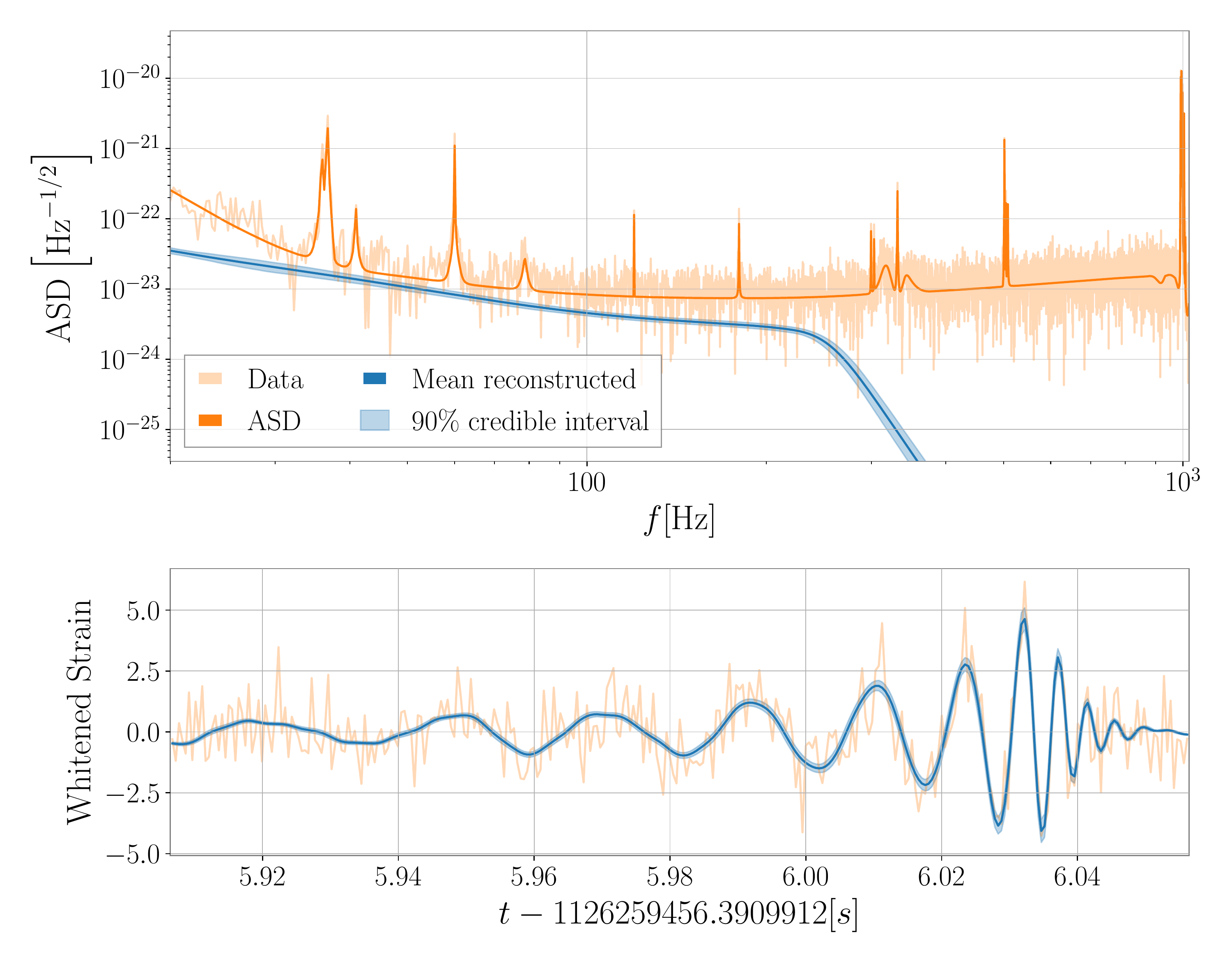}
    \caption{
        Reconstructed waveform for GW150914 for LIGO Hanford.
        The top panel shows the amplitude spectral density of the signal (blue), data (light orange), and estimated noise amplitude spectral density (dark orange).
        The bottom panel shows the time domain data (light orange) and waveform estimate (blue).
        The dark blue curves show the mean recovered waveform and the light blue shaded region the 90\% confidence interval.
    }
    \label{fig:waveform-posterior}
\end{figure}

\bilby users can produce sets of posterior plots specific to gravitational-wave transient analysis.
We use the \skymap~\citep{singer16a, singer16b} package to produce sky maps in both the {\tt fits} format commonly used for electromagnetic observation and standard image formats.
We are also able to produce plots showing our inferred posterior on the detector calibration and waveform models, in addition to the parameters describing these models.
We present examples of these plots for GW150914 in Figures~\ref{fig:calibration-posterior} and ~\ref{fig:waveform-posterior} respectively.
In such plots, we show the mean reconstructed model and symmetric $90\%$ credible intervals.

\subsection{Validation of \bilby}\label{subsec:validation}

A common consistency test of the performance of sampling algorithms is to check that the correct proportion of true parameter values are found within a given probability interval for simulated systems ~\citep{cook06,talts18}---i.e.
that $10\%$ of events are found within the $0.1$ probability credible interval, $50\%$ are found within the $0.5$ probability credible interval, etc.
We generate a set of CBC signals with true parameter values drawn from our prior probability distributions and inject these into simulated noise. 
Parameter estimation is then performed on each signal to determine the credible level at which the true value of each parameter is found. 
This test is traditionally used in validating gravitational-wave inference codes~\citep{sidery14,veitch15,berry15,pankow15,singer16a,biwer18,delpozzo18}.

To test \bilby's parameter estimation, we simulate $100$ synthetic CBC signals for a two-detector Hanford--Livingston network and add the signals to Gaussian noise colored to the anticipated Advanced LIGO design sensitivity~\citep{abbott_19_observing_scenarios}.
The parameters of the simulated events are drawn from the default $4\,\mathrm{s}$ prior set, detailed in Section~\ref{subsec:default_priors}.

\begin{figure}
    \centering
    \includegraphics[width=\linewidth]{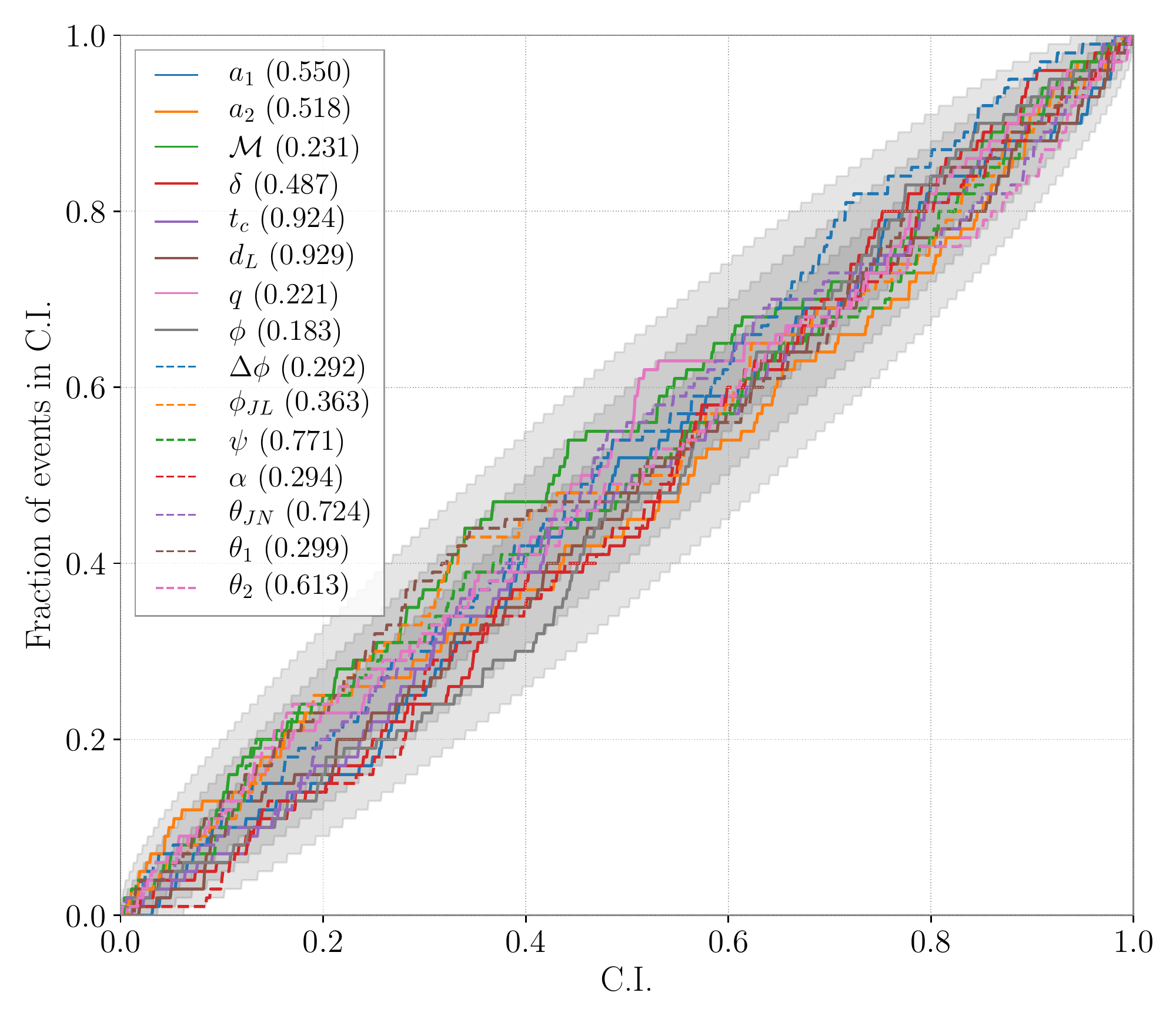}
    \caption{Results of 100 injections drawn from the four-second prior defined in Section \ref{subsec:default_priors}. The gray regions cover the cumulative $1$-, $2$- and $3$-$\sigma$ confidence intervals in order of decreasing opacity.
    Each colored line tracks the cumulative fraction of events within this confidence interval for a different parameter. The combined $p$-value for all parameters, over all tests, is \pptestpvalue, consistent with the individual $p$-values being drawn from a uniform distribution. Individual parameter $p$-values are displayed in parentheses in the plot legend. The marginalised parameters---geocenter time $t_\mathrm{c}$, luminosity distance $d_\mathrm{L}$ and phase $\phi$--are reconstructed in post-processing. Other parameters provided in the plot legend are defined in Appendix \ref{sec:definitions}.}
    \label{fig:pp-plot}
\end{figure}

Parameter estimation is performed using the \dynesty sampler with the distance, time, and phase-marginalized likelihood. Analysis of the performance of other samplers is left to future work.
Results of the test are shown in Figure~\ref{fig:pp-plot}, where the fraction of events for which the true parameter is found at a particular confidence level is plotted against that particular confidence interval.\footnote{These plots are referred to as P--P plots, where P could stand for probability, percent or proportion. Instructions for generating P--P plots are provided in the \bilby documentation at \href{https://git.ligo.org/lscsoft/bilby_pipe/wikis/pp/howto}{git.ligo.org/lscsoft/bilby\_pipe/wikis/pp/howto}.}
We also show the individual parameter $p$-values representing the probability that the fraction of events in a particular confidence interval is drawn from a uniform distribution, as expected for a Gaussian likelihood, and the combined $p$-value quantifying the probability that the individual $p$-values are drawn from a uniform distribution.
The combined $p$-value obtained with the latest version of \bilby is \pptestpvalue and the minimum is 0.183 for $\phi$, which is entirely consistent with chance for the set of 15 parameters, 
indicating that the posterior probability distributions produced by \bilby are well-calibrated.
The grey regions show the 1, 2, and $3\sigma$ confidence intervals so we expect the lines to deviate from this region approximately 0.3\% of the time, which is consistent with what we see.

In addition to the procedure described above, we verify the suitability of the sampler settings for the problem of sampling the CBC parameter space using a series of review tests. 
These are described in detail in Appendix~\ref{appendix:review-tests}.
The settings used for each of the tests described here are provided in Appendix~\ref{appendix:run-settings}.
In addition to these review tests, \bilby has an extensive set of unit tests, which scrutinize the behaviour of the software in high detail every time a change is made to the code;
these unit tests can be found within the \bilby package.\footnote{\href{https://git.ligo.org/lscsoft/bilby/tree/master/test}{git.ligo.org/lscsoft/bilby/tree/master/test}}

\subsection{Automation of \bilby for gravitational-wave inference}\label{sec:automation}
\label{sec:bilbypipe}

With the improvement in sensitivity and expansion of the gravitational-wave observatory network comes an increasing rate of detections. Streamlining the deployment of \bilby analysis is therefore vital.
We introduce \bilbypipe, a Python package providing a set of command-line tools designed to allow performance of parameter estimation on gravitational-wave data with all settings either passed in a configuration file or via the command line.\footnote{\label{footnote:bilby_pipe}The source-code is available on the git repository \href{https://git.ligo.org/lscsoft/bilby_pipe}{git.ligo.org/lscsoft/bilby\_pipe}.
Specifics about the installation, functionality and user examples are also provided \href{https://lscsoft.docs.ligo.org/bilby_pipe}{lscsoft.docs.ligo.org/bilby\_pipe}.} 
This tool was used to perform the analyses of the GWTC-1 catalogue events presented in Section~\ref{sec:event-catalog}, and is integral to the automatic online parameter estimation that is triggered by potential gravitational-wave events.

The \bilbypipe workflow consists of two key stages: data generation, and data analysis.
These steps are outlined in Section~\ref{sec:daq_analysis}.
The pipelines provided by \bilbypipe can be utilized to distribute analysis of a single event over multiple CPUs using \pbilby~\citep{Smith:2019ucc}, which is described in Section~\ref{sec:parallel_bilby}.
The workflow for the automated running of \bilby on gravitational-wave candidates is detailed in Section~\ref{sec:online_bilby}.

\subsubsection{Data generation and analysis}
\label{sec:daq_analysis}

Gravitational-wave detectors record and store time-domain strain data and information about the behavior internal to the detectors, as well as data from a suite of environmental sensors. To obtain gravitational-wave strain data, we recommend using the \gwpy library~\citep{gwpy}. \gwpy can retrieve both public data from the Gravitational Wave Open Science Center~\citep{abbott_19_gwosc}, and proprietary data using the Network Data Server protocol (NDS2) to acquire data from LIGO servers. 
Given a GPS trigger time and a required data duration, \bilbypipe uses \gwpy to extract an analysis segment of strain data around the trigger, as well as a segment of strain data used to estimate the noise PSD.
The default duration for the analysis segment is $T=\unit[4]{s}$, which is considered adequate for sources with detector-frame chirp masses $\mathcal{M} \gtrsim \unit[15]{\text{M}_\odot}$.
Sources with lower $\mathcal{M}$ have longer signals, so longer analysis segments should be used.
A portion of data following the trigger time is required to encompass the remaining merger and post-coalescence ringdown signal; this is $\unit[2]{s}$ by default.

A \bilbypipe user can provide pre-generated PSDs, and a range of design-sensitivity noise spectra for current and future detectors are available as part of the \bilby package.
For the analyses we present in Section~\ref{sec:event-catalog}, we use event-specific PSDs produced using \bayeswave~\citep{Cornish14}. 
When a PSD is not provided, \bilbypipe uses the median-average power spectrum method described by \citet{findchirp}, and implemented in \gwpy, to calculate the PSD; this method has the advantage of downweighting outliers in the off-source data~\citep{findchirp, veitch15}.
In order to avoid including any signal in the PSD calculation, \bilbypipe uses a stretch of data preceding the analysis segment.
Following \citet{veitch15} and \citet{chatziioannou2019noise}, we use data stretches of length $\min(\unit[32T, 1024]{s})$ by default, although both of these values can be altered by the user.
The upper limit of $\unit[1024]{s}$ is required because the PSD of gravitational-wave detectors is non-stationary over long time-periods~\citep{chatziioannou2019noise}. 
To further mitigate this issue, the data is divided into segments of length $T$, with each segment overlapping $50\%$ of the previous segment; this allows a shorter total stretch of data to be used to calculate the PSD.
Following \citet{findchirp}, segments are Tukey windowed with a $\unit[0.4]{s}$ roll-off to suppress spectral leakage \citep{ligo2019guide}, before computing their one-sided power spectra.

The priors for the analysis can be specified by the user, either by providing a path to a file containing the priors in \bilby syntax, or by giving the name of one of the default \bilbypipe priors described in Section~\ref{subsec:default_priors}.
By default, the \bilby \texttt{GravitationalWaveTransient} likelihood is used with the waveform template generated by \lalsimulation~\citep{LALSuite}.
However, users can specify their own source models and modified likelihoods in the configuration file.
After saving the necessary data, \bilbypipe launches parameter estimation on the analysis segment in accordance with the procedure outlined in Section~\ref{subsec:bayesian_inference}.

\subsubsection{Parallel \bilby}
 \label{sec:parallel_bilby}

Parallel \bilby~\citep{Smith:2019ucc} is a parallel implementation of \bilby which uses Message Passing Interface \citep[MPI;][]{farah19} to distribute the \dynesty nested sampling package over a pool of CPUs.
Nested sampling requires drawing successive samples satisfying a likelihood constraint from the prior.
Faithfully drawing samples from this constrained prior requires many likelihood evaluations.
We use a CPU pool to draw prior samples in parallel at each iteration of the algorithm to reduce the wall-time needed to complete an analysis.

Qualitatively, \pbilby works by using a pool of $n_{\mathrm{cores}}$ CPUs to draw $n_{\mathrm{cores}}-1$ samples from the prior in parallel at each iteration of the sampling algorithm. The $n_{\mathrm{cores}}-1$ proposed samples are ranked by likelihood and the lowest-likelihood live point is replaced. The prior volume is then updated on all $n_{\mathrm{cores}}$ processes and the sampling step is repeated until the algorithm is converged. The speedup $S$ of the parallel implementation is a function of the number of live points $n_{\mathrm{live}}$ and the number of parallel processes \citep{Smith:2019ucc}:
\begin{equation}
    S = n_{\mathrm{live}}\ln\left( 1+\frac{n_{\mathrm{cores}}}{n_{\mathrm{live}}}\right).
\end{equation}

Currently, \pbilby only supports the \dynesty and \ptemcee sampling packages.
All of the functionality of \bilby, as described in Section~\ref{sec:changes}, is supported by \pbilby.

\pbilby is highly scalable, and is thus well suited to accelerating applications in which the gravitational-wave signal or noise models are computationally expensive to evaluate, e.g., time-domain signal models such as spin-precessing effective-one-body models with higher-order modes \citep{PhysRevD.95.044028, Ossokine20}, numerical-relativity surrogate models \citep{PhysRevD.96.024058} and models including tidal effects \citep{Nagar18, Lackey19}.
Other well-suited applications include those where sampling convergence can be slow due to high dimensionality of the parameter space, e.g., when calibration~\citep{farr14b} or beyond-general-relativity parameters are used~\citep{abbott16_gw150914_testingGR,abbott19_TGR}, or when a large number of live points is required to effectively estimate the evidence.

In order to facilitate efficient inter-CPU communication with MPI, \pbilby is a stand-alone package,
though it still uses the underlying \bilby modules.

In addition to the hugely parallel \pbilby, many of the implemented sampling packages support parallelization through a user specified pool of processes.
For these samplers \bilby natively supports local parallelization using the \python~{\sc multiprocessing} package.
When available, the number of parallel computational threads to use is specified using the {\tt nthreads} argument.

\subsubsection{Online \bilby} \label{sec:online_bilby}
The gravitational-wave candidate event database GraceDB\footnote{\href{https://gracedb.ligo.org}{gracedb.ligo.org}} provides a centralized location for collecting and distributing gravitational-wave triggers uploaded in real time from search pipelines.
Once uploaded, each trigger is assigned a unique identifier, and LIGO--Virgo users are notified via an \lvalert (LIGO--Virgo Alert Network).
\gwcelery~\citep{GWCelery}, a Python-based package designed to facilitate interactions with GraceDB, responds to an alert by first creating a Superevent, which groups triggers from multiple search pipelines and then chooses a preferred event based on the signal-to-noise ratio of the triggers.
If the preferred candidate has a false-alarm-rate (FAR) below a given threshold, \gwcelery automatically launches multiple parameter estimation jobs.
For the case of \bilby, this involves making a call to the \bilbypipegdb executable.

\begin{figure}
    \centering
    \includegraphics[width=0.45\textwidth]{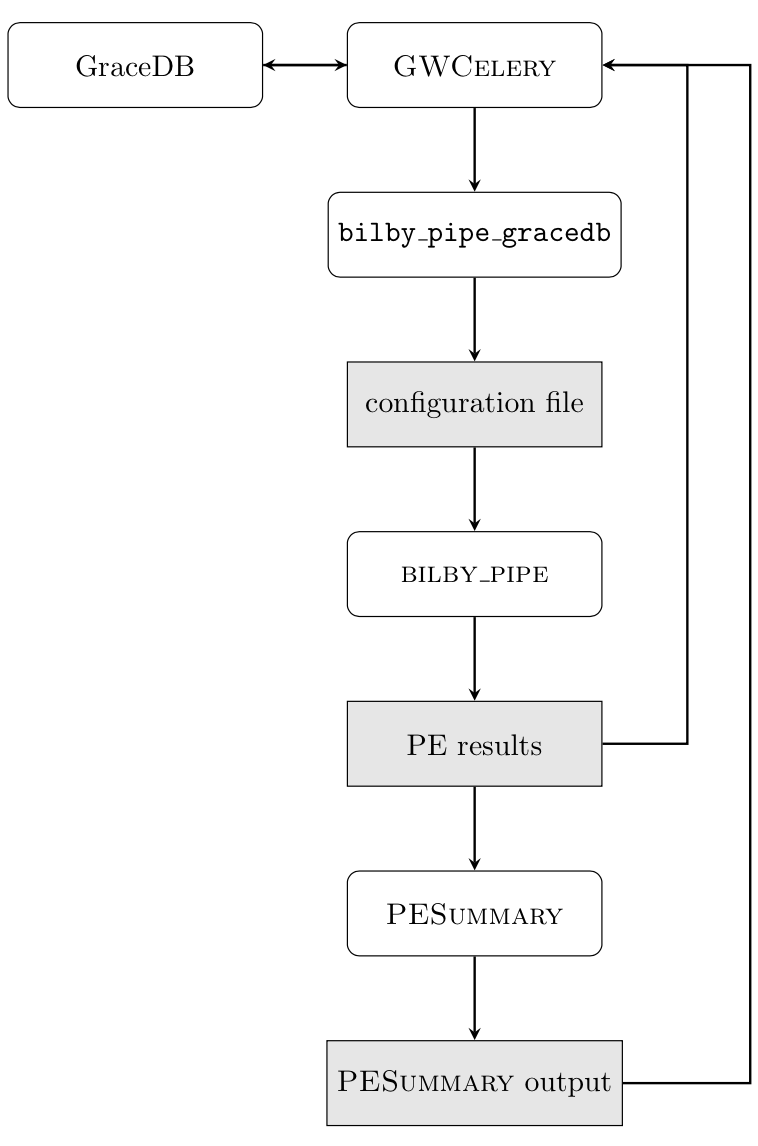}
     \caption{Workflow for online \bilby parameter estimation.}
     \label{fig:flowchart}
\end{figure}

The \bilbypipegdb executable takes the GraceDB event ID as input and generates a configuration file based on the trigger time of the candidate.
A prior file is selected from the set of default priors using the chirp mass of the gravitational-wave signal template that triggered the \lvalert.
Further details about the default priors can be found in Section~\ref{subsec:default_priors}.
These files are then passed to the \texttt{bilby\_pipe} executable, which runs parameter estimation on the event. \pesummary~\citep{pesummary}, a Python-based package designed to post-process inference package output in a number of formats, then generates updated source classification probabilities and webpages displaying diagnostic plots.
Once this step is complete, \gwcelery uploads the posterior samples, post-processing pages and updated source classification probabilities to GraceDB.
Figure~\ref{fig:flowchart} illustrates the process of automated parameter estimation from the trigger of a gravitational-wave event to the upload of \bilby parameter estimation results to GraceDB.

\subsubsection{Run times}

The overall run time of a \bilby parameter estimation job depends on the specific input data and can vary considerably based on the chosen sampler settings and signal-to-noise ratio.
The overall wall time can be reduced by allowing for marginalization over certain parameters, as described in Section~\ref{sec:marginalisation}, or by using the parallelization methods described in Section~\ref{sec:parallel_bilby}.
For a GW150914-like binary black hole merger, the expected run time for a time, distance  and phase marginalized \bilby analysis using the default waveform model \imrphenomp~\citep{IMRPhenomP} is $\mathcal{O}(10)$ hours. The waveform models needed to analyse binary neutron star merger events are much longer than those required for binary black holes, and therefore are more computationally expensive. Hence, for a GW170817-like binary neutron star merger event, we use \pbilby to distribute the analysis over a pool of CPUs, as described in Section~\ref{sec:parallel_bilby}; the expected run time in this case is $\mathcal{O}(10)$ hours.

\section{Gravitational-wave Transient catalogue}
\label{sec:event-catalog}
This section contains our run settings for performing parameter estimation on GWTC-1 events using \bilby, in addition to the results we obtain from this analysis.
We describe our default priors and sampler settings in Sections \ref{subsec:default_priors}--\ref{subsec:data}. 
Further details about these settings are given in Appendix \ref{appendix:run-settings}. 
We provide our results in Section \ref{subsec:results}, where we assess their statistical similarity to those published in GWTC-1~\citep{abbott2019_GWTC1}.\footnote{The \lalinference posterior samples that we show in this section are taken from the Parameter Estimation Sample Release for GWTC-1~\citep{GWTC1-samples}. The posterior samples from \lalinference are obtained using a mixture of the nested sampling algorithm of \texttt{LALInferenceNest} and the Markov-chain Monte Carlo algorithm of \texttt{LALInferenceMCMC}~\citep{veitch15}.}
All \bilbypipe configuration files, posterior samples and \bilby results files are made available online~\citep{Bilby-GWTC-1-Analysis-and-Verification}.

\subsection{Default priors}\label{subsec:default_priors}

The default prior distributions contained in \bilbypipe are predominantly tailored to specific signal durations, with the exception of a high-mass prior tailored to particularly heavy sources with detector-frame chirp mass $\mathcal{M}$ up to $175 M_\odot$.
For each event in GWTC-1, we choose the default prior that best covers the prior volume studied using \lalinference for the original samples release.
This means that two events (GW150914 and GW151012) are analysed using priors suited to signals of duration $T = \unit[4]{s}$, even though we match the data duration to that used in the original \lalinference analysis ($T = \unit[8]{s}$).
The prior on $\mathcal{M}$ is uniform in the detector frame, while the prior on $d_\mathrm{L}$ is uniform in comoving volume and source frame time, as implemented in the \texttt{UniformSourceFrame} prior class described in Section~\ref{sec:cosmological_priors}. 
The $\mathcal{M}$, $d_\mathrm{L}$ and spin magnitude prior limits vary between prior sets, while the other source parameters are assigned priors that are consistent between sets.
The shapes and limits of all priors are defined in Appendix~\ref{appendix:priors}.
The prior files can be found in the \bilbypipe git repository.\textsuperscript{\ref{footnote:bilby_pipe}}

\begin{table*}
    \centering
    \caption{
    Summary statistics for each event in GWTC-1, as recovered by \bilby. 
    We quote median values along with the symmetric $90\%$ credible interval range around the median. 
    For mass ratio $q$, we quote the $90\%$ lower limit ($10\%$ quantile), with all events being consistent with equal mass ($q=1$).
    We use a fixed-sky prior on source location for GW170817, the binary neutron star merger, fixing the source at the right ascension and declination of its electromagnetic counterpart~\citep{abbott17_gw170817_multimessenger}. 
    The $90\%$ credible areas for sky location are computed using $3000$ samples from each posterior.     
    The final column lists the maximum Jensen--Shannon (JS) divergence statistic (a measure of the similarity between two distributions) between the \bilby GTWC1 samples, and the \lalinference GWTC-1 posterior samples across the model parameters. We consider JS divergence values greater than \unit[0.002]{nat} to be statistically significant. \label{tab:events}}
    \begin{tabular}{l c c c c c c c D{=}{\,=\,}{-1} }
    \hline
     Event & Prior & $\mathcal{M}/M_\odot$ & $\mathcal{M}^\mathrm{source}/M_\odot$ & $q$ lower limit  & $d_\mathrm{L}/\mathrm{Mpc}$ & $\chi_\mathrm{eff}$ & $\Delta\Omega/ \mathrm{deg}^2$ & \multicolumn{1}{c}{Max-JS$/$nat} \\
    \hline
 GW150914 & \unit[4]{s}  & $31^{+1}_{-1}$          & $28^{+2}_{-1}$                          & $0.72$      & $420^{+160}_{-165}$     & $-0.0^{+0.1}_{-0.1}$ & $169$ & \mathrm{JS}_{\theta_{JN}}=0.0019 \\
 GW151012 & \unit[4]{s}  & $18^{+2}_{-1}$          & $15^{+2}_{-1}$                          & $0.41$      & $1015^{+498}_{-472}$    & $0.0^{+0.2}_{-0.2}$    & $1457$ &  \mathrm{JS}_{\mathcal{M}}=0.0014 \\
 GW151226 & \unit[8]{s}  & $9.7^{+0.1}_{-0.1}$     & $8.9^{+0.3}_{-0.3}$                     & $0.38$      & $428^{+196}_{-189}$ & $0.2^{+0.1}_{-0.1}$   & $1022$  &  \mathrm{JS}_{q}=0.0017 \\
 GW170104 & \unit[4]{s}  & $26^{+2}_{-2}$          & $22^{+2}_{-2}$                          & $0.48$      & $935^{+441}_{-411}$    & $-0.0^{+0.2}_{-0.2}$   & $900$ &  \mathrm{JS}_{\mathcal{M}}=0.0007 \\
 GW170608 & \unit[16]{s} & $8.5^{+0.0}_{-0.0}$     & $7.9^{+0.2}_{-0.2}$                     & $0.49$      & $317^{+122}_{-115}$ & $0.0^{+0.1}_{-0.0}$    & $1462$    &  \mathrm{JS}_{q}=0.0011 \\
 GW170729 & High-mass    & $51^{+8}_{-9}$          & $35^{+6}_{-5}$                          & $0.43$      & $2548^{+1369}_{-1235}$    & $0.3^{+0.2}_{-0.3}$    & $1050$  & \mathrm{JS}_{\alpha}=0.0026 \\
 GW170809 & \unit[4]{s}  & $30^{+2}_{-2}$          & $25^{+2}_{-2}$                          & $0.51$      & $995^{+311}_{-411}$       & $0.1^{+0.2}_{-0.2}$    & $300$ &  \mathrm{JS}_{\mathcal{M}}=0.0010 \\
 GW170814 & \unit[4]{s}  & $27^{+1}_{-1}$          & $24^{+1}_{-1}$                          & $0.69$      & $572^{+154}_{-212}$       & $0.1^{+0.1}_{-0.1}$    & $77$ &  \mathrm{JS}_{\theta_1}=0.0009 \\
 GW170817 & Custom       & $1.1975^{+0.0001}_{-0.0001}$ & $1.187^{+0.004}_{-0.002}$            & $0.74$ & $40^{+8}_{-16}$        & $0.00^{+0.02}_{-0.01}$ & N/A  &  \mathrm{JS}_{\tilde{\Lambda}}=0.0019 \\
 GW170818 & \unit[4]{s}  & $32^{+2}_{-2}$          & $27^{+2}_{-2}$                          & $0.58$      & $1017^{+407}_{-348}$      & $-0.1^{+0.2}_{-0.2}$   & $29$      &  \mathrm{JS}_{\alpha}=0.0064 \\
 GW170823 & High-mass  & $39^{+5}_{-4}$          & $29^{+4}_{-3}$                          & $0.54$      & $1771^{+857}_{-831}$      & $0.0^{+0.2}_{-0.2}$    & $1570$ &  \mathrm{JS}_{\theta_{N}}=0.0009 \\
    \hline
    \end{tabular}
\end{table*}

\subsection{Likelihood}\label{subsec:likelihood}

Our likelihood is marginalized over reference phase and source luminosity distance, as described in Section~\ref{sec:marginalisation}.
For binary black hole merger analyses, we use the waveform model \imrphenomp~\citep{IMRPhenomP,hannam2014,khan16,Bohe:PhenomPv2} as our signal template.
For the binary neutron star GW170817, we use the \imrphenompvtwonrtidal waveform model with tidal effects \citep{Dietrich19}.

\subsection{Sampling}\label{subsubsec:sampler_settings}

We use \dynesty~\citep{dynesty} as our sampler; see Appedix \ref{appendix:sampler} for the detailed sampler settings.
We use the static version of  \dynesty, as is default for \bilbypipe.
For each event, we run five analyses in parallel, merging the resultant posterior samples in post-processing. 
When combining results, care must be taken to weight each set of samples appropriately by its relative evidence.
The weight applied to the $i$th component of $N$ sets of posterior samples is given by
\begin{equation}
    w_i = \frac{\mathcal{Z}_i}{\sum_{j=i}^{N} \mathcal{Z}_j},
\end{equation}
where $\mathcal{Z}_i$ is the evidence of the $i$th set of samples.

\subsection{Data used}\label{subsec:data}

We use detector noise PSDs and calibration envelopes data from the data releases accompanying GWTC-1~\citep{abbott2019_GWTC1, GWTC1-PSDS, GWTC1-CAL}.
The data for each event are obtained through \bilbypipe using methods from the \gwpy~\citep{gwpy} package as outlined in Section~\ref{sec:daq_analysis}.
Appendix~\ref{appendix:run-settings} contains details of the trigger times and data segment durations specified for each event, which we choose to match those used in the original \lalinference analysis.

\subsection{Analysis of binary neutron star merger GW170817}
\label{subsec:neutron_star}

The first observation of a binary neutron star coalescence, GW170817, by LIGO--Virgo \citep{abbott17_gw170817_detection} presented a new challenge for gravitational-wave transient inference. The longer signal durations increase the typical computing requirements, and for systems containing a neutron star, tidal effects become important in the waveform models.
The original discovery \citep{abbott17_gw170817_detection} and subsequent follow-up studies \citep{abbott18_GW170817_properties} analysed the data with a variety of waveform models and under differing assumptions.

We employ \pbilby for this analysis, with \bilbypipe default sampler settings. We use priors chosen to match those of the LVC analysis \citep{abbott18_GW170817_properties}, but sample in chirp mass and mass ratio rather than component masses. Our likelihood is computed using the tidal waveform model \imrphenompvtwonrtidal~\citep{Dietrich19}. This \pbilby analysis took approximately \unit[11]{hours} on 560 cores.

\subsection{Results}\label{subsec:results}

We make posterior samples and \bilbypipe configuration settings files available online~\citep{Bilby-GWTC-1-Analysis-and-Verification, Bilby_samples}.
To directly compare \bilby posterior samples to those obtained using \lalinference, we reweight the \lalinference posterior distributions by \bilbypipe default priors. Appendix~\ref{appendix:reweighting} contains the details of this reweighting procedure.

To quantitatively assess the similarity between \bilby and \lalinference posterior samples, we measure their Jensen--Shannon~\citep[JS;][]{61115} divergence.
This is a symmetrized extension of the Kullback--Leibler divergence~\citep{kullback1951} that is used to quantify the information gain going between two distributions.
The JS divergence is defined to be between \unit[0]{nat} and \unit[1]{nat}, where \unit[0]{nat} represents no additional information going from one distribution to the other (the two distributions are identical) and $\unit[ln(2)]{nat} = \unit[0.69]{nat}$ represents maximal divergence.\footnote{In v1 of this paper, we stated JS divergence values with incorrect units (bits). These units have now been corrected.} For different sets of samples drawn from the same Gaussian distribution, we find JS divergence values of $\lesssim\unit[0.0010]{nat}$ while the number of samples $N\gtrsim2000$, and JS divergence values of $\lesssim\unit[0.0004]{nat}$ when $N \gtrsim5000$. To compare \bilby and \lalinference results, we use $N = \text{min}(N_{\rm LI}, 10000)$, where $N_{\rm LI}$ is the number of samples left in the \lalinference posterior after the reweighting procedure.

Our goal is to use the JS divergence as a quantitative indicator that the \bilby GWTC-1 samples are in agreement with those produced by \lalinference. 
To investigate the typical distributions of JS divergence values due to sampling error, we calculated JS values for posteriors from two distinct \lalinference runs on GW150914 with identical configurations. Bootstrapping was used to generate 100 posterior realizations from each run, which were used to obtain a distribution of JS divergences for each of the binary parameters included in the public \lalinference GWTC-1 posterior sample release. Across different parameters, we typically found mean values of \unit[0.0007]{nat}, with a maximum of \unit[0.0015]{nat}.
As such, we determined the following naive criteria for evaluating the JS divergence values when comparing the \bilby and \lalinference GWTC-1 posteriors.
For a JS divergence value less than \unit[0.0015]{nat}, we conclude the samples are, to within statistical uncertainties, drawn from the same distribution, and values larger than \unit[0.0015]{nat} require manual inspection.

\begin{figure}
    \centering
    \includegraphics[width=\linewidth]{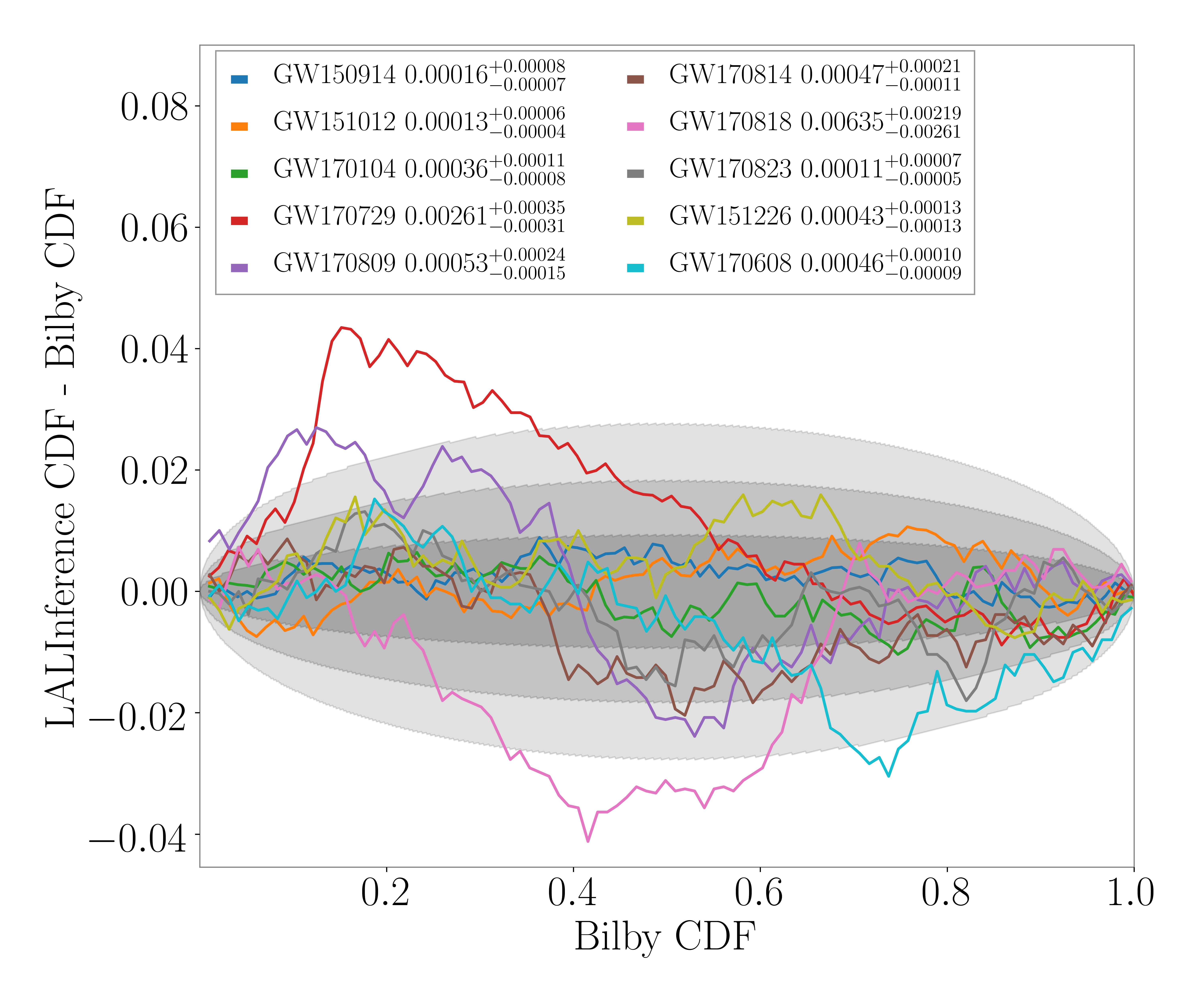}
    \caption{Difference between the right-ascension ($\alpha$) samples recovered by \bilby and \lalinference for all BBH events. This is the worst recovered parameter according to the JS-divergence. Labels show the mean JS-divergence between $\alpha$ samples, evaluated by random re-sampling over 100 iterations.}
    \label{fig:ra_pp_plot}
\end{figure}

\begin{figure}
    \centering
    \includegraphics[width=\linewidth]{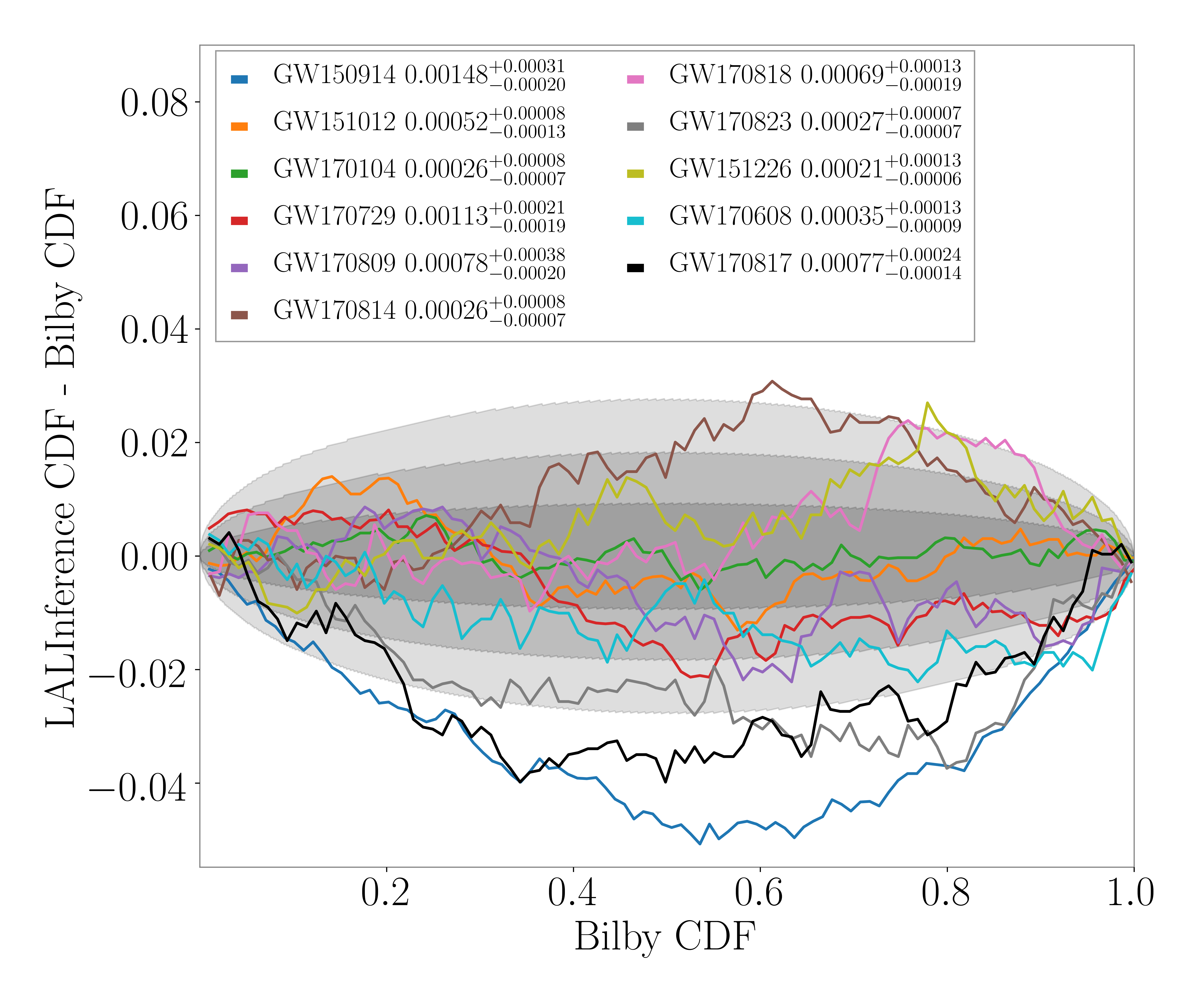}
    \caption{Difference between the luminosity distance ($d_\mathrm{L}$) samples recovered by \bilby and \lalinference for all events. Labels show the mean JS-divergence between $d_\mathrm{L}$ samples, evaluated by random re-sampling over 100 iterations.}
    \label{fig:dl_pp_plot}
\end{figure}

In Table~\ref{tab:events}, we list the maximum JS divergence for the model parameters for each event.
Of these, six pass our naive criterion described above.
For the remaining events, we manually inspect the posterior distributions to look for discrepancies. 
The parameter with the largest JS divergence value across all BBH events is the right ascension, $\alpha$.
Events with large sky areas, such as GW170729, suffer from large deviations between the \bilby and \lalinference posteriors in the sky position parameters.
The sky position was fixed to the location of the EM counterpart for GW170817.
We show the difference between the \bilby and \lalinference posterior cumulative density functions (CDFs) for $\alpha$ in Figure~\ref{fig:ra_pp_plot} and for the luminosity distance $d_{\rm L}$, which passes the naive criterion on the JS divergence for all events, in Figure~\ref{fig:dl_pp_plot}.
For GW170818, $\alpha$ has the largest JS divergence value ($\unit[0.006]{nat}$) despite the fact that the \bilby and \lalinference CDFs match at the $2\sigma$ level.
This is because the distribution is approximated using a kernel density estimate (KDE) in order to compute the JS divergence, and the posterior for this particular event has a sharp drop-off, which is difficult to model faithfully using the KDE.

Upon manual inspection, we find that the posteriors with JS divergence values up to $\sim\unit[0.002]{nat}$ are consistent between the \lalinference and \bilby samples.
The remaining parameters with significant deviations between the two samplers are the sky position parameters for GW170729.
Investigations into the source of these discrepancies are ongoing.
The differences between the \bilby and \lalinference CDFs for all events and all parameters are shown in Appendix~\ref{appendix:event_cdf_plots}.
A similar comparison was made in~\cite{abbott2019_GWTC1} analyzing the posterior distributions obtained using two different waveform approximants for each event.
The maximum difference between the posteriors assuming the two different waveform models in that work is typically $\sim \unit[0.02]{nat}$, an order of magnitude larger than the differences here. 

\begin{figure}
    \centering
    \includegraphics[width=\linewidth]{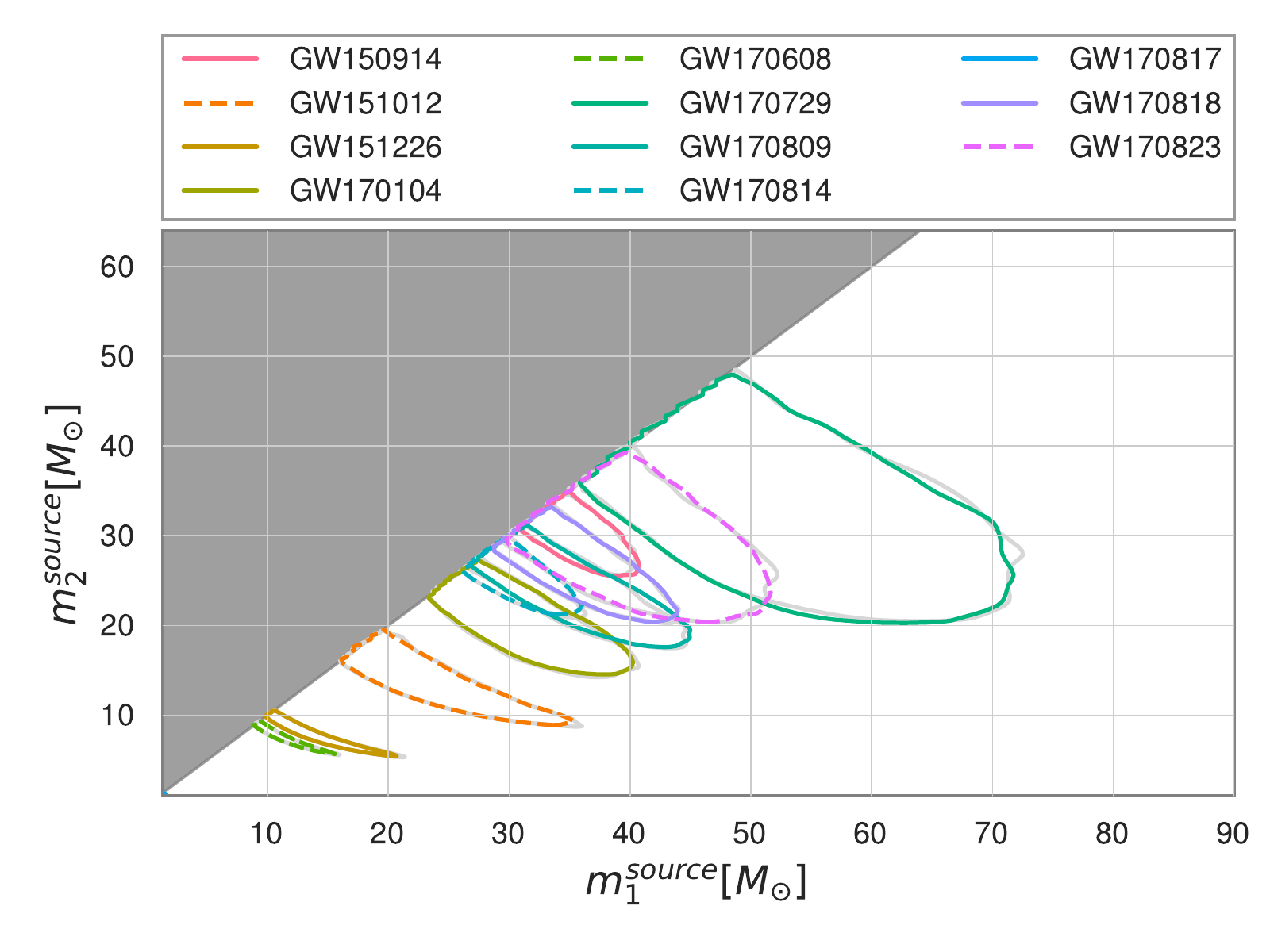}
    \caption{Comparison of the posterior distributions between the \lalinference (gray) and \bilby (colored) packages over the source primary mass $m_{1}^\mathrm{source}$ and source secondary mass $m_{2}^\mathrm{source}$ parameter space. Each contour shows the $90\%$ credible area, with the \lalinference posterior samples reweighted to the \bilby priors.}
    \label{fig:posterior_comparison}
\end{figure}

As another way to visualize the differences between the \bilby and \lalinference samples, in Figure~\ref{fig:posterior_comparison}, we compare the 90\% credible areas of the  two posteriors on the source-frame primary mass $m_{1}^\mathrm{source}$ and secondary mass $m_{2}^\mathrm{source}$ for all GWTC-1 events. As indicated by the low JS divergence values for the mass parameters, the two samplers produce posteriors on these parameters that agree within expected statistical fluctuations.

\begin{figure}
    \centering
    \includegraphics[width=\linewidth]{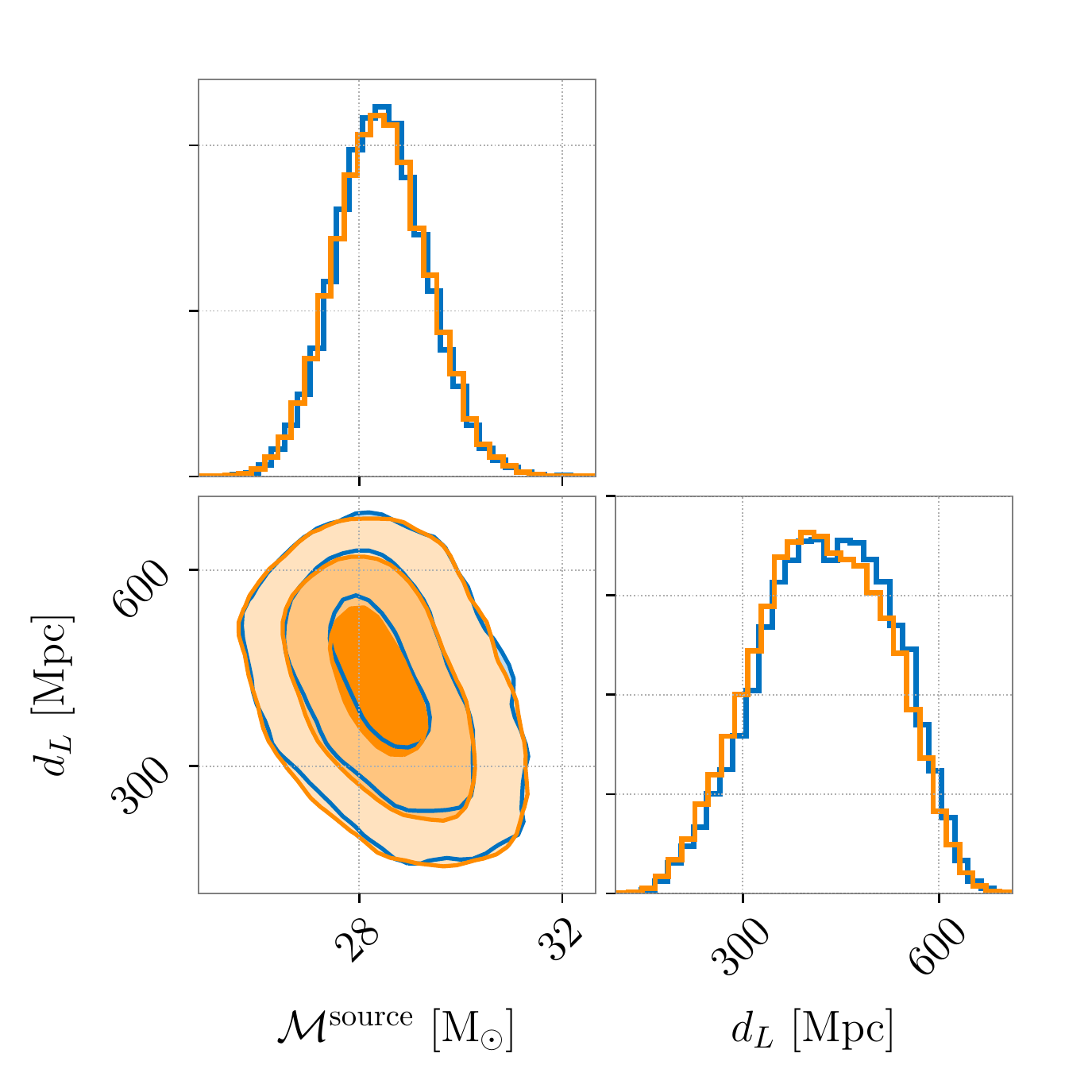}
    \caption{
    Posterior probability distributions for source-frame chirp mass $\mathcal{M}^\mathrm{source}$ and luminosity distance $d_\mathrm{L}$ for GW150914. We display posteriors obtained using \bilby in orange, and \lalinference posteriors in blue. We reweight the \lalinference posteriors to the \bilby default priors using the procedure outlined in Appendix~\ref{appendix:reweighting}. The one-dimensional JS divergence on chirp mass $\mathcal{M}$ and luminosity distance $d_\mathrm{L}$ for this event are JS$_\mathcal{M}=\unit[0.0017]{nat}$ and JS$_{d_\mathrm{L}}=\unit[0.0015]{nat}$.}
    \label{fig:source_frame_chirp_mass_vs_distance}
\end{figure}

We compare \bilby posteriors on source-frame chirp mass $\mathcal{M}^{\mathrm{source}}$ and luminosity distance $d_\mathrm{L}$ for the first observed gravitational-wave event, GW150914~\citep{abbott16_gw150914_detection}, in Figure~\ref{fig:source_frame_chirp_mass_vs_distance}.
The \lalinference distance posterior here matches the \bilby posterior more closely than was demonstrated in Figure~2 of \citet{ashton19}.
This is due to an issue in the application of the time-domain window being fixed in \lalinference, which had affected the distance posterior~\citep{talbot20}.

\begin{figure*}
    \centering
    \begin{subfigure}
        \centering
        \includegraphics[width=0.45\linewidth]{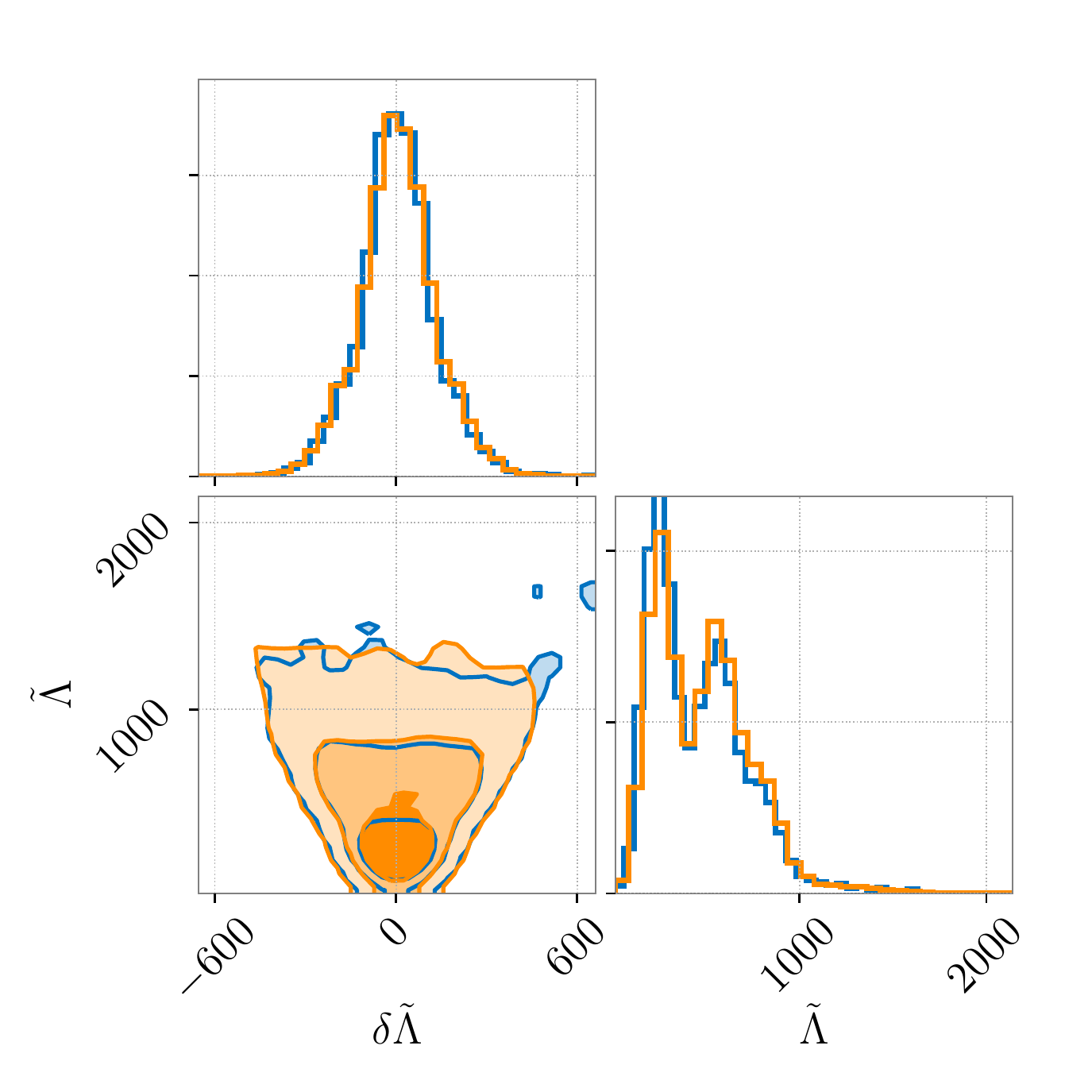}
    \end{subfigure}
    ~ 
    \begin{subfigure}
        \centering
        \includegraphics[width=0.45\linewidth]{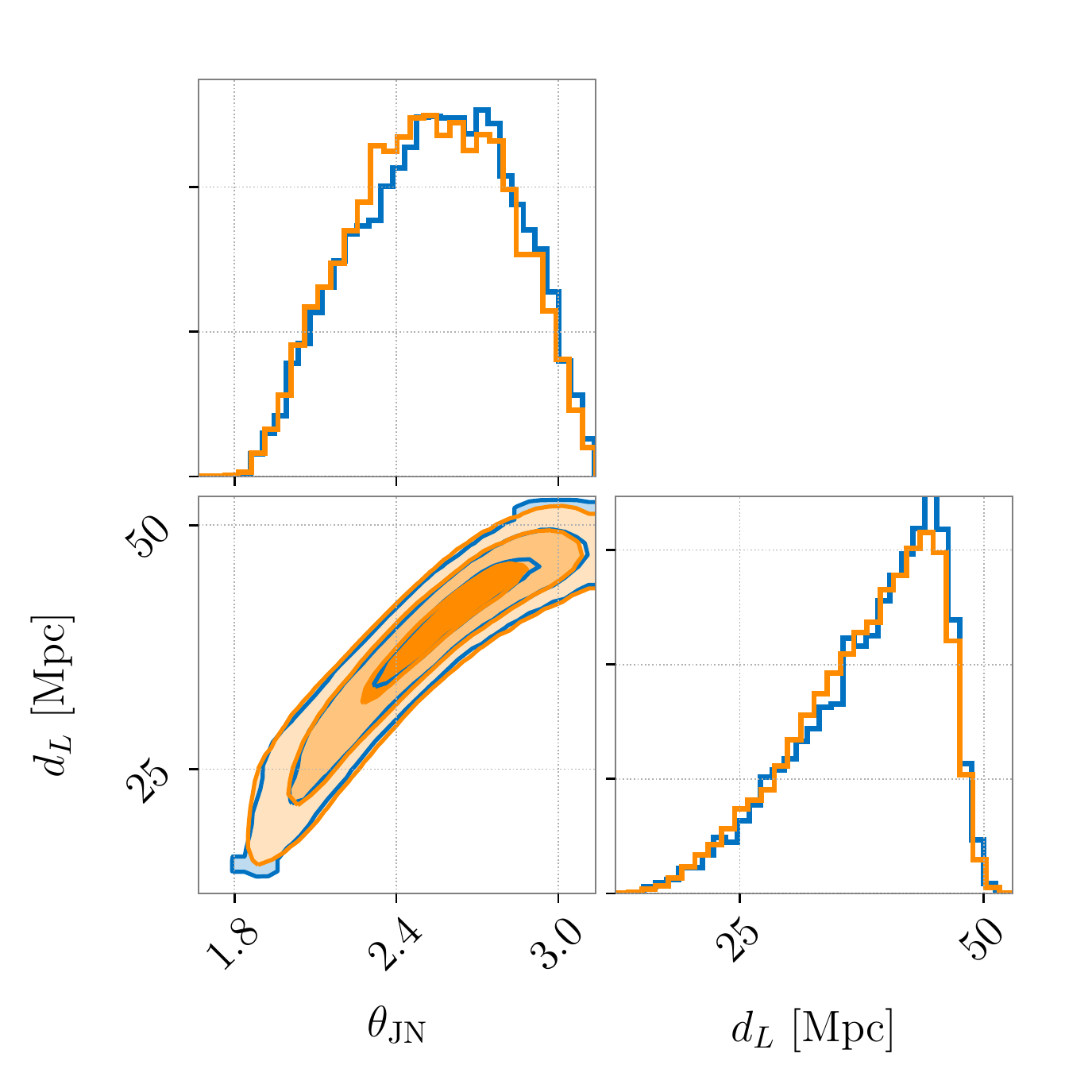}
    \end{subfigure}
    \caption{Joint posterior distributions for parameters of GW170817, comparing \pbilby posteriors in orange and \lalinference posteriors in blue. 
    Left: Posterior probability distributions for tidal parameters $\tilde{\Lambda}$ ($\text{JS}_{\tilde{\Lambda}} = \unit[0.0019]{nat}$) and $\delta\tilde{\Lambda}$ ($\text{JS}_{\delta\tilde{\Lambda}} = \unit[0.0008]{nat}$). 
    Right: Posterior probability distributions for inclination angle $\theta_\mathrm{JN}$ ($\text{JS}_{\theta_\mathrm{JN}} = \unit[0.0009]{nat}$) and luminosity distance $d_\mathrm{L}$ ($\text{JS}_{d_\mathrm{L}} = \unit[0.0008]{nat}$).
    \label{fig:GW170817-distance-inlicination}}
\end{figure*}

For the first observed binary neutron-star merger event, GW170817, we compare the \bilby posterior distributions on tidal parameters $\tilde{\Lambda}$ and $\delta\tilde{\Lambda}$, as well as $\theta_{JN}$ and $d_\mathrm{L}$, to those obtained using \lalinference in Figure~\ref{fig:GW170817-distance-inlicination}. 
The maximum JS divergence for this event is JS$_{q}=\unit[0.0017]{nat}$.
Additional posterior probability plots for all parameters of all eleven CBC events can be found within the online resources that accompany this paper~\citep{Bilby-GWTC-1-Analysis-and-Verification}.

Based on these results, we conclude that \bilby and \lalinference produce statistically indistinguishable results for all parameters and all events reported in GWTC-1 with the exception of the sky area for GW170729 and GW151226. We emphasize that the differences in the CDFs for these parameters are still small compared to other sources of error such as waveform systematics~\citep{abbott2019_GWTC1} and uncertainty in the power spectral density~\citep{biscoveanu2020}.
We provide \pesummary comparison pages between \bilby and reweighted \lalinference posteriors for all GWTC-1 events online.\footnote{\href{https://bilby-gwtc1.github.io}{bilby-gwtc1.github.io}}

\section{Summary}
\bilby is a modern and versatile Bayesian inference library, and has been primed for analysis of gravitational-wave observations.
\bilby performs reliably, producing accurate and unbiased parameter estimation results when analysing simulated signals.
We validate \bilby results for GWTC-1 using the JS divergence statistic between posterior distributions obtained using \bilby and the previously published \lalinference results, finding a maximum JS value of JS$_{\alpha} = \unit[0.0026]{nat}$ for GW170729.
The similarity between the two results indicate that both the \bilby samples obtained with \dynesty and the \lalinference samples are well-converged, and efforts to further validate these results using alternative samplers within \bilby are ongoing. 
Posterior probability distributions generated by \bilby and \lalinference, when run on the same GWTC-1 data and using identical analysis settings, are consistent to the level of sampling noise.
The \bilby posterior samples for events in GWTC-1 are available online~\citep{Bilby_samples}. 
We conclude that \bilby is well-suited to meet the challenges of gravitational-wave parameter estimation in the era of frequent detections.

\section*{Acknowledgements}
{
We thank Stephen Green for helpful insight into the calculation of JS divergence values.
This work is supported through Australian Research Council (ARC) Centre of Excellence CE170100004. 
PDL is supported through ARC Future Fellowship FT160100112 and ARC Discovery Project DP180103155.  
ET is supported through ARC Future Fellowship FT150100281 and CE170100004. 
This work is partially supported by the National Research Foundation of Korea (NRF) grant funded by the Korea government (MEST) (No. 2019R1A2C2006787).
NB acknowledges Inspire division, DST, Government of India for the fellowship support.
This work is partially supported by the National Science Foundation under Grant No.\ PHY-1912648. 
SB, C-JH., and CT acknowledge support of the National Science Foundation, and the LIGO Laboratory. 
SB is also supported by the Paul and Daisy Soros Fellowship for New Americans and the NSF Graduate Research Fellowship under Grant No. DGE-1122374.
This work was partially supported by European Union FEDER funds, the Spanish Ministry of Science and Innovation and the Spanish Agencia Estatal de Investigaci{\'o}n grants FPA2016-76821-P and PID2019-106416GB-I00/AEI/10.13039/501100011033, the Comunitat Autonoma de les Illes Balears through the Direcció General de Pol{\'i}tica Universitaria i Recerca with funds from the Tourist Stay Tax Law ITS 2017-006 (PRD2018/24), the Vicepresid{\`e}ncia i Conselleria d'Innovaci{\'o}, Recerca i Turisme, Conselleria d'Educaci{\'o}, i Universitats del Govern de les Illes Balears and Fons Social Europeu. M.C.~acknowledges funding from the European Union's Horizon 2020 research and innovation programme, under the Marie Sk{\l}odowska-Curie grant agreement No. 751492. D.K.~is supported by the Spanish Ministerio de Ciencia, Innovaci{\'o}n y Universidades (ref.~BEAGAL 18/00148) and cofinanced by the Universitat de les Illes Balears.
This work used \bilby= v0.6.9, \bilbypipe= v0.3.12, \dynesty= v1.0.1, \lalsuite=v6.49, \pesummary= v0.5.6

This research has made use of data, software and/or web tools obtained from the Gravitational Wave Open Science Center~\citep{abbott_19_gwosc}, a service of LIGO Laboratory, the LIGO Scientific Collaboration and the Virgo Collaboration.
Computing was performed on the OzSTAR Australian national facility at Swinburne University of Technology, which receives funding in part from the Astronomy National Collaborative Research Infrastructure Strategy (NCRIS) allocation provided by the Australian Government, LIGO Laboratory computing clusters at California Institute of Technology and LIGO Hanford Observatory supported by National Science Foundation Grants PHY-0757058 and PHY-0823459, and the Quest computing cluster, which is jointly supported by the Office of the Provost, the Office for Research and Northwestern University Information Technology, and funded by the National Science Foundation under Grant No.\ PHY-1726951.
LIGO was constructed by the California Institute of Technology and Massachusetts Institute of Technology with funding from the National Science Foundation and operates under cooperative agreement PHY-1764464. Virgo is funded by the French Centre National de Recherche Scientifique (CNRS), the Italian Istituto Nazionale della Fisica Nucleare (INFN) and the Dutch Nikhef, with contributions by Polish and Hungarian institutes.
}

\section*{Data Availability Statement}
{
We analyse publicly-available data~\citep{abbott_19_gwosc}, and make use of publicly-available PSDs~\citep{GWTC1-PSDS} and calibration envelopes~\citep{GWTC1-CAL}. We compare our results against publicly-available posterior samples~\citep{GWTC1-samples}. We make our own results publicly accessible online~\citep{Bilby_samples}.
}

\appendix
\section{Additional \bilby validation tests}
\label{appendix:review-tests}

In addition to the tests described in the main body of the paper, we performed several additional validation tests which are standard benchmarks for stochastic sampling codes.

\subsection{Prior sampling}
The initial distribution of samples drawn from the prior must faithfully represent the shape of the prior function. In addition to being used for review, the prior sampling test also forms part of \bilby's unit test suite.
Prior samples can be obtained using \bilby via two different methods.
The first is to use the \texttt{sample} method of each \texttt{Prior} object, which generates samples by rescaling from a unit cube.
The second is to run the sampler with a null likelihood using the \texttt{ZeroLikelihood} object so that the returned posterior samples actually reflect the prior. 
To test the consistency of the two methods, we generate prior samples via both methods for a standard 15-dimensional binary black hole signal injected into simulated Gaussian noise.
We perform a Kolmogorov--Smirnov test~\citep{kolmogorov1933, smirnov1948} to evaluate the similarity of the two sets of samples, calculating a $p$-value for each parameter, which quantifies the probability that the two sets of samples are drawn from identical distributions. A combined p-value is then computed, representing the probability that the ensemble of individual-parameter p-values is drawn from a unit uniform distribution. We consider the test to pass if this combined $p$-value is greater than $0.01$.
For a representative run with the latest version of \bilby, we obtain a combined $p$-value of $0.017$.

\subsection{15-dimensional Gaussian}
\begin{figure*}
    \centering
    \begin{subfigure}
        \centering
        \includegraphics[width=0.45\linewidth]{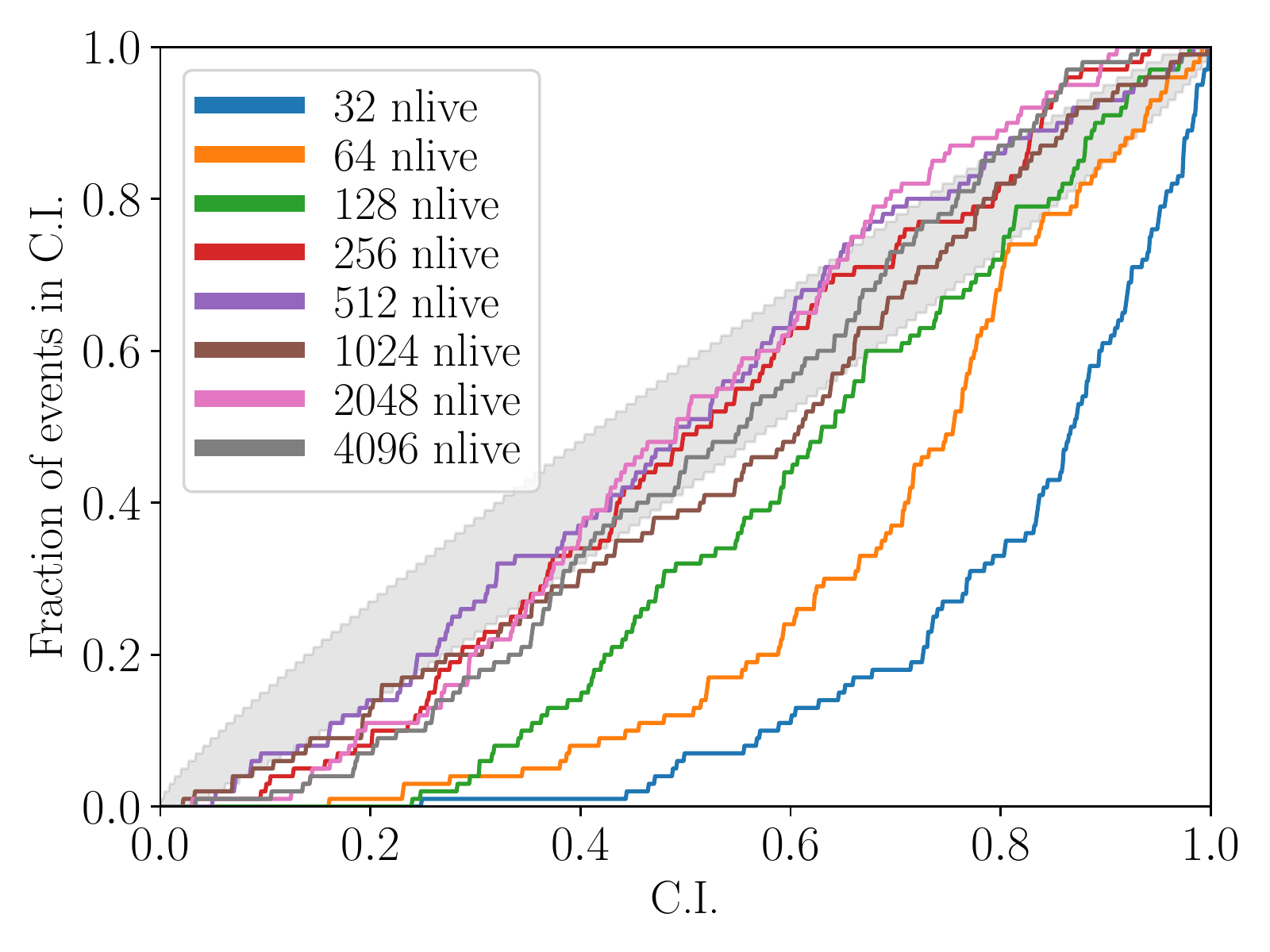}
    \end{subfigure}
    ~ 
    \begin{subfigure}
        \centering
        \includegraphics[width=0.45\linewidth]{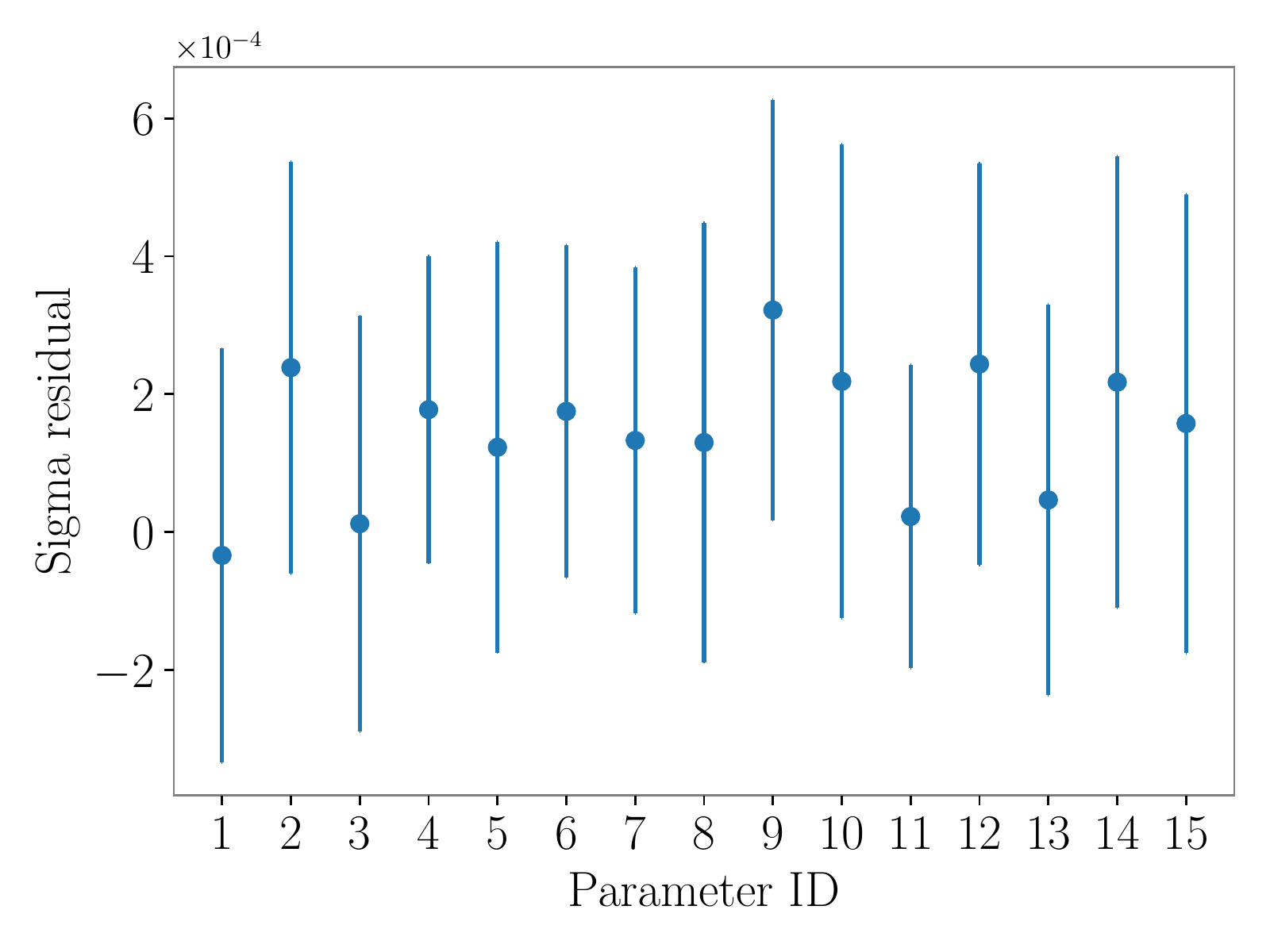}
    \end{subfigure}
    \caption{
    Left: Illustration of the frequency with which the true evidence is within a given credible interval for the unimodal Gaussian-shaped likelihood.
    The legend shows how many live points are used to produce the individual curves.
    For lower number of live points, systematic errors in the evidence estimation cause significant underestimates of the error.
    Starting at 1024 live points, the evidence error reasonably reflects the true uncertainty.
    The grey band shows the $90 \%$ confidence interval.
    Right: Residuals of the true width of the analytical likelihood minus the average recovered one for 1024 live points in each dimension based on 100 independent runs.
    The error bars show the $90 \%$ confidence interval of the average mean of the distribution.
    There is a small $\mathcal{O}(0.1\%)$ systematic bias to underestimate the width, i.e. the parameter is on average slighty overconstrained.
    However, this bias is negligibly small compared to stochastic sampling uncertainties for individual runs.}
    \label{fig:15_d}
\end{figure*}

Sampling an analytically-known likelihood distribution is an important test to verify that we can recover the correct posterior.
For this test, we choose the \scipy implementation of a multivariate normal distribution (\texttt{scipy.stats.multivariate\_normal}) as our likelihood.
We choose the distribution to be $15$-dimensional since this reflects the typical number of dimensions we encounter in binary black hole problems.
We set the means of all parameters to be zero, and choose a covariance matrix $\mathrm{COV}_{ij}$ with standard deviations for each of the parameters ranging between $0.15$ and $0.25$ to match past tests done with \lalinference.
Using the \bilby default sampler settings for a $15$-dimensional problem, we test if we correctly recover the posterior distribution by drawing samples from this $15$-dimensional likelihood and comparing the obtained means and standard deviations to the true values.
Additionally, we verify that we recover the expected evidence within the estimated error. 
Since the likelihood distribution is normalized and we use uniform priors for each parameter in the range $[-5, 5]$, the evidence can be approximated by the prior volume, since the standard deviations are small enough that the value of the likelihood evaluated at the edges of the prior is negligible: 
\begin{equation}
    \ln \mathcal{Z} \approx - \ln X \, ,
\end{equation}
where $X$ is the prior volume.
In Figure~\ref{fig:15_d} on the left hand side we find the measured standard deviations and the evidence to be in broad agreement with analytical expectations.
While the evidence errors quoted by \dynesty are not truly Gaussian, the one-sigma credible interval is consistent with covering the true evidence $68\%$ of the time if one uses more than $1000$ live points. 
Additionally, the overshoot at high values of the credible interval indicates that there are fewer outliers than we would for a Gaussian distribution.
The right hand side of Figure~\ref{fig:15_d} demonstrates that the width of the posterior distribution is correctly recovered.
We have thus shown that the \dynesty implementation in \bilby has no significant issues in recovering the shape of posterior distributions and the correct evidence for this  fundamental problem.

We performed the same test using a bimodal Gaussian distribution, with means separated by $8$ standard deviations in each dimension.
While it is more difficult to correctly sample a degenerate likelihood surface, we still find 1000 live points sufficient to reasonably recover the evidence.
Individual runs of the bimodal likelihood may produce a biased set posterior samples in favour of one of the modes over the other, which is why multiple runs should be combined.
We verified that none of the modes is preferred if we use all 100 runs.
Thus, there are also no substantial issues that arise in sampling multimodal distributions with \bilby.

\subsection{Fiducial event simulations}

We analyse two fiducial simulated signals; one binary black hole merger, and one binary neutron star merger with tides. We use a LIGO Hanford--Livingston detector network and add the simulated signals into design sensitivity Gaussian noise. For the binary black hole, we use the \imrphenomp waveform and the default \unit[4]{s} prior described in Table~\ref{tab:mass_distance_priors}. For the binary neutron star, we use the ROQ implementation of the \imrphenompvtwonrtidal waveform~\citep{amanda_baylor_2019_3478659} with the 128~s tidal low-spin prior. The binary black hole and neutron star systems have network optimal SNRs of $8.8$ and $27.9$, respectively.\footnote{The binary black hole analysis was performed using \bilby version 0.6.3, while the neutron star analysis used \bilby 1.0.0. The default Advanced LIGO design PSD changed between these two versions of \bilby to reflect the updated detector sensitivity predictions~\citep{abbott_19_observing_scenarios}. Parameter estimation is performed using \dynesty with the default settings.} In Table \ref{tab:fiducial_injection_recovery}, we show the true values along with the recovered median and 90\% credible interval values for each parameter. Nearly all the true parameter values for both systems are recovered within the 90\% credible interval, and those that are not are consistent with deviations due to the Gaussian noise realization. Full corner plots for both simulated signals are available online~\citep{Bilby-GWTC-1-Analysis-and-Verification}.

\begin{table}
    \centering
    \caption{Our injected and recovered values for the two fiducial event analyses. Recovered median values are quoted with the symmetric 90\% credible interval around the median.}
    \begin{tabular}{c c c@{\extracolsep{4pt}} c c}
    \hline
      & \multicolumn{2}{c} {{ BBH }} & \multicolumn{2}{c} {{ BNS }} \\
     \cline{2-3} \cline{4-5}
     { Parameter } & { Inject } & { Recover } & { Inject } & { Recover } \\
     \hline
    $\mathcal{M}/\mathrm{M}_{\odot}$ & $15.53$ & $15.4_{-0.4}^{+0.3}$ & $1.486$ & $1.486_{-0.0001}^{+0.0001}$ \\
    $q$ & $0.52$ & $0.7_{-0.4}^{+0.3}$ & $0.9$ & $0.9_{-0.2}^{+0.1}$ \\
    $a_{1}$ & $0.65$ & $0.6_{-0.5}^{+0.3}$ & $0.04$ & $0.02_{-0.02}^{+0.02}$ \\
    $a_{2}$ & $0.65$ & $0.5_{-0.4}^{+0.4}$ & $0.01$ & $0.02_{-0.02}^{+0.02}$ \\
    $\theta_{1}$ & $1.24$ & $1.1_{-0.6}^{+0.8}$ & $1.03$ & $1.5_{-0.9}^{+1.0}$ \\
    $\theta_{2}$ & $0.80$ & $1.3_{-0.9}^{+1.1}$ & $2.17$ & $1.6_{-1.0}^{+1.0}$ \\
    $\phi_{12}$ & $1.5$ & $3.1_{-2.8}^{+2.9}$ & $5.10$ & $3.2_{-2.9}^{+2.8}$ \\
    $\phi_{\rm JL}$ & $3.01$ & $3.2_{-2.9}^{+2.8}$ & $2.52$ & $3.1_{-2.8}^{+2.9}$ \\ 
    $d_{\mathrm{L}}/\mathrm{Mpc}$ & $614$ & $1018_{-623}^{+1147}$ & $100$ & $86_{-26}^{+17}$ \\
    $\delta$ & $1.00$ & $0.7_{-1.6}^{+0.4}$ & $0.2$ & $0.3_{-0.1}^{+0.1}$ \\
    $\alpha$ & $2.00$ & $4.6_{-2.7}^{+1.0}$ & $3.95$ & $3.9_{-0.1}^{+0.1}$ \\
    $\theta_{J N}$ & $1.65$ & $1.8_{-0.8}^{+1.0}$ & $0.25$ & $0.6_{-0.4}^{+0.7}$ \\
    $\psi$ & $1.50$ & $1.6_{-1.4}^{+1.4}$ & $2.70$ & $1.5_{-1.4}^{+1.5}$ \\
    $\phi$ & $2.00$ & $3.1_{-2.8}^{+2.8}$ & $3.69$ & $3.1_{-2.8}^{+2.8}$ \\
    $t_{\mathrm{geo}}/\mathrm{s}$ & $0.04$ & $0.04_{-0.02}^{+0.00}$ & $-0.01$ & $-0.01_{-0.00}^{+0.00}$ \\
    $\Lambda_{1}$ & $-$ & $-$ & $1500$ & $752_{-657}^{+915}$ \\
    $\Lambda_{2}$ & $-$ & $-$ & $750$ & $1437_{-1216}^{+1294}$ \\
    \hline
    \end{tabular}
    \label{tab:fiducial_injection_recovery}
\end{table}

\section{Run setting details}
\label{appendix:run-settings}

\subsection{Sampler settings}
\label{appendix:sampler}

The default sampler used by \bilby is \dynesty~\citep{dynesty}, an off-the-shelf nested sampling~\citep{Skilling06} package. 
The first step in nested sampling is to draw $N$ random live points from the prior.
At each iteration, the lowest-likelihood sample from the initial $N$ points is discarded in favour of a higher-likelihood point, again randomly chosen from the prior.
After every step, the actively-sampled region of the prior shrinks to the volume contained by the hyperplane of constant minimum likelihood for the current population of live points. 
When the live domain has reduced sufficiently, it becomes inefficient to select higher-likelihood points uniformly from the restricted prior space.

After the uniform sampling becomes sufficiently inefficient, new points are selected by randomly walking using a custom Markov-chain Monte Carlo algorithm starting from the sample being replaced.
The transition probability is determined by the distribution of the set of current live points.
The number of steps taken in the chain is determined such that the length of the chain is at least some multiple $n_\mathrm{act}$ of the auto-correlation length of the chain \citep{sokal94}.
For the analysis in this paper, we require $n_\mathrm{act}=10$.
A Markov-chain Monte Carlo walker algorithm then takes at least $n$ steps to draw a new sample from the restricted prior.
In order to reduce bottlenecks while using multiprocessing we impose a maximum length of the chain.
If no point with a higher likelihood than the original point is found within this number of steps, we return a random point from the prior distribution.
Nested sampling is able to well-resolve multimodal distributions, making it useful for exploring complicated parameter spaces.
For all events in GWTC-1, we give the sampler $N=2000$ live points and $n=100$ steps. 

\subsection{Priors}
\label{appendix:priors}
\begin{table}
    \centering
    \caption{
    Lower and upper limits on chirp mass $\mathcal{M}$, luminosity distance $d_\mathrm{L}$ and dimensionless spin magnitude $a_1, a_2$ priors for each of the default prior sets contained in \bilbypipe. \label{tab:mass_distance_priors}}
    \begin{tabular}{c c c c}
    \hline
    Prior & $\mathcal{M}/$M$_\odot$ & $d_\mathrm{L}/\mathrm{Mpc}$ & $a_1, a_2$ \\
    \hline
        High-mass & 25--175 & 100--7000  & 0--0.99 \\
        $4~\mathrm{s}$ & 12.299703--45 & 100--5000 & 0--0.88 \\
        $8~\mathrm{s}$ & 7.932707--14.759644 & 100--5000 & 0--0.8 \\
        $16~\mathrm{s}$ & 5.141979--9.519249 & 100--4000 & 0--0.8 \\
        $32~\mathrm{s}$ & 3.346569--6.170374 & 100--3000 & 0--0.8 \\
        $64~\mathrm{s}$ & 2.184345--4.015883 & 20--2000 & 0--0.8 \\
        $128~\mathrm{s}$ & 1.420599--2.602169 & 1--500 & 0--0.8 \\
        $128~\mathrm{s}$ tidal & 1.485--1.49 & 1--300 & 0--0.89 \\
        $128~\mathrm{s}$ tidal low-spin & 1.485--1.49 & 1--300 & 0--0.05 \\
    \hline
    \end{tabular}
\end{table}
~
\begin{table}
    \centering
    \caption{
    Default prior settings for $10$ of the $17$ parameters studied for CBCs observed with gravitational waves.
    The settings given in this table are consistent between all default prior sets contained in \bilbypipe.\label{tab:consistent_priors}}
    \begin{tabular}{c c c c}
    \hline
    Parameter & Shape & Limits & Boundary \\
    \hline
          $q$ & Uniform & 0.125--1 & -- \\
          $\theta_1$, $\theta_2$ & Sinusoidal & 0--$\pi$ & -- \\
          $\phi_{12}$, $\phi_{JL}$ & Uniform & 0--$2\pi$ & Periodic \\
          $\theta_{JN}$ & Sinusoidal & 0--$\pi$ &  -- \\
          $\psi$ & Uniform & 0--$\pi$ &  Periodic \\
          $\phi$ & Uniform & 0--$2\pi$ &  Periodic \\
          $\alpha$ & Uniform & 0--$2\pi$ &  Periodic \\
          $\delta$ & Cosinusoidal & $-\pi / 2$--$\pi / 2$ &  -- \\
    \hline
    \end{tabular}
\end{table}

We sample directly in $\mathcal{M}$ and $q$ to avoid issues associated with sampling extremely thin regions of parameter space, which occurs when sampling in component masses 
(\bilby and \bilbypipe can easily be made to sample in other parameters such as component masses; here we only discuss default parameters and priors used for analysis of the eleven events in GWTC-1).
Our prior on mass ratio is uniform in the range $0.125 \leq q \leq 1.0$, with the lower limit determined due to limitations of the \imrphenomp ROQ.

Prior limits used for $\mathcal{M}$, $d_\mathrm{L}$, $a_1$ and $a_2$ are provided in Table \ref{tab:mass_distance_priors}.
The chirp mass prior limits are based on those stated in the ROQ git repository.\textsuperscript{\ref{footnote:ROQ}}
We use a luminosity distance prior that is uniform in the source frame, with limits motivated by the scaling of gravitational-wave amplitude with both chirp mass and distance. 
The uniform-in-source-frame prior, which indicates a uniform distribution of mergers in our Universe~\citep{ade2016}, differs from the $d_\mathrm{L}^{2}$ power-law prior used in the \lalinference analyses, which indicates a uniform distribution in a Euclidean, non-expanding universe.
We use dimensionless component spin priors that are uniform between $0$ and an upper limit that is determined by the mass range assumed.
For non-tidal waveform models, we use an upper limit that is either $0.8$, $0.88$ or $0.99$.
For tidal approximants, both a low-spin and a high-spin prior are available. Our component spin prior upper limits are $0.05$ (low-spin) and $0.89$ (high-spin) in these cases.
The upper limits on spin magnitude are determined by the training range of the ROQ basis \citep[e.g.,][]{smith16}.
For analysis of binary neutron star coalescence signal GW170817, we sample in the dimensionless tidal parameters $\Lambda_1$ and $\Lambda_2$, which describe the deformability of the primary and secondary masses.
If  $\Lambda_{i} = 0$, the neutron star is non-deformable and thus has no tides. We set our priors on $\Lambda_1$ and $\Lambda_2$ to be uniform between $0$ and $5000$ to reflect our ignorance of the neutron star equation of state. 
The remainder of our priors are standard and geometrically motivated.

\subsection{Data}

    \begin{table}
    \centering
    \caption{GPS trigger time and data segment duration used for each event. By default, the data segment is positioned such that there are $2~\mathrm{s}$ of data after the trigger time.  \label{tab:triggers_and_durations}}
    \begin{tabular}{l c c}
    \hline
    Event & GPS trigger time $t_\mathrm{trig}/\mathrm{s}$ & Data duration $T/\mathrm{s}$ \\
    \hline
        GW150914 & 1126259462.391 & 8 \\
        GW151012 & 1128678900.400 & 8 \\
        GW151226 & 1135136350.600 & 8 \\
        GW170104 & 1167559936.600 & 4 \\
        GW170608 & 1180922494.500 & 16 \\
        GW170729 & 1185389807.300 & 4 \\
        GW170809 & 1186302519.700 & 4 \\
        GW170814 & 1186741861.500 & 4 \\
        GW170817 & 1187008882.430 & 128 \\
        GW170818 & 1187058327.100 & 4 \\
        GW170823 & 1187529256.500 & 4 \\
    \hline
    \end{tabular}
\end{table}

The data segments we use are accessed using the \gwpy~\citep{gwpy} method \texttt{TimeSeries.get(channel\_name, start\_time, end\_time)}. The \texttt{start\_time} $t_\mathrm{start}$ and \texttt{end\_time} $t_\mathrm{end}$ are defined relative to the \texttt{trigger\_time} $t_\mathrm{trig}$ of each event, such that
\begin{align}
    t_\mathrm{end} &= t_\mathrm{trig} + t_\mathrm{post-trig}; \quad
    t_\mathrm{start} = t_\mathrm{end} - T.
\end{align}
Here $T$ is the total duration of the data segment and $t_\mathrm{post-tri}$ is the post-trigger duration, which is $2~\mathrm{s}$ in \bilby by default. We provide the trigger times and data segment durations for all GWTC-1 events in Table \ref{tab:triggers_and_durations}. The \texttt{channel\_name} used to obtain strain data from both the LIGO Hanford and LIGO Livingston detectors is \texttt{DCS-CALIB\_STRAIN\_C02} for all events, with the exception of GW170817, for which we use the \texttt{channel\_name} of \texttt{DCH-CLEAN\_STRAIN\_C02\_T1700406\_v3} to obtain glitch-subtracted strain data from LIGO Livingston. We also obtain Virgo data for events that occurred from July until mid-August 2017 (GW170729, GW170809, GW170814, GW170817 and GW170818) using the \texttt{channel\_name} of \texttt{Hrec\_hoft\_V1O2Repro2A\_16384Hz}.

Strain data is available from the Gravitational Wave Open Science Centre~\citep{abbott_19_gwosc} sampled at both $\unit[16384]{Hz}$ (the native sampling frequency of advanced LIGO and advanced Virgo) and down-sampled to $\unit[4096]{Hz}$.
We download the data sampled at $\unit[16384]{Hz}$.
The \lalinference~\citep{LALSuite} analysis of binary black holes in~\cite{abbott2019_GWTC1} was performed with data down-sampled to $\unit[2048]{Hz}$ using a \lal down-sampling function and integrated to the Nyquist frequency ($\unit[1024]{Hz}$).

In \bilbypipe the user can choose to either not down-sample, down-sample using the same \lal routine as done in \lalinference and \bayeswave~\citep{Cornish14}, or down-sample using the \gwpy method.
In general, we recommend users do not down-sample the time domain data, but rather apply cuts directly in the frequency domain.
However, since the PSDs used in this analysis were made with \bayeswave and the \lalinference analysis we compare with use the \lal down-sampling, we also use this method.

The default method implemented in \lal and used by \lalinference and \bayeswave is done in the time domain and consists of two stages.
First the data are low-passed using a 20th-order zero-phase Butterworth filter.
The filter is customised such that the power at the low-pass frequency $f_{c}$ is reduced by a factor of ten.
The frequency response of the filter is given by
 \begin{align}
     R(f; f_{c}, n, a_{c}) &= \left[1 + \left(a_{c}^{-1/2} - 1\right) \left(\frac{f}{f_{c}}\right)^{2n}\right]^{-1}.
 \end{align}
The data are then down-sampled by a factor of $N$ by taking every $N$th sample, this aliases the data.
This aliasing means that any signal close to the new Nyquist frequency will be suppressed and aliased which may introduce a bias in our inference.
The final frequency domain strain after downsampling by a factor of $N$ is given by
\begin{align}
    \bar{h}(f; f_c, n, a_{c}) &= h(f) R(f; f_c, n, a_c) \nonumber\\
    & + \sum_{i = {\rm odd}}^{N} h((i + 1) f_c - f) R((i + 1) f_c - f; f_c, n, a_c)  \nonumber\\
    & + \sum_{i = {\rm even}}^{N} h(i f_c + f) R(i f_c + f; f_c, n, a_c).
\end{align}
Here $h(f)$ is the frequency-domain data without low-pass filtering or downsampling.
Of the events analysed in this work, the lowest mass events (GW151226, GW170608, and GW170817) have frequency content close to or above the down-sampled Nyquist frequency.
We expect the bias introduced by this to be small.

\begin{figure}
    \centering
    \includegraphics[width=\linewidth]{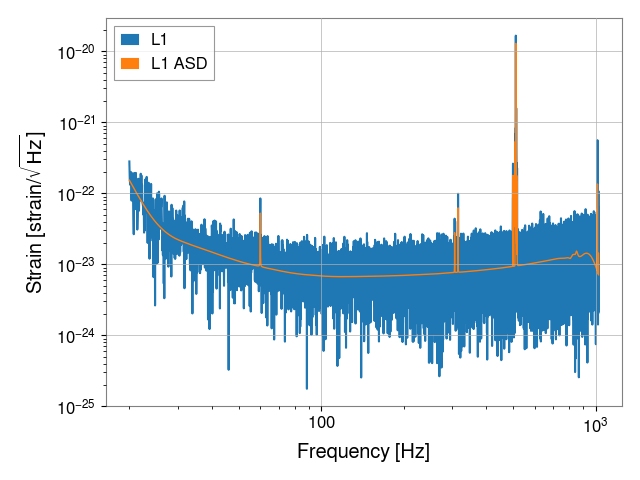}
    \includegraphics[width=\linewidth]{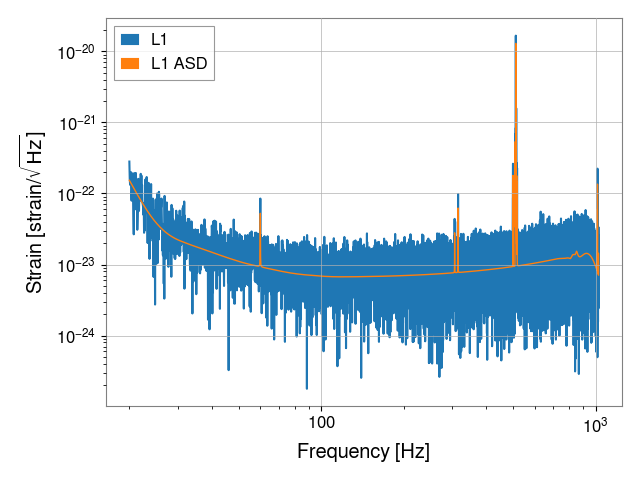}
    \caption{
    The data and PSD in the LIGO Livingston interferometer at the time of GW170608.
    In the upper/lower panel we show the data with/without being low-pass filtered and down-sampled to $\unit[2048]{Hz}$.
    We can see the effect of the low-pass filter in suppressing the data above $\sim\unit[900]{Hz}$.
    The filtering and down-sampling was applied when computing the PSD and so the data on the left better matches the PSD.
    }
    \label{fig:sample_rate_data_comparison}
\end{figure}

In Figure~\ref{fig:sample_rate_data_comparison} we show the data containing GW170608 along with the PSD produced by \bayeswave with (left) and without (right) downsampling the data to a new sampling rate of $\unit[2048]{Hz}$ for the LIGO Livingston observatory.
On the right we can see the turnover in the data and the PSD close to the new Nyquist frequency $\unit[1024]{Hz}$.

\section{Prior Reweighting}
\label{appendix:reweighting}
 In order to compare posterior samples that are unbiased by differing prior choices, we reweight samples obtained using \lalinference priors $\pi_\mathrm{LI}$ by \bilby default priors $\pi_\mathrm{B}$, with weights expressed as
\begin{equation}
    \mathcal{W} = \frac{\pi_\mathrm{B}}{\pi_\mathrm{LI}}.
\end{equation}
We must also account for the fact that \bilbypipe uses default priors that are flat in $\mathcal{M}$ and $q$, whereas \lalinference uses priors that are uniform in component masses.
We therefore rejection sample from the released posterior samples with weights given by the inverse of the Jacobian given in Eq.~(21) of \citet{veitch15},
\begin{equation}
    \mathcal{J} = \frac{\mathcal{M}}{m_1^2}.
\end{equation}
The complete reweighting procedure can be written
\begin{equation}
p_{\pi_\mathrm{B}} = \mathcal{W}\mathcal{J}p_{\pi_\mathrm{LI}},
\end{equation}
where $p_{\pi_\mathrm{B}}$ and $p_{\pi_\mathrm{LI}}$ are the posterior probabilities computed using \bilby and \lalinference priors, respectively. In practice, we reweight by rejection sampling in order to preserve the independence of samples. We also account for a difference in the definition of the Solar mass $\text{M}_\odot$ between the current version of \bilby and the version of \lalinference used to produce the public GWTC-1 samples that we compare against. 

\section{CDF Comparisons for GWTC-1 Events}
\label{appendix:event_cdf_plots}
In this Appendix we present the comparisons of the CDFs obtained using \bilby and \lalinference for all parameters and for all events. The legend shows the JS divergence and uncertainty for each parameter, and the shaded regions represent the $1$-, $2$-, and $3$-$\sigma$ confidence intervals.
\begin{figure*}
    \centering
    \includegraphics[width=0.8\linewidth]{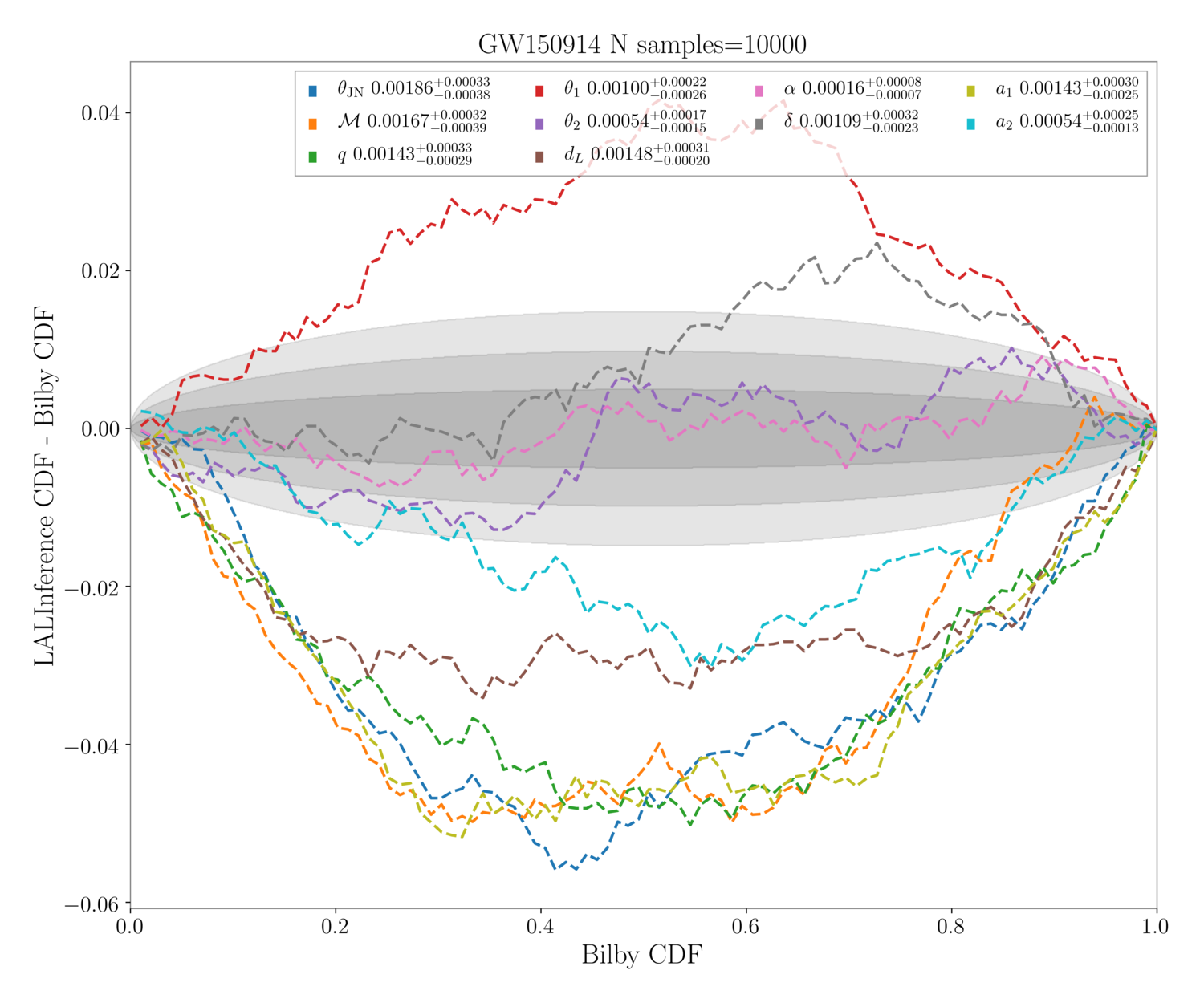}
    \includegraphics[width=0.8\linewidth]{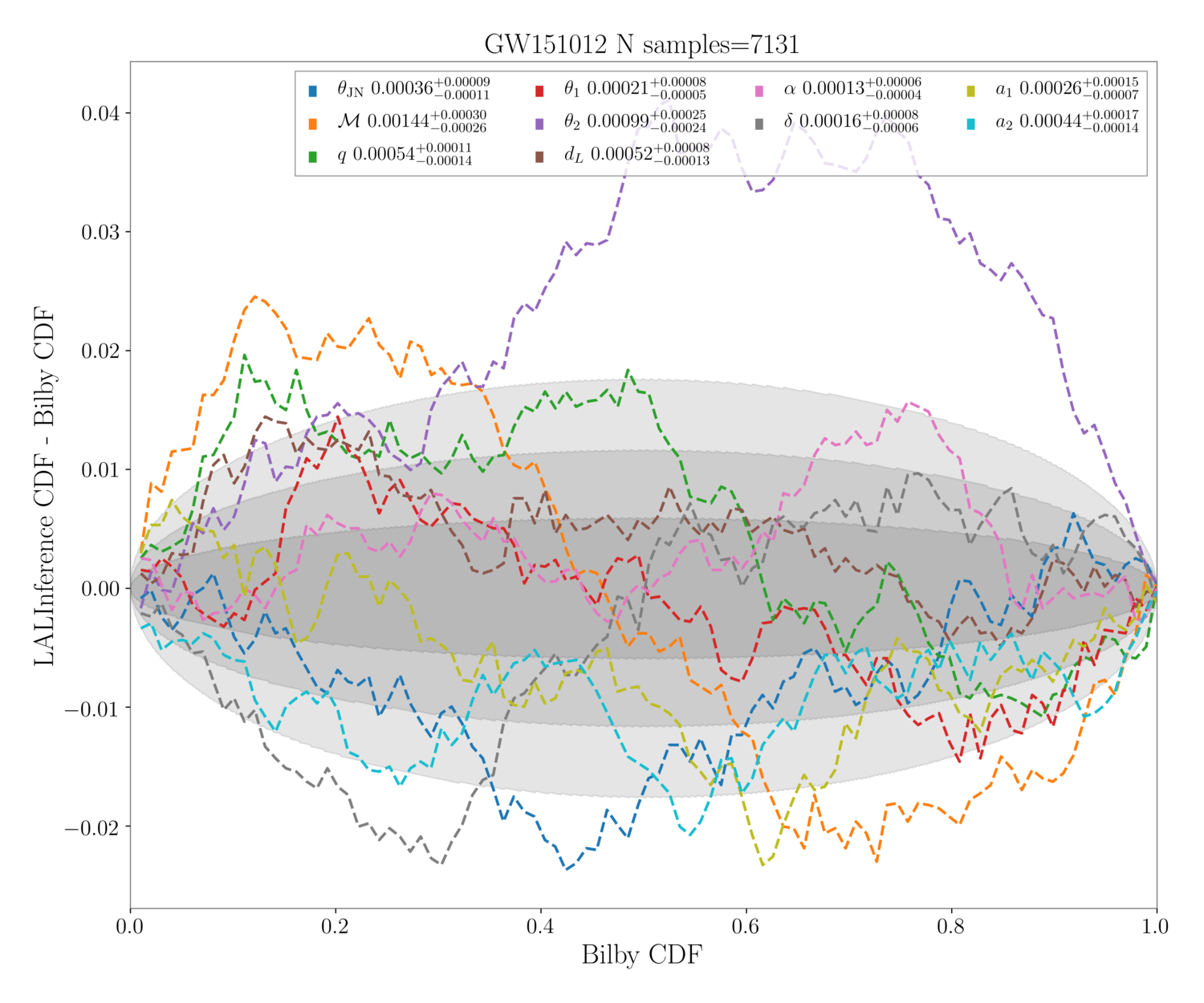}
    \caption{CDF comparison between \bilby and \lalinference for GW150914 and GW151012.}
\end{figure*}

\begin{figure*}
    \centering
    \includegraphics[width=0.8\linewidth]{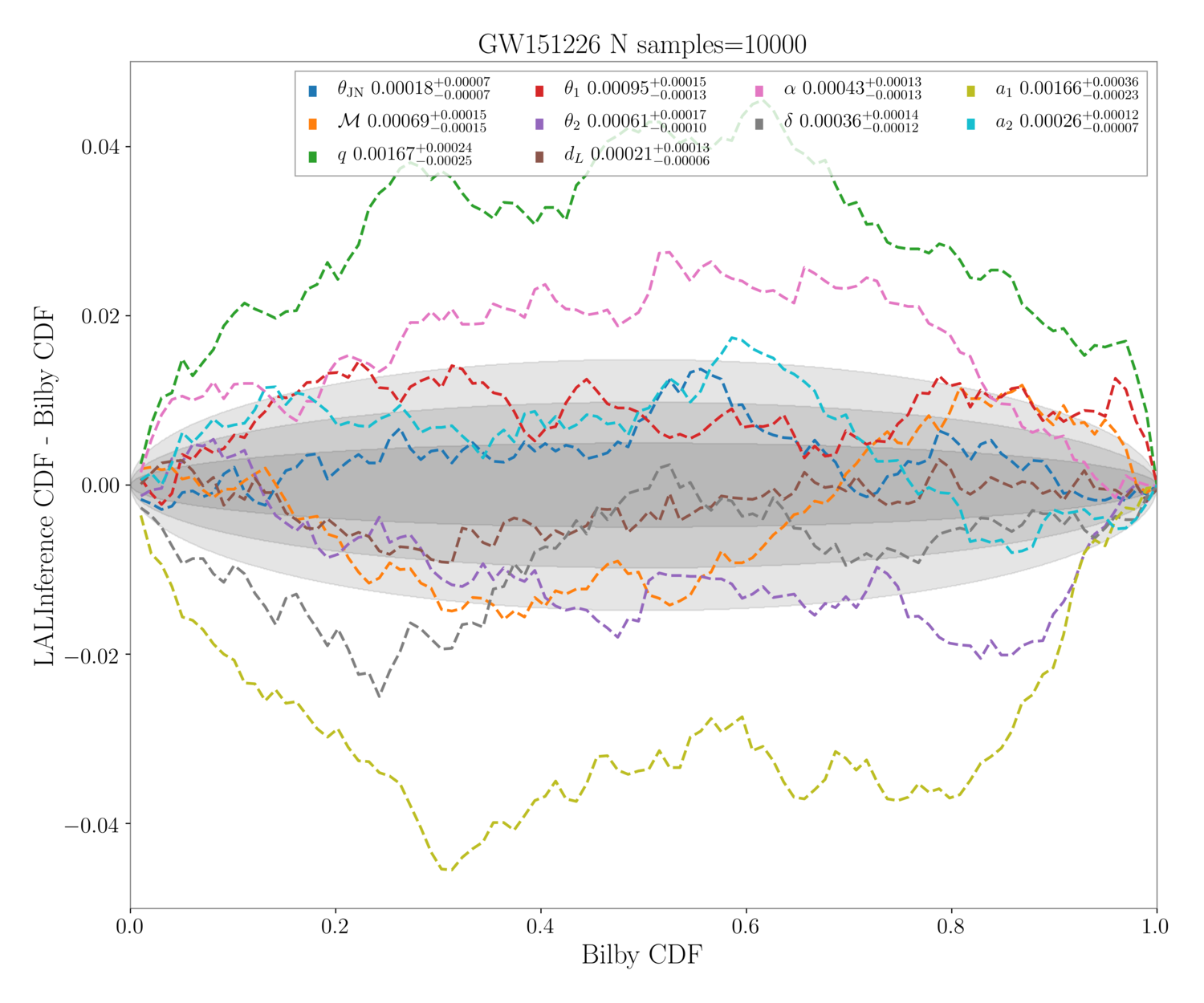}
    \includegraphics[width=0.8\linewidth]{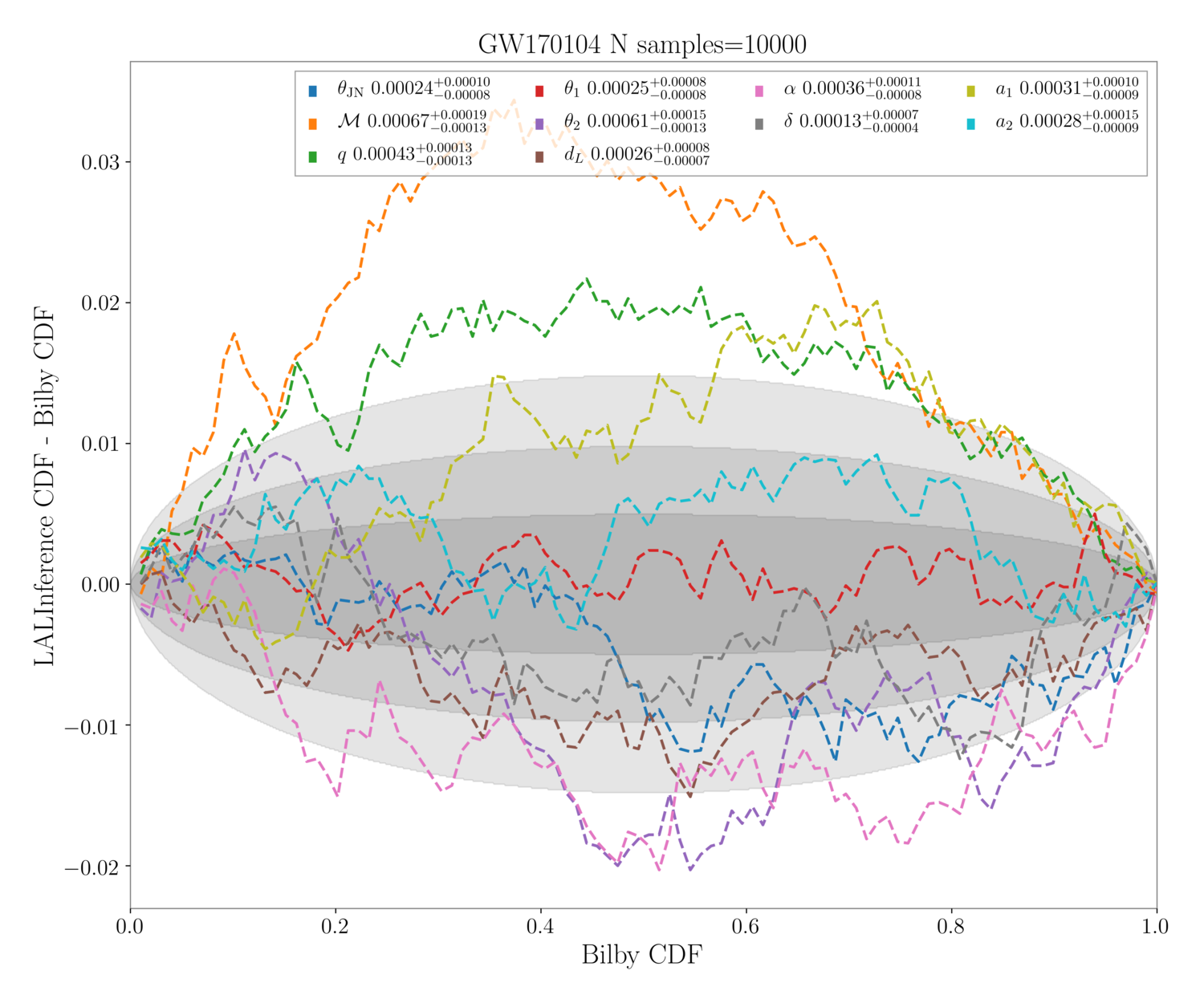}
    \caption{CDF comparison between \bilby and \lalinference for GW151226 and GW170104.}
\end{figure*}

\begin{figure*}
    \centering
    \includegraphics[width=0.8\linewidth]{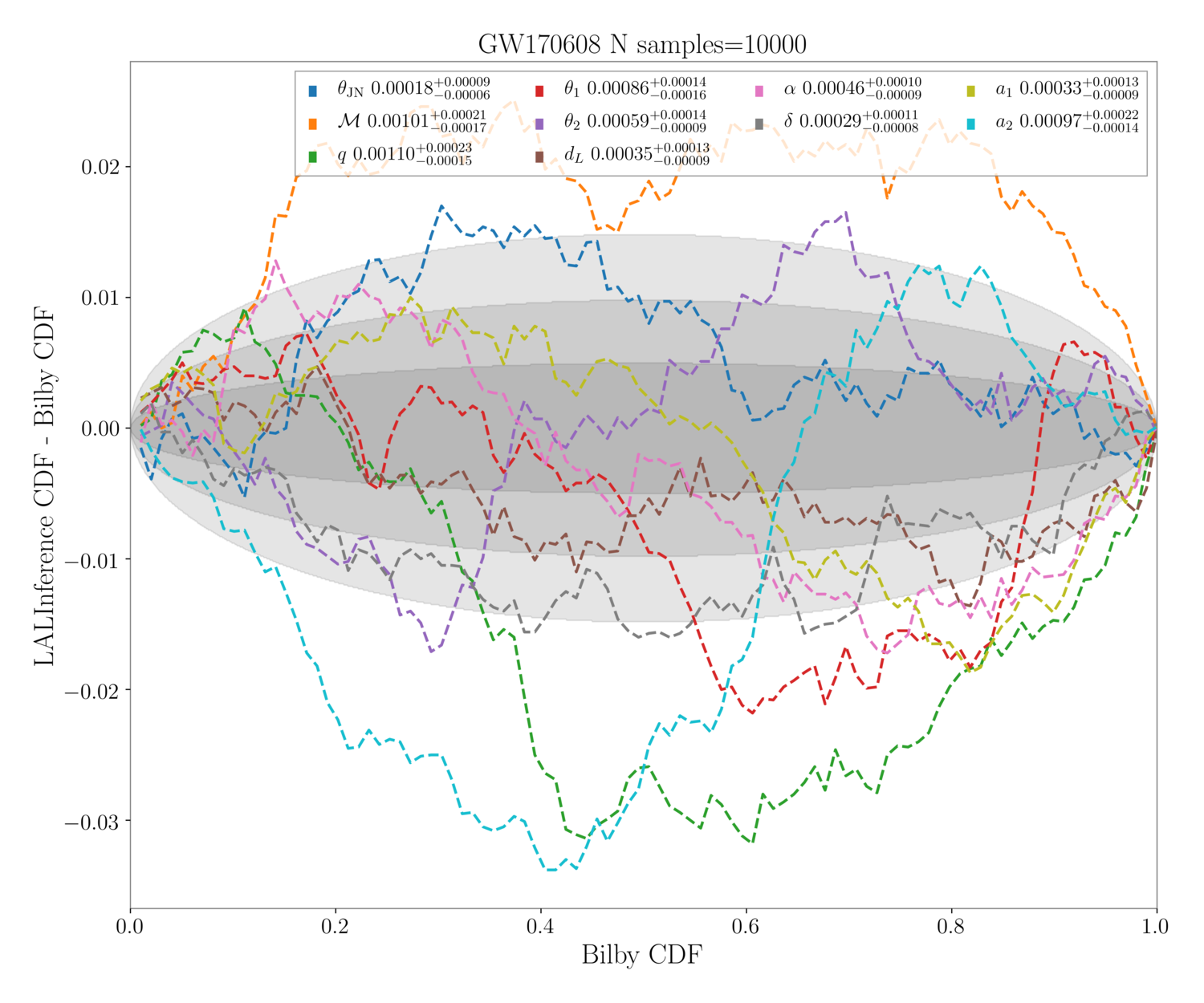}
    \includegraphics[width=0.8\linewidth]{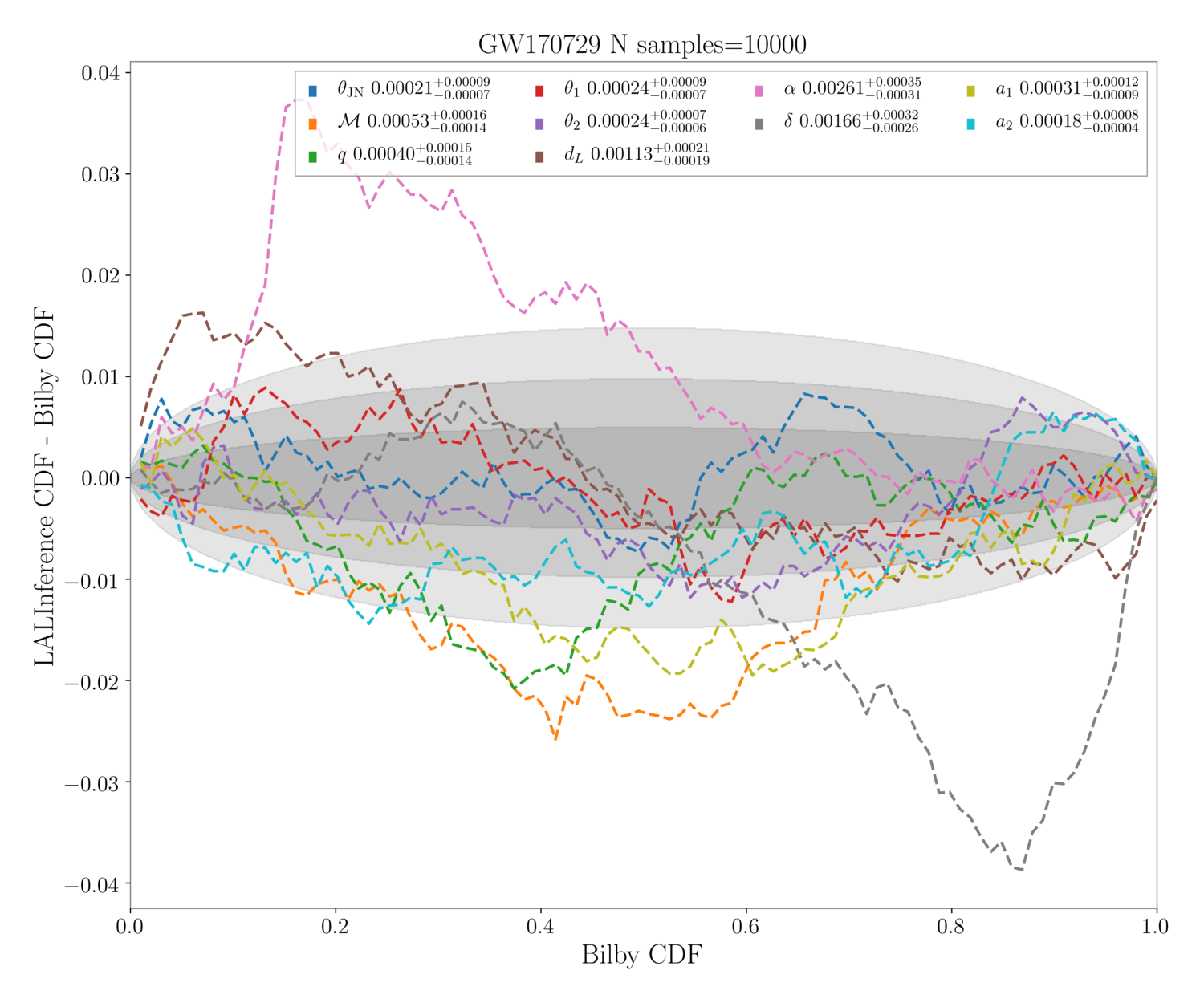}
    \caption{CDF comparison between \bilby and \lalinference for GW170608 and GW170729.}
\end{figure*}

\begin{figure*}
    \centering
    \includegraphics[width=0.8\linewidth]{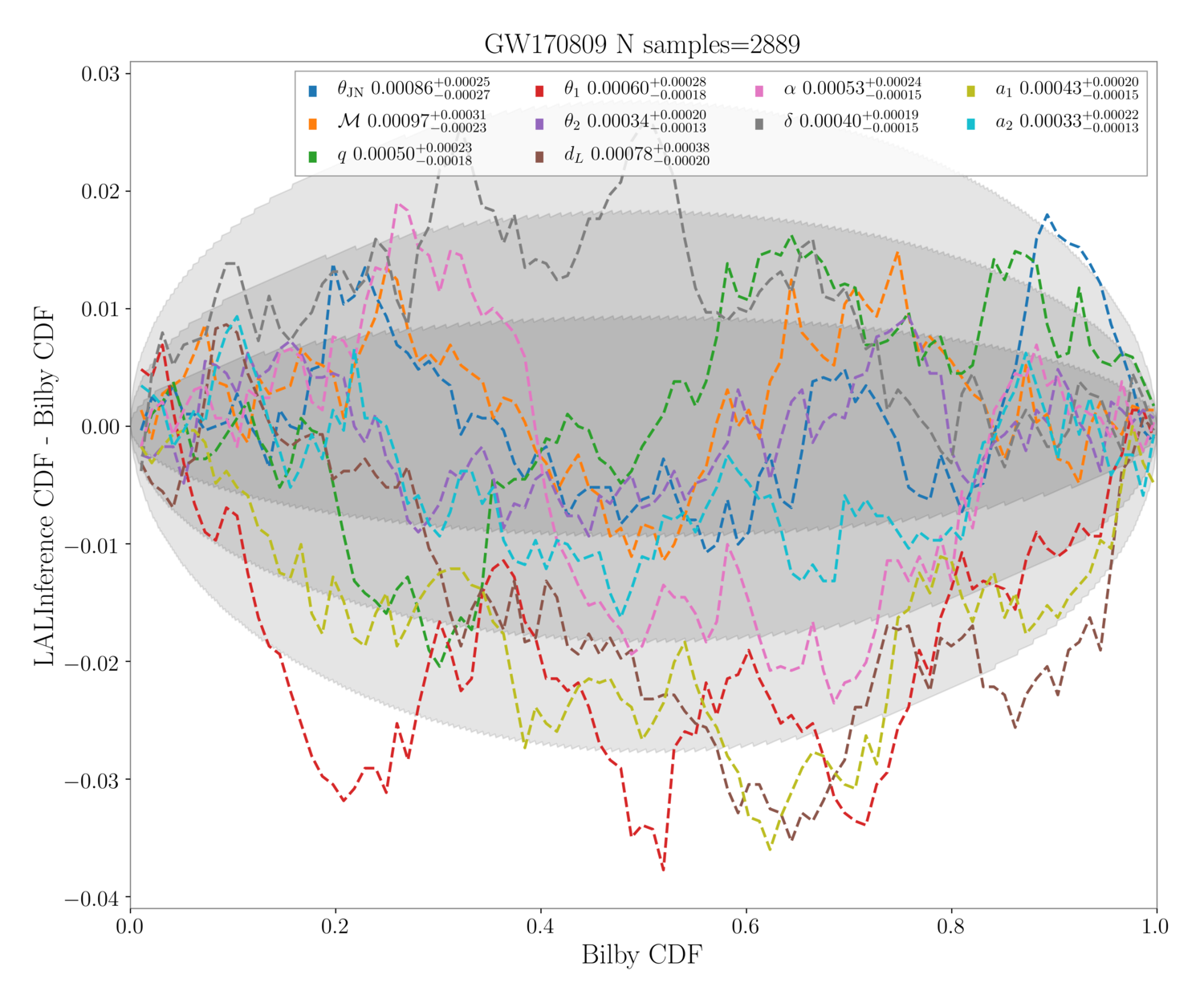}
    \includegraphics[width=0.8\linewidth]{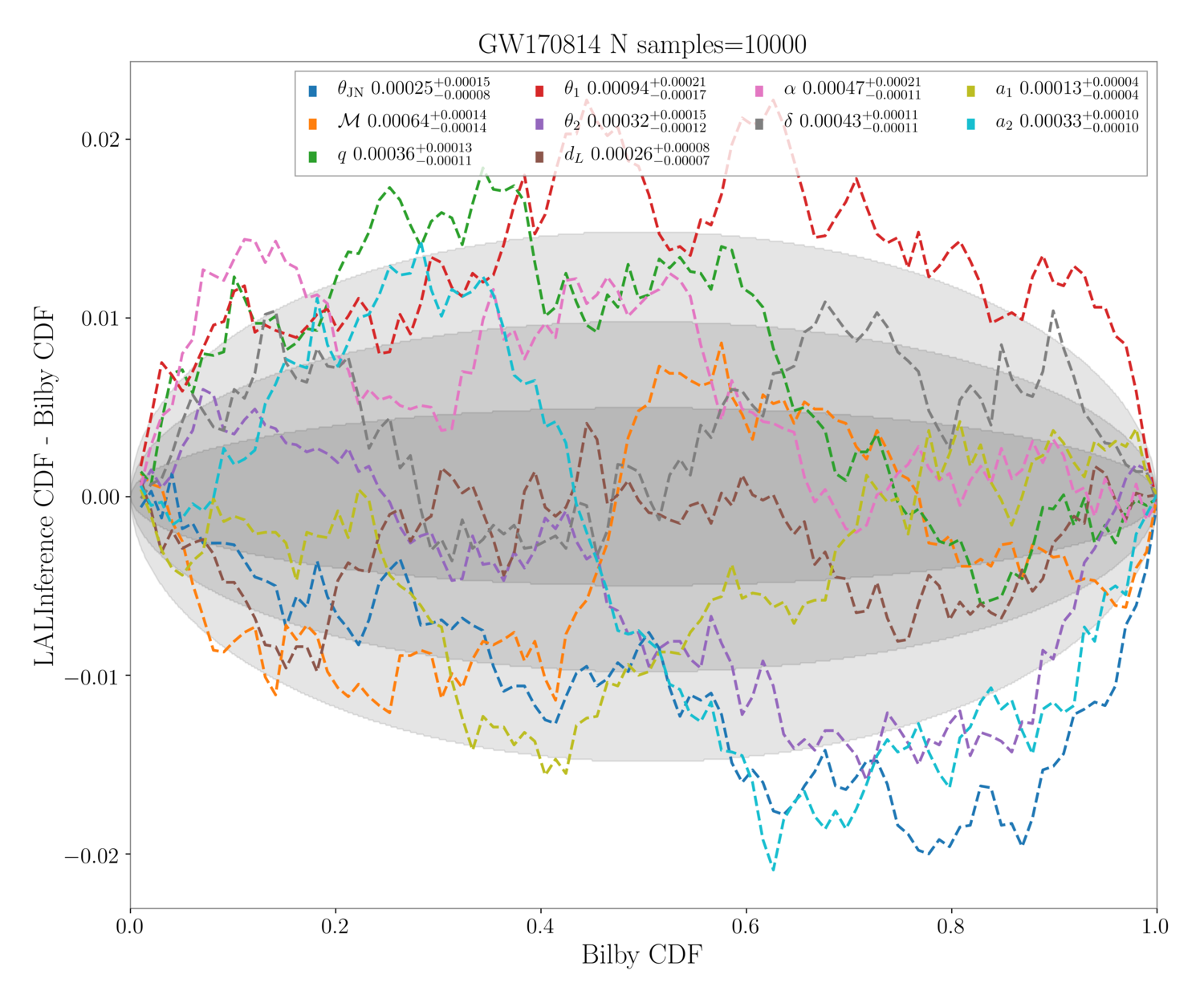}
    \caption{CDF comparison between \bilby and \lalinference for GW170809 and GW170814.}
\end{figure*}

\begin{figure*}
    \centering
    \includegraphics[width=0.8\linewidth]{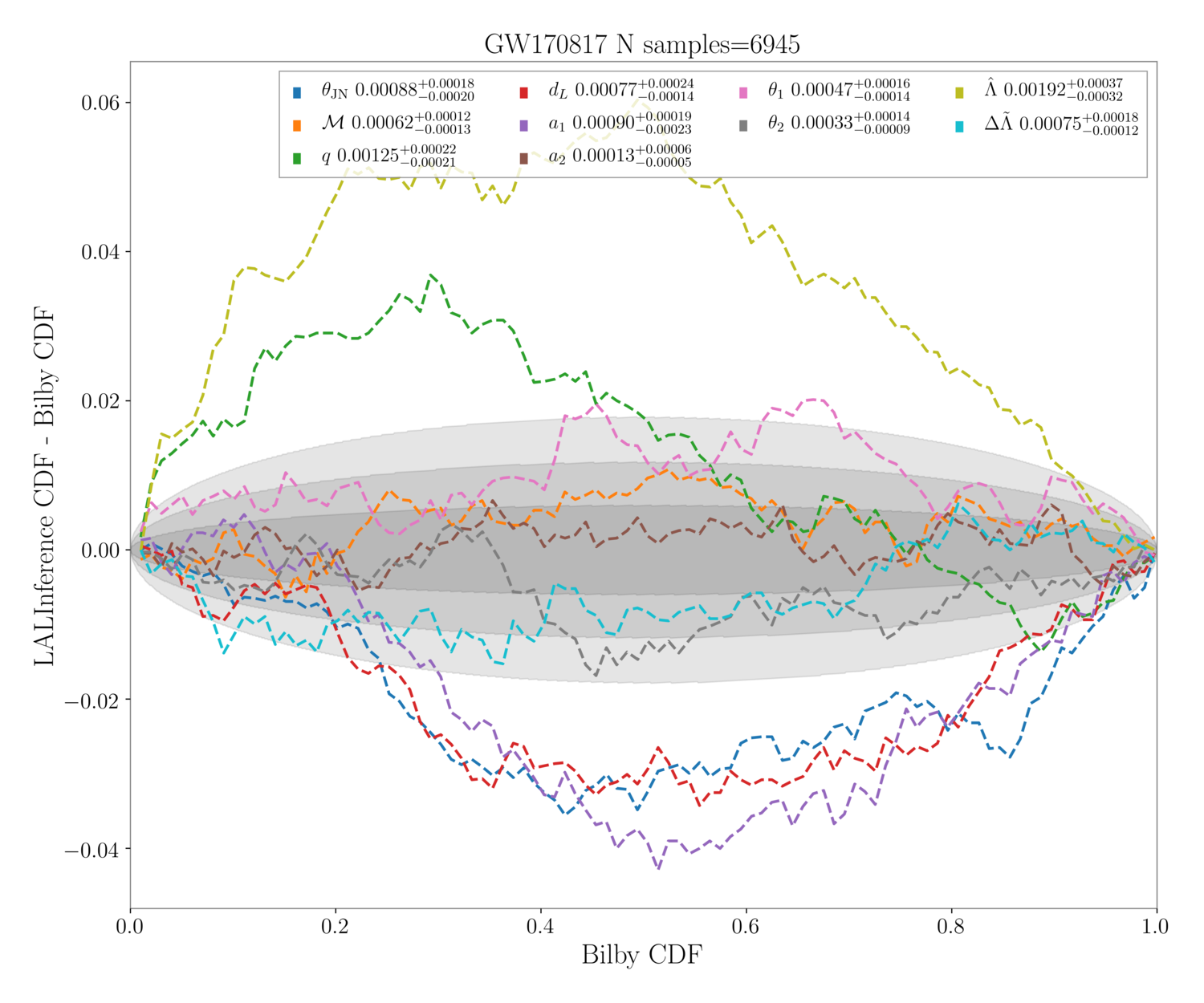}
    \includegraphics[width=0.8\linewidth]{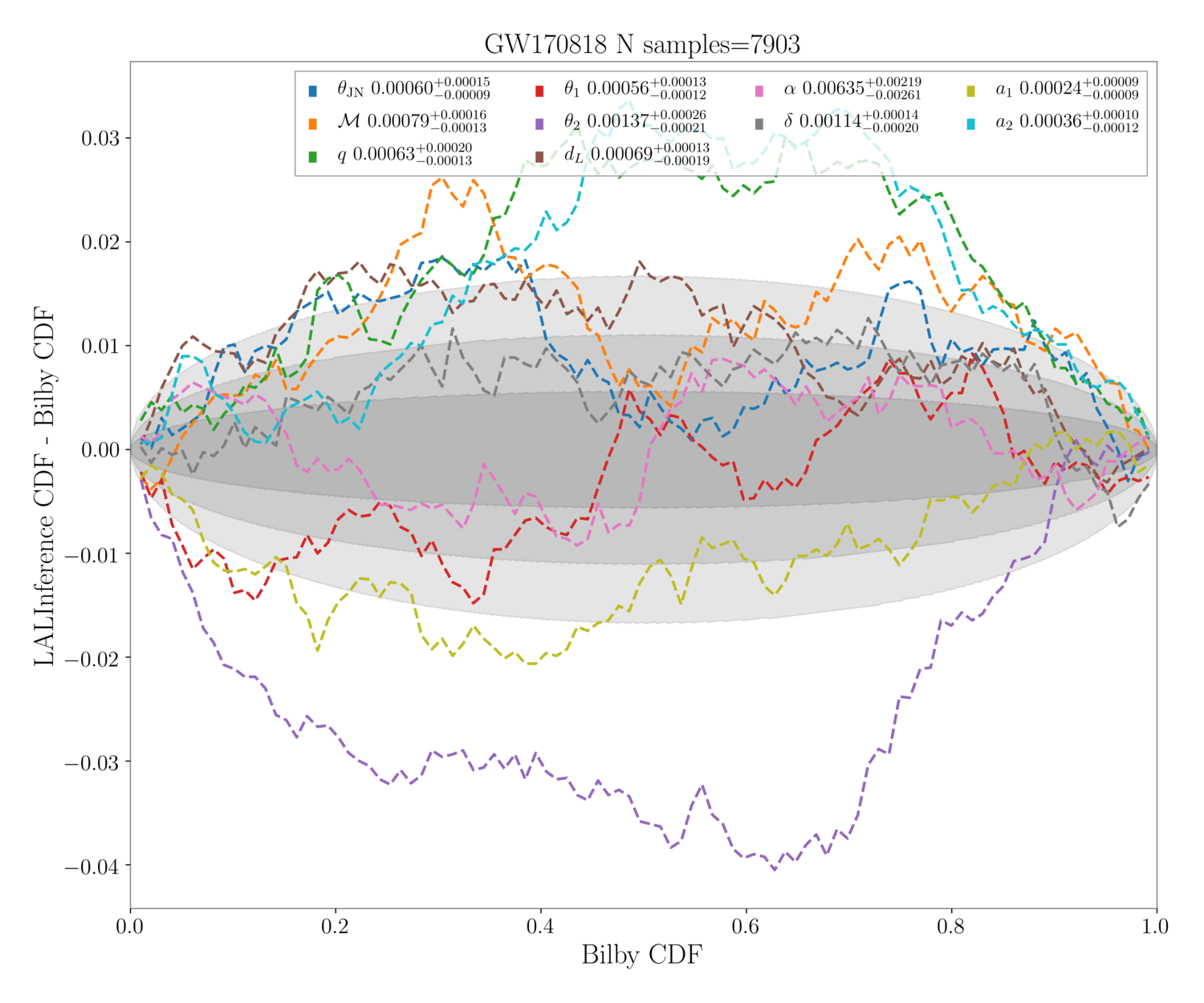}
    \caption{CDF comparison between \bilby and \lalinference for GW170817 and GW170818.}
\end{figure*}

\begin{figure*}
    \centering
    \includegraphics[width=0.8\linewidth]{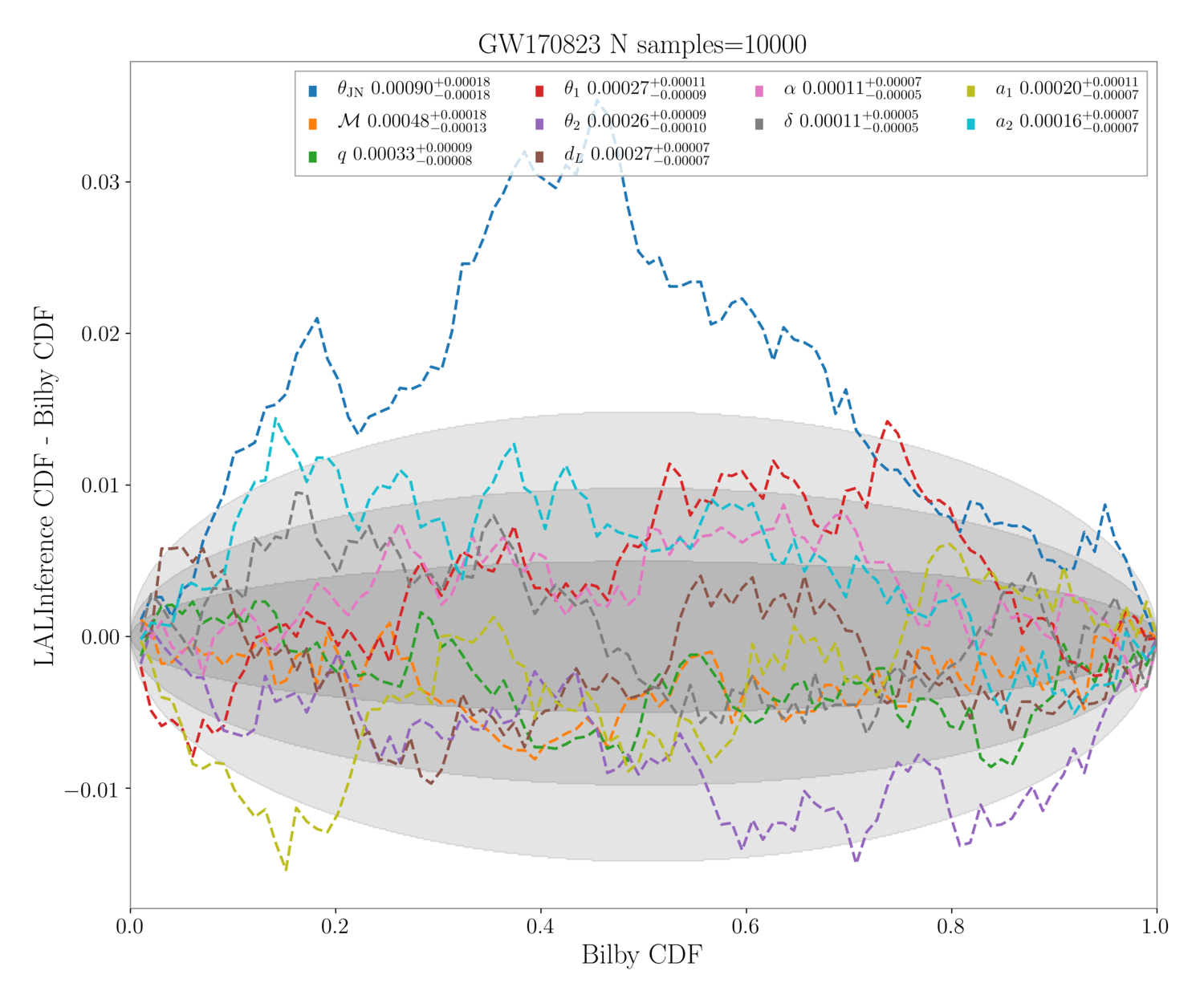}
    \caption{CDF comparison between \bilby and \lalinference for GW170823.}
\end{figure*}

\clearpage
\section{Parameter definitions}
\label{sec:definitions}
\bilby is able to sample in a range of different parameterisations of compact binaries.
In Table~\ref{tab:parameter-definitions}, we describe the definitions of these parameters as implemented in \bilby.
Unless otherwise specified all of these parameters can be sampled in, using the standard waveform model, likelihood, and conversion functions.

Currently, there is a relative lack of support for sampling parameters describing eccentric orbits: the eccentricity $e$ and the argument of periapsis $\omega$. This is because the frequency-domain eccentric waveforms available in \lalsimulation are less complete than their quasi-circular counterparts, containing only the inspiral section of the signal.

\begin{table*}
    \centering
    \caption{
    Definition of parameters typically considered for CBC inference.
    Subscript $i = 1, 2$ indicates whether the parameter pertains to the primary (1) or secondary (2) binary object. Subscript $k = x, y, z$ refers to a quantity measured in the $\hat{x}$, $\hat{y}$ or $\hat{z}$ direction; $\hat{z}$ points along the binary axis of rotation, while the $\hat{x}$, $\hat{y}$ directions are orthogonal to each other and $\hat{z}$, defined at reference phase $\phi$, and differ by phase offset $\phi_{12}$ between the two objects. Additional subscripts: 
    $^*$ - defined at a reference frequency,
    $^\dagger$ - parameter cannot be sampled, only generated in post-processing, 
    $^\times$ - parameter cannot yet be sampled or generated in post-processing. \label{tab:parameter-definitions}
    }
    \begin{tabular}{l p{10cm} c c}
    \hline
    Name & Description & \LaTeX{} label & Units \\
    \hline
    {\tt mass\_i} & Detector-frame (redshifted) mass of the $i$th object & $m_i$ & M$_\odot$ \\
    {\tt chirp\_mass} & Detector-frame chirp mass $\mathcal{M} = (m_1 m_2)^{3/5}/(m_1 + m_2)^{1/5}$ \citep{finn1993,poisson1995,blanchet1995} & $\mathcal{M}$ & M$_\odot$ \\
    {\tt total\_mass} & Detector-frame combined mass of the primary and secondary masses & $M$ & M$_\odot$ \\
    {\tt mass\_ratio} & The ratio of the secondary and primary masses $q = m_2 / m_1 \leq 1$ & $q$ & -- \\
    {\tt symmetric\_mass\_ratio} & A definition of mass ratio which is independent of the identity of the primary/secondary $\eta = q / (1 + q)^2$ & $\eta$ & -- \\
    {\tt mass\_i\_source} & Source-frame mass of the $i$th object $m^\mathrm{source}_i = m_i / (1 + z)$ \citep{krolak1987} & $m^\mathrm{source}_i$ & M$_\odot$ \\
    {\tt chirp\_mass\_source} & Source-frame chirp mass $\mathcal{M}^\mathrm{source} =\mathcal{M} / (1 + z)$ & $\mathcal{M}^\mathrm{source}$ & M$_\odot$ \\
    {\tt total\_mass\_source} & Source-frame total mass $M^\mathrm{source}= M / (1 + z)$ & $M^\mathrm{source}$ & M$_\odot$ \\
    {\tt a\_i} & Dimensionless spin magnitude of the $i$th object & $a_i$ & -- \\
    {\tt tilt\_i}$^*$ & Zenith angle between the spin and orbital angular momenta for the $i$th object & $\theta_i$ & rad \\
    {\tt cos\_tilt\_i}$^*$ & Cosine of the zenith angle between the spin and orbital angular momenta for the $i$th object & $\cos\theta_i$ & -- \\
    {\tt phi\_jl}$^*$ & Difference between total and orbital angular momentum azimuthal angles & $\phi_\mathrm{JL}$ & rad \\
    {\tt phi\_12}$^*$ & Difference between the azimuthal angles of the individual spin vector projections onto the orbital plane & $\phi_{12}$ & rad \\
    {\tt chi\_i}$^*$ (a.k.a.\ {\tt spin\_i\_z})& $i$th object aligned spin: projection of the $i$th object spin onto the orbital angular momentum $\chi_i =a_i\cos(\theta_i)$ & $\chi_i$ & -- \\
    {\tt chi\_i\_in\_plane}$^{*\dagger}$ & $i$th object in-plane spin: magnitude of the projection of the $i$th object spin onto the orbital plane $\chi^{\perp}_i=|a_i\sin(\theta_i)|$ & $\chi^{\perp}_i$ & -- \\
    {\tt chi\_eff}$^{*\dagger}$ & Effective inspiral spin parameter $\chi_\mathrm{eff}=(\chi_1 + q\chi_2)/(1 + q)$ \citep{santamaria2010,ajith2011} & $\chi_\mathrm{eff}$  & -- \\
    {\tt chi\_p}$^{*\dagger}$ & Effective precession spin parameter $\chi_{p} =\max\{\chi^\perp_1, q (3q + 4) / (4q + 3) \chi^\perp_2\}$ \citep{hannam2014,schmidt2015} & $\chi_{p}$ & -- \\
    {\tt spin\_i\_k}$^{*\dagger}$ & $k$th component of $i$th object spin in Euclidean coordinates & $S_{i, k}$ & -- \\
    {\tt lambda\_i} & Dimensionless tidal deformability of the $i$th object& $\Lambda_i$ & -- \\
    {\tt lambda\_tilde} & Combined dimensionless tidal deformability \citep{flanagan2008,favata2014} & $\tilde{\Lambda}$  & -- \\
    {\tt delta\_lambda\_tilde} & Relative difference in the combined tidal deformability  \citep{favata2014,wade2014} & $\delta\tilde{\Lambda}$ & -- \\
    {\tt eccentricity}$^{*(\dagger)}$ & Orbital eccentricity defined at a reference frequency & $e$ & -- \\
    {\tt argument\_of\_periapsis}$^*\times$ & The angle between the secondary mass and the ascending node of the orbit when the secondary mass is at periapsis & $\omega$ & rad \\
    {\tt ra} & Right ascension & $\alpha$ & rad \\
    {\tt dec} & Declination & $\delta$ & rad \\
    {\tt zenith} & Zenith angle in the detector-based sky parameterisation & $\kappa$ & rad \\
    {\tt azimuth} & Azimuthal angle in the detector-based sky parameterisation & $\epsilon$ & rad \\
    {\tt luminosity\_distance} & Luminosity distance to the source & $d_\mathrm{L}$ & Mpc \\
    {\tt comoving\_distance} & Comoving distance depending on specified cosmology & $d_\mathrm{C}$ & Mpc \\
    {\tt redshift} & Redshift depending on specified cosmology & $z$ & -- \\
    {\tt geocent\_time} & GPS reference time at the geocenter, typically merger time & $t_\mathrm{c}$ & $\mathrm{s}$ \\
    {\tt IFO\_time} & GPS reference time at the detector with name {\tt IFO}, e.g., {\tt H1\_time}, typically merger time & $t_\mathrm{IFO}$ & $\mathrm{s}$ \\
    {\tt time\_jitter} & Shift to apply for time array used in time marginalization & $\delta t$ & $\mathrm{s}$ \\
    {\tt psi} & Polarization angle of the source & $\psi$ & rad \\
    {\tt phase}$^*$ & Binary phase at a reference frequency & $\phi$ & rad \\
    {\tt theta\_jn} & Zenith angle between the total angular momentum and the line of sight & $\theta_{JN}$ & rad \\
    {\tt cos\_theta\_jn} & Cosine of the zenith angle between the total angular momentum and the line of sight & $\cos\theta_{JN}$ & -- \\
    {\tt iota}$^*$ & Zenith angle between the orbital angular momentum and the line of sight & $\iota$ & rad \\
    {\tt cos\_iota}$^*$ & Cosine of the zenith angle between the orbital angular momentum and the line of sight & $\cos\iota$ & -- \\
    \hline
    \end{tabular}
\end{table*}

\bibliographystyle{mnras}
\bibliography{BilbyGWTC}

\begin{thebibliography}{}
\makeatletter
\relax
\def\mn@urlcharsother{\let\do\@makeother \do\$\do\&\do\#\do\^\do\_\do\%\do\~}
\def\mn@doi{\begingroup\mn@urlcharsother \@ifnextchar [ {\mn@doi@}
  {\mn@doi@[]}}
\def\mn@doi@[#1]#2{\def\@tempa{#1}\ifx\@tempa\@empty \href
  {http://dx.doi.org/#2} {doi:#2}\else \href {http://dx.doi.org/#2} {#1}\fi
  \endgroup}
\def\mn@eprint#1#2{\mn@eprint@#1:#2::\@nil}
\def\mn@eprint@arXiv#1{\href {http://arxiv.org/abs/#1} {{\tt arXiv:#1}}}
\def\mn@eprint@dblp#1{\href {http://dblp.uni-trier.de/rec/bibtex/#1.xml}
  {dblp:#1}}
\def\mn@eprint@#1:#2:#3:#4\@nil{\def\@tempa {#1}\def\@tempb {#2}\def\@tempc
  {#3}\ifx \@tempc \@empty \let \@tempc \@tempb \let \@tempb \@tempa \fi \ifx
  \@tempb \@empty \def\@tempb {arXiv}\fi \@ifundefined
  {mn@eprint@\@tempb}{\@tempb:\@tempc}{\expandafter \expandafter \csname
  mn@eprint@\@tempb\endcsname \expandafter{\@tempc}}}

\bibitem[\protect\citeauthoryear{{Aasi}, {Abadie}, {Abbott}  et~al.}{{Aasi}
  et~al.}{2013}]{aasi13_big_dog}
{Aasi} J.,  {Abadie} J.,  {Abbott} B.~P.,   et~al., 2013, \mn@doi [\prd]
  {10.1103/PhysRevD.88.062001}, \href
  {https://ui.adsabs.harvard.edu/abs/2013PhRvD..88f2001A} {88, 062001}

\bibitem[\protect\citeauthoryear{{Aasi} et~al.}{{Aasi} et~al.}{2015}]{ligo}
{Aasi} J.,  et~al., 2015, \mn@doi [\cqg] {10.1088/0264-9381/32/7/074001}, \href
  {http://adsabs.harvard.edu/abs/2015CQGra..32g4001T} {32, 074001}

\bibitem[\protect\citeauthoryear{{Abbott}, {Abbott}, {Abbott}  et~al.}{{Abbott}
  et~al.}{2016a}]{abbott16_gw150914_detection}
{Abbott} B.~P.,  {Abbott} R.,  {Abbott} T.~D.,   et~al., 2016a, \mn@doi [\prl]
  {10.1103/PhysRevLett.116.061102}, \href
  {http://adsabs.harvard.edu/abs/2016PhRvL.116f1102A} {116, 061102}

\bibitem[\protect\citeauthoryear{{Abbott}, {Abbott}, {Abbott}  et~al.}{{Abbott}
  et~al.}{2016b}]{abbott16_gw150914_testingGR}
{Abbott} B.~P.,  {Abbott} R.,  {Abbott} T.~D.,   et~al., 2016b, \mn@doi [\prl]
  {10.1103/PhysRevLett.116.221101}, \href
  {http://adsabs.harvard.edu/abs/2016PhRvL.116v1101A} {116, 221101}

\bibitem[\protect\citeauthoryear{{Abbott}, {Abbott}, {Abbott}  et~al.}{{Abbott}
  et~al.}{2016c}]{abbott16_gw150914_pe}
{Abbott} B.~P.,  {Abbott} R.,  {Abbott} T.~D.,   et~al., 2016c, \mn@doi [\prl]
  {10.1103/PhysRevLett.116.241102}, \href
  {http://adsabs.harvard.edu/abs/2016PhRvL.116x1102A} {116, 241102}

\bibitem[\protect\citeauthoryear{{Abbott}, {Abbott}, {Abbott}  et~al.}{{Abbott}
  et~al.}{2016d}]{abbott16_gw150914_astro}
{Abbott} B.~P.,  {Abbott} R.,  {Abbott} T.~D.,   et~al., 2016d, \mn@doi [\apjl]
  {10.3847/2041-8205/818/2/L22}, \href
  {https://ui.adsabs.harvard.edu/abs/2016ApJ...818L..22A} {818, L22}

\bibitem[\protect\citeauthoryear{{Abbott}, {Abbott}, {Abbott}  et~al.}{{Abbott}
  et~al.}{2017a}]{abbott17_gw150914_systematics}
{Abbott} B.~P.,  {Abbott} R.,  {Abbott} T.~D.,   et~al., 2017a, \mn@doi
  [Classical and Quantum Gravity] {10.1088/1361-6382/aa6854}, \href
  {https://ui.adsabs.harvard.edu/abs/2017CQGra..34j4002A} {34, 104002}

\bibitem[\protect\citeauthoryear{{Abbott}, {Abbott}, {Abbott}  et~al.}{{Abbott}
  et~al.}{2017b}]{abbott17_gw170104_detection}
{Abbott} B.~P.,  {Abbott} R.,  {Abbott} T.~D.,   et~al., 2017b, \mn@doi [\prl]
  {10.1103/PhysRevLett.118.221101}, \href
  {https://ui.adsabs.harvard.edu/#abs/2017PhRvL.118v1101A} {118, 221101}

\bibitem[\protect\citeauthoryear{{Abbott}, {Abbott}, {Abbott}  et~al.}{{Abbott}
  et~al.}{2017c}]{abbott17_gw170817_detection}
{Abbott} B.~P.,  {Abbott} R.,  {Abbott} T.~D.,   et~al., 2017c, \mn@doi [\prl]
  {10.1103/PhysRevLett.119.161101}, \href
  {http://adsabs.harvard.edu/abs/2017PhRvL.119p1101A} {119, 161101}

\bibitem[\protect\citeauthoryear{{Abbott}, {Abbott}, {Abbott}  et~al.}{{Abbott}
  et~al.}{2017d}]{abbott17_gw170817_Hubble}
{Abbott} B.~P.,  {Abbott} R.,  {Abbott} T.~D.,   et~al., 2017d, \mn@doi [\nat]
  {10.1038/nature24471}, \href
  {http://adsabs.harvard.edu/abs/2017Natur.551...85A} {551, 85}

\bibitem[\protect\citeauthoryear{{Abbott}, {Abbott}, {Abbott}  et~al.}{{Abbott}
  et~al.}{2017e}]{abbott17_gw170817_multimessenger}
{Abbott} B.~P.,  {Abbott} R.,  {Abbott} T.~D.,   et~al., 2017e, \mn@doi [\apj]
  {10.3847/2041-8213/aa91c9}, \href
  {http://adsabs.harvard.edu/abs/2017ApJ...848L..12A} {848, L12}

\bibitem[\protect\citeauthoryear{{Abbott}, {Abbott}, {Abbott}  et~al.}{{Abbott}
  et~al.}{2017f}]{abbott17_gw170817_gwgrb}
{Abbott} B.~P.,  {Abbott} R.,  {Abbott} T.~D.,   et~al., 2017f, \mn@doi [\apj]
  {10.3847/2041-8213/aa920c}, \href
  {http://adsabs.harvard.edu/abs/2017ApJ...848L..13A} {848, L13}

\bibitem[\protect\citeauthoryear{{Abbott}, {Abbott}, {Abbott}  et~al.}{{Abbott}
  et~al.}{2017g}]{abbott17_gw170817_ejecta}
{Abbott} B.~P.,  {Abbott} R.,  {Abbott} T.~D.,   et~al., 2017g, \mn@doi [\apjl]
  {10.3847/2041-8213/aa9478}, \href
  {https://ui.adsabs.harvard.edu/abs/2017ApJ...850L..39A} {850, L39}

\bibitem[\protect\citeauthoryear{{Abbott}, {Abbott}, {Abbott}  et~al.}{{Abbott}
  et~al.}{2017h}]{abbott17_gw170817_progenitor}
{Abbott} B.~P.,  {Abbott} R.,  {Abbott} T.~D.,   et~al., 2017h, \mn@doi [\apjl]
  {10.3847/2041-8213/aa93fc}, \href
  {https://ui.adsabs.harvard.edu/abs/2017ApJ...850L..40A} {850, L40}

\bibitem[\protect\citeauthoryear{{Abbott}, {Abbott}, {Abbott}  et~al.}{{Abbott}
  et~al.}{2018a}]{GWTC1-samples}
{Abbott} B.~P.,  {Abbott} R.,  {Abbott} T.~D.,   et~al., 2018a,
  \url{https://dcc.ligo.org/LIGO-P1800370/public}

\bibitem[\protect\citeauthoryear{{Abbott}, {Abbott}, {Abbott}  et~al.}{{Abbott}
  et~al.}{2018b}]{abbott_19_observing_scenarios}
{Abbott} B.~P.,  {Abbott} R.,  {Abbott} T.~D.,   et~al., 2018b, \mn@doi [Living
  Reviews in Relativity] {10.1007/s41114-018-0012-9}, \href
  {https://ui.adsabs.harvard.edu/abs/2018LRR....21....3A} {21, 3}

\bibitem[\protect\citeauthoryear{{Abbott}, {Abbott}, {Abbott}  et~al.}{{Abbott}
  et~al.}{2018c}]{abbott18_GW170817_NS_parameters}
{Abbott} B.~P.,  {Abbott} R.,  {Abbott} T.~D.,   et~al., 2018c, \mn@doi [\prl]
  {10.1103/PhysRevLett.121.161101}, \href
  {https://ui.adsabs.harvard.edu/abs/2018PhRvL.121p1101A} {121, 161101}

\bibitem[\protect\citeauthoryear{{Abbott}, {Abbott}, {Abbott}  et~al.}{{Abbott}
  et~al.}{2019b}]{GWTC1-PSDS}
{Abbott} B.~P.,  {Abbott} R.,  {Abbott} T.~D.,   et~al., 2019b,
  \url{https://dcc.ligo.org/LIGO-P1900011/public}

\bibitem[\protect\citeauthoryear{{Abbott}, {Abbott}, {Abbott}  et~al.}{{Abbott}
  et~al.}{2019c}]{GWTC1-CAL}
{Abbott} B.~P.,  {Abbott} R.,  {Abbott} T.~D.,   et~al., 2019c,
  \url{https://dcc.ligo.org/LIGO-P1900040/public}

\bibitem[\protect\citeauthoryear{{Abbott}, {Abbott}, {Abbott}  et~al.}{{Abbott}
  et~al.}{2019a}]{ligo2019guide}
{Abbott} B.~P.,  {Abbott} R.,  {Abbott} T.~D.,   et~al., 2019a, arXiv preprint
  arXiv:1908.11170

\bibitem[\protect\citeauthoryear{{Abbott}, {Abbott}, {Abbott}  et~al.}{{Abbott}
  et~al.}{2019d}]{abbott19_O2_cosmo}
{Abbott} B.~P.,  {Abbott} R.,  {Abbott} T.~D.,   et~al., 2019d, arXiv e-prints,
  \href {https://ui.adsabs.harvard.edu/abs/2019arXiv190806060T} {p.
  arXiv:1908.06060}

\bibitem[\protect\citeauthoryear{{Abbott}, {Abbott}, {Abraham}
  et~al.}{{Abbott} et~al.}{2019e}]{abbott_19_gwosc}
{Abbott} R.,  {Abbott} T.~D.,  {Abraham} S.,   et~al., 2019e, arXiv e-prints,
  \href {https://ui.adsabs.harvard.edu/abs/2019arXiv191211716T} {p.
  arXiv:1912.11716}

\bibitem[\protect\citeauthoryear{{Abbott}, {Abbott}, {Abbott}  et~al.}{{Abbott}
  et~al.}{2019f}]{abbott2019_GWTC1}
{Abbott} B.~P.,  {Abbott} R.,  {Abbott} T.~D.,   et~al., 2019f, \mn@doi [\prx]
  {10.1103/PhysRevX.9.031040}, 9, 031040

\bibitem[\protect\citeauthoryear{{Abbott}, {Abbott}, {Abbott}  et~al.}{{Abbott}
  et~al.}{2019g}]{abbott18_GW170817_properties}
{Abbott} B.~P.,  {Abbott} R.,  {Abbott} T.~D.,   et~al., 2019g, \mn@doi [\prx]
  {10.1103/PhysRevX.9.011001}, \href
  {https://ui.adsabs.harvard.edu/abs/2019PhRvX...9a1001A} {9, 011001}

\bibitem[\protect\citeauthoryear{{Abbott}, {Abbott}, {Abbott}  et~al.}{{Abbott}
  et~al.}{2019h}]{abbott19_TGR}
{Abbott} B.~P.,  {Abbott} R.,  {Abbott} T.~D.,   et~al., 2019h, \mn@doi [\prd]
  {10.1103/PhysRevD.100.104036}, \href
  {https://ui.adsabs.harvard.edu/abs/2019PhRvD.100j4036A} {100, 104036}

\bibitem[\protect\citeauthoryear{{Abbott}, {Abbott}, {Abbott}  et~al.}{{Abbott}
  et~al.}{2019i}]{abbott19_O2_pops}
{Abbott} B.~P.,  {Abbott} R.,  {Abbott} T.~D.,   et~al., 2019i, \mn@doi [\apjl]
  {10.3847/2041-8213/ab3800}, \href
  {https://ui.adsabs.harvard.edu/abs/2019ApJ...882L..24A} {882, L24}

\bibitem[\protect\citeauthoryear{{Abbott}, {Abbott}, {Abbott}  et~al.}{{Abbott}
  et~al.}{2020a}]{abbott20_GW190412}
{Abbott} B.~P.,  {Abbott} R.,  {Abbott} T.~D.,   et~al., 2020a, arXiv e-prints,
  \href {https://ui.adsabs.harvard.edu/abs/2020arXiv200408342T} {p.
  arXiv:2004.08342}

\bibitem[\protect\citeauthoryear{{Abbott}, {Abbott}, {Abbott}  et~al.}{{Abbott}
  et~al.}{2020b}]{abbott19_EOS_model-select}
{Abbott} B.~P.,  {Abbott} R.,  {Abbott} T.~D.,   et~al., 2020b, \mn@doi
  [Classical and Quantum Gravity] {10.1088/1361-6382/ab5f7c}, \href
  {https://ui.adsabs.harvard.edu/abs/2020CQGra..37d5006A} {37, 045006}

\bibitem[\protect\citeauthoryear{Abbott, Abbott, Abraham, Acernese, Ackley
  et~al.}{Abbott et~al.}{2020c}]{abbott20_GW190521}
Abbott R.,  Abbott T.~D.,  Abraham S.,  Acernese F.,  Ackley K.,   et~al.,
  2020c, \mn@doi [Phys. Rev. Lett.] {10.1103/PhysRevLett.125.101102}, 125,
  101102

\bibitem[\protect\citeauthoryear{{Abbott}, {Abbott}, {Abbott}  et~al.}{{Abbott}
  et~al.}{2020d}]{abbott20_GW190425}
{Abbott} B.~P.,  {Abbott} R.,  {Abbott} T.~D.,   et~al., 2020d, \mn@doi [\apjl]
  {10.3847/2041-8213/ab75f5}, \href
  {https://ui.adsabs.harvard.edu/abs/2020ApJ...892L...3A} {892, L3}

\bibitem[\protect\citeauthoryear{{Acernese} et~al.}{{Acernese}
  et~al.}{2015}]{virgo}
{Acernese} F.,  et~al., 2015, \mn@doi [Classical Quantum Gravity]
  {10.1088/0264-9381/32/2/024001}, \href
  {http://adsabs.harvard.edu/abs/2015CQGra..32b4001A} {32, 024001}

\bibitem[\protect\citeauthoryear{Ade et~al.,}{Ade et~al.}{2016}]{ade2016}
Ade P.~A.,  et~al., 2016, Astronomy \& Astrophysics, 594, A13

\bibitem[\protect\citeauthoryear{{Ajith} et~al.,}{{Ajith}
  et~al.}{2011}]{ajith2011}
{Ajith} P.,  et~al., 2011, \mn@doi [\prl] {10.1103/PhysRevLett.106.241101},
  \href {https://ui.adsabs.harvard.edu/abs/2011PhRvL.106x1101A} {106, 241101}

\bibitem[\protect\citeauthoryear{{Allen}, {Anderson}, {Brady}, {Brown}  \&
  {Creighton}}{{Allen} et~al.}{2012}]{findchirp}
{Allen} B.,  {Anderson} W.~G.,  {Brady} P.~R.,  {Brown} D.~A.,   {Creighton} J.
  D.~E.,  2012, \mn@doi [\prd] {10.1103/PhysRevD.85.122006}, \href
  {https://ui.adsabs.harvard.edu/abs/2012PhRvD..85l2006A} {85, 122006}

\bibitem[\protect\citeauthoryear{{Ashton} \& {Khan}}{{Ashton} \&
  {Khan}}{2020}]{ashton2019b}
{Ashton} G.,  {Khan} S.,  2020, \mn@doi [\prd] {10.1103/PhysRevD.101.064037},
  \href {https://ui.adsabs.harvard.edu/abs/2020PhRvD.101f4037A} {101, 064037}

\bibitem[\protect\citeauthoryear{{Ashton} et~al.,}{{Ashton}
  et~al.}{2019}]{ashton19}
{Ashton} G.,  et~al., 2019, \apjs, 241, 27

\bibitem[\protect\citeauthoryear{{Babak} et~al.,}{{Babak}
  et~al.}{2008}]{Babak08}
{Babak} S.,  et~al., 2008, \mn@doi [Classical and Quantum Gravity]
  {10.1088/0264-9381/25/18/184026}, \href
  {https://ui.adsabs.harvard.edu/abs/2008CQGra..25r4026B} {25, 184026}

\bibitem[\protect\citeauthoryear{{Babak} et~al.,}{{Babak}
  et~al.}{2010}]{Babak10}
{Babak} S.,  et~al., 2010, \mn@doi [Classical and Quantum Gravity]
  {10.1088/0264-9381/27/8/084009}, \href
  {https://ui.adsabs.harvard.edu/abs/2010CQGra..27h4009B} {27, 084009}

\bibitem[\protect\citeauthoryear{Barrett, Gaebel, Neijssel, Vigna-G{\'{o}}mez,
  Stevenson, Berry, Farr  \& Mandel}{Barrett et~al.}{2018}]{Barrett2018}
Barrett J.~W.,  Gaebel S.~M.,  Neijssel C.~J.,  Vigna-G{\'{o}}mez A.,
  Stevenson S.,  Berry C. P.~L.,  Farr W.~M.,   Mandel I.,  2018, \mn@doi [Mon.
  Not. R. Astron. Soc.] {10.1093/mnras/sty908}, 477, 4685

\bibitem[\protect\citeauthoryear{{Bavera} et~al.,}{{Bavera}
  et~al.}{2020}]{bavera19}
{Bavera} S.~S.,  et~al., 2020, \mn@doi [\aap] {10.1051/0004-6361/201936204},
  \href {https://ui.adsabs.harvard.edu/abs/2020A&A...635A..97B} {635, A97}

\bibitem[\protect\citeauthoryear{{Bayes}}{{Bayes}}{1763}]{Bayes}
{Bayes} T.,  1763, \mn@doi [Philosophical Transactions of The Royal Society]
  {10.1098/rstl.1763.0053}

\bibitem[\protect\citeauthoryear{Baylor, Smith  \& Chase}{Baylor
  et~al.}{2019}]{amanda_baylor_2019_3478659}
Baylor A.,  Smith R.,   Chase E.,  2019,
  IMRPhenomPv2\_NRTidal\_GW190425\_narrow\_Mc, \mn@doi{10.5281/zenodo.3478659},
  \url {https://doi.org/10.5281/zenodo.3478659}

\bibitem[\protect\citeauthoryear{{Belczynski} et~al.,}{{Belczynski}
  et~al.}{2018}]{belczynski18}
{Belczynski} K.,  et~al., 2018, arXiv e-prints, \href
  {https://ui.adsabs.harvard.edu/abs/2018arXiv181210065B} {p. arXiv:1812.10065}

\bibitem[\protect\citeauthoryear{{Berry} et~al.,}{{Berry}
  et~al.}{2015}]{berry15}
{Berry} C. P.~L.,  et~al., 2015, \mn@doi [\apj] {10.1088/0004-637X/804/2/114},
  \href {https://ui.adsabs.harvard.edu/abs/2015ApJ...804..114B} {804, 114}

\bibitem[\protect\citeauthoryear{{Biscoveanu}, {Vitale}  \&
  {Haster}}{{Biscoveanu} et~al.}{2019}]{biscoveanu2019b}
{Biscoveanu} S.,  {Vitale} S.,   {Haster} C.-J.,  2019, \mn@doi [\apjl]
  {10.3847/2041-8213/ab479e}, \href
  {https://ui.adsabs.harvard.edu/abs/2019ApJ...884L..32B} {884, L32}

\bibitem[\protect\citeauthoryear{{Biscoveanu}, {Haster}, {Vitale}  \&
  {Davies}}{{Biscoveanu} et~al.}{2020a}]{biscoveanu2020}
{Biscoveanu} S.,  {Haster} C.-J.,  {Vitale} S.,   {Davies} J.,  2020a, \mn@doi
  [\prd] {10.1103/PhysRevD.102.023008}, \href
  {https://ui.adsabs.harvard.edu/abs/2020PhRvD.102b3008B} {102, 023008}

\bibitem[\protect\citeauthoryear{{Biscoveanu}, {Thrane}  \&
  {Vitale}}{{Biscoveanu} et~al.}{2020b}]{biscoveanu2019a}
{Biscoveanu} S.,  {Thrane} E.,   {Vitale} S.,  2020b, \mn@doi [\apj]
  {10.3847/1538-4357/ab7eaf}, \href
  {https://ui.adsabs.harvard.edu/abs/2020ApJ...893...38B} {893, 38}

\bibitem[\protect\citeauthoryear{{Biwer}, {Capano}, {De}, {Cabero}, {Brown},
  {Nitz}  \& {Raymond}}{{Biwer} et~al.}{2019}]{biwer18}
{Biwer} C.~M.,  {Capano} C.~D.,  {De} S.,  {Cabero} M.,  {Brown} D.~A.,  {Nitz}
  A.~H.,   {Raymond} V.,  2019, \mn@doi [\pasp] {10.1088/1538-3873/aaef0b},
  \href {https://ui.adsabs.harvard.edu/abs/2019PASP..131b4503B} {131, 024503}

\bibitem[\protect\citeauthoryear{Blackman et~al.,}{Blackman
  et~al.}{2017}]{PhysRevD.96.024058}
Blackman J.,  et~al., 2017, \mn@doi [Phys. Rev. D]
  {10.1103/PhysRevD.96.024058}, 96, 024058

\bibitem[\protect\citeauthoryear{{Blanchet}, {Damour}, {Iyer}, {Will}  \&
  {Wiseman}}{{Blanchet} et~al.}{1995}]{blanchet1995}
{Blanchet} L.,  {Damour} T.,  {Iyer} B.~R.,  {Will} C.~M.,   {Wiseman} A.~G.,
  1995, \mn@doi [\prl] {10.1103/PhysRevLett.74.3515}, \href
  {https://ui.adsabs.harvard.edu/abs/1995PhRvL..74.3515B} {74, 3515}

\bibitem[\protect\citeauthoryear{Boh{\'e}, Hannam, Husa, Ohme, Puerrer  \&
  Schmidt}{Boh{\'e} et~al.}{2016}]{Bohe:PhenomPv2}
Boh{\'e} A.,  Hannam M.,  Husa S.,  Ohme F.,  Puerrer M.,   Schmidt P.,  2016,
  Technical Report {LIGO}-T1500602, PhenomPv2 - Technical Notes for LAL
  Implementation, \url {https://dcc.ligo.org/LIGO-T1500602}.
{LIGO} Project, \url {https://dcc.ligo.org/LIGO-T1500602}

\bibitem[\protect\citeauthoryear{Boh\'e et~al.,}{Boh\'e
  et~al.}{2017}]{PhysRevD.95.044028}
Boh\'e A.,  et~al., 2017, \mn@doi [Phys. Rev. D] {10.1103/PhysRevD.95.044028},
  95, 044028

\bibitem[\protect\citeauthoryear{Buchner}{Buchner}{2016}]{ultranest1}
Buchner J.,  2016, \mn@doi [Statistics and Computing]
  {10.1007/s11222-014-9512-y}, 26, 383–392

\bibitem[\protect\citeauthoryear{Buchner}{Buchner}{2019}]{ultranest2}
Buchner J.,  2019, \mn@doi [Publications of the Astronomical Society of the
  Pacific] {10.1088/1538-3873/aae7fc}, 131, 108005

\bibitem[\protect\citeauthoryear{{Buchner} et~al.}{{Buchner}
  et~al.}{2014}]{pymultinest}
{Buchner} J.,  et~al., 2014, \mn@doi [\aap] {10.1051/0004-6361/201322971},
  \href {http://adsabs.harvard.edu/abs/2014A%26A...564A.125B} {564, A125}

\bibitem[\protect\citeauthoryear{{Cahillane} et~al.,}{{Cahillane}
  et~al.}{2017}]{cahillane17}
{Cahillane} C.,  et~al., 2017, \mn@doi [\prd] {10.1103/PhysRevD.96.102001},
  \href {https://ui.adsabs.harvard.edu/abs/2017PhRvD..96j2001C} {96, 102001}

\bibitem[\protect\citeauthoryear{{Cantiello} et~al.,}{{Cantiello}
  et~al.}{2018}]{cantiello18}
{Cantiello} M.,  et~al., 2018, \mn@doi [\apjl] {10.3847/2041-8213/aaad64},
  \href {https://ui.adsabs.harvard.edu/abs/2018ApJ...854L..31C} {854, L31}

\bibitem[\protect\citeauthoryear{{Chatziioannou}, {Haster}, {Littenberg},
  {Farr}, {Ghonge}, {Millhouse}, {Clark}  \& {Cornish}}{{Chatziioannou}
  et~al.}{2019}]{chatziioannou2019noise}
{Chatziioannou} K.,  {Haster} C.-J.,  {Littenberg} T.~B.,  {Farr} W.~M.,
  {Ghonge} S.,  {Millhouse} M.,  {Clark} J.~A.,   {Cornish} N.,  2019, \mn@doi
  [\prd] {10.1103/PhysRevD.100.104004}, \href
  {https://ui.adsabs.harvard.edu/abs/2019PhRvD.100j4004C} {100, 104004}

\bibitem[\protect\citeauthoryear{{Chen}, {Fishbach}  \& {Holz}}{{Chen}
  et~al.}{2018}]{chen18}
{Chen} H.-Y.,  {Fishbach} M.,   {Holz} D.~E.,  2018, \mn@doi [\nat]
  {10.1038/s41586-018-0606-0}, \href
  {https://ui.adsabs.harvard.edu/abs/2018Natur.562..545C} {562, 545}

\bibitem[\protect\citeauthoryear{{Chornock} et~al.,}{{Chornock}
  et~al.}{2017}]{chornock17}
{Chornock} R.,  et~al., 2017, \mn@doi [\apjl] {10.3847/2041-8213/aa905c}, \href
  {https://ui.adsabs.harvard.edu/abs/2017ApJ...848L..19C} {848, L19}

\bibitem[\protect\citeauthoryear{{Christensen} \& {Meyer}}{{Christensen} \&
  {Meyer}}{1998}]{christensen98}
{Christensen} N.,  {Meyer} R.,  1998, \mn@doi [\prd]
  {10.1103/PhysRevD.58.082001}, \href
  {https://ui.adsabs.harvard.edu/abs/1998PhRvD..58h2001C} {58, 082001}

\bibitem[\protect\citeauthoryear{{Christensen} \& {Meyer}}{{Christensen} \&
  {Meyer}}{2001}]{christensen01}
{Christensen} N.,  {Meyer} R.,  2001, \mn@doi [\prd]
  {10.1103/PhysRevD.64.022001}, \href
  {https://ui.adsabs.harvard.edu/abs/2001PhRvD..64b2001C} {64, 022001}

\bibitem[\protect\citeauthoryear{Cook, Gelman  \& Rubin}{Cook
  et~al.}{2006}]{cook06}
Cook S.~R.,  Gelman A.,   Rubin D.~B.,  2006, \mn@doi [Journal of Computational
  and Graphical Statistics] {10.1198/106186006X136976}, 15, 675

\bibitem[\protect\citeauthoryear{{Cornish}}{{Cornish}}{2010}]{Cornish10}
{Cornish} N.~J.,  2010, arXiv e-prints, \href
  {https://ui.adsabs.harvard.edu/abs/2010arXiv1007.4820C} {p. arXiv:1007.4820}

\bibitem[\protect\citeauthoryear{Cornish \& Littenberg}{Cornish \&
  Littenberg}{2015}]{Cornish14}
Cornish N.~J.,  Littenberg T.~B.,  2015, \mn@doi [\cqg]
  {10.1088/0264-9381/32/13/135012}, 32, 135012

\bibitem[\protect\citeauthoryear{{Cornish} \& {Shuman}}{{Cornish} \&
  {Shuman}}{2020}]{Cornish20}
{Cornish} N.~J.,  {Shuman} K.,  2020, arXiv e-prints, \href
  {https://ui.adsabs.harvard.edu/abs/2020arXiv200503610C} {p. arXiv:2005.03610}

\bibitem[\protect\citeauthoryear{{Coughlin} \& {Dietrich}}{{Coughlin} \&
  {Dietrich}}{2019}]{coughlin19}
{Coughlin} M.~W.,  {Dietrich} T.,  2019, \mn@doi [\prd]
  {10.1103/PhysRevD.100.043011}, \href
  {https://ui.adsabs.harvard.edu/abs/2019PhRvD.100d3011C} {100, 043011}

\bibitem[\protect\citeauthoryear{{Del Pozzo}, {Berry}, {Ghosh}, {Haines},
  {Singer}  \& {Vecchio}}{{Del Pozzo} et~al.}{2018}]{delpozzo18}
{Del Pozzo} W.,  {Berry} C.~P.~L.,  {Ghosh} A.,  {Haines} T.~S.~F.,  {Singer}
  L.~P.,   {Vecchio} A.,  2018, \mn@doi [\mnras] {10.1093/mnras/sty1485}, \href
  {https://ui.adsabs.harvard.edu/abs/2018MNRAS.479..601D} {479, 601}

\bibitem[\protect\citeauthoryear{{Dhawan}, {Bulla}, {Goobar}, {Sagu{\'e}s
  Carracedo}  \& {Setzer}}{{Dhawan} et~al.}{2020}]{dhawan19}
{Dhawan} S.,  {Bulla} M.,  {Goobar} A.,  {Sagu{\'e}s Carracedo} A.,   {Setzer}
  C.~N.,  2020, \mn@doi [\apj] {10.3847/1538-4357/ab5799}, \href
  {https://ui.adsabs.harvard.edu/abs/2020ApJ...888...67D} {888, 67}

\bibitem[\protect\citeauthoryear{{Dietrich}, {Samajdar}, {Khan},
  {Johnson-McDaniel}, {Dudi}  \& {Tichy}}{{Dietrich} et~al.}{2019}]{Dietrich19}
{Dietrich} T.,  {Samajdar} A.,  {Khan} S.,  {Johnson-McDaniel} N.~K.,  {Dudi}
  R.,   {Tichy} W.,  2019, \mn@doi [\prd] {10.1103/PhysRevD.100.044003}, \href
  {https://ui.adsabs.harvard.edu/abs/2019PhRvD.100d4003D} {100, 044003}

\bibitem[\protect\citeauthoryear{Ellis \& van Haasteren}{Ellis \& van
  Haasteren}{2017}]{ptmcmc}
Ellis J.,  van Haasteren R.,  2017, jellis18/PTMCMCSampler: Official Release,
  \mn@doi{10.5281/zenodo.1037579}, \url
  {https://doi.org/10.5281/zenodo.1037579}

\bibitem[\protect\citeauthoryear{{Essick}, {Landry}  \& {Holz}}{{Essick}
  et~al.}{2020}]{essick19}
{Essick} R.,  {Landry} P.,   {Holz} D.~E.,  2020, \mn@doi [\prd]
  {10.1103/PhysRevD.101.063007}, \href
  {https://ui.adsabs.harvard.edu/abs/2020PhRvD.101f3007E} {101, 063007}

\bibitem[\protect\citeauthoryear{{Farah} et~al.,}{{Farah}
  et~al.}{2019}]{farah19}
{Farah} W.,  et~al., 2019, \mn@doi [\mnras] {10.1093/mnras/stz1748}, \href
  {https://ui.adsabs.harvard.edu/abs/2019MNRAS.488.2989F} {488, 2989}

\bibitem[\protect\citeauthoryear{{Farr}}{{Farr}}{2014}]{farr14}
{Farr} W.~M.,  2014, Technical Report LIGO-T1400460, Marginalisation of the
  time and phase parameters in CBC parameter estimation, \url
  {https://dcc.ligo.org/LIGO-T1400460/public}.
\url {https://dcc.ligo.org/LIGO-T1400460/public}

\bibitem[\protect\citeauthoryear{{Farr} \& {Farr}}{{Farr} \&
  {Farr}}{2015}]{kombine}
{Farr} B.,  {Farr} W.~M.,  2015, kombine: a kernel-density-based,
  embarrassingly parallel ensemble sampler

\bibitem[\protect\citeauthoryear{{Farr}, Farr  \& Littenberg}{{Farr}
  et~al.}{2014}]{farr14b}
{Farr} W.~M.,  Farr B.,   Littenberg T.,  2014, Technical Report LIGO-T1400682,
  Modelling calibration errors in CBC waveforms, \url
  {https://dcc.ligo.org/LIGO-T1400682/public}.
\url {https://dcc.ligo.org/LIGO-T1400682/public}

\bibitem[\protect\citeauthoryear{{Favata}}{{Favata}}{2014}]{favata2014}
{Favata} M.,  2014, \mn@doi [\prl] {10.1103/PhysRevLett.112.101101}, \href
  {https://ui.adsabs.harvard.edu/abs/2014PhRvL.112j1101F} {112, 101101}

\bibitem[\protect\citeauthoryear{{Feroz} \& {Hobson}}{{Feroz} \&
  {Hobson}}{2008}]{multinest1}
{Feroz} F.,  {Hobson} M.~P.,  2008, \mn@doi [\mnras]
  {10.1111/j.1365-2966.2007.12353.x}, \href
  {http://adsabs.harvard.edu/abs/2008MNRAS.384..449F} {384, 449}

\bibitem[\protect\citeauthoryear{{Feroz}, {Hobson}  \& {Bridges}}{{Feroz}
  et~al.}{2009}]{multinest2}
{Feroz} F.,  {Hobson} M.~P.,   {Bridges} M.,  2009, \mn@doi [\mnras]
  {10.1111/j.1365-2966.2009.14548.x}, \href
  {http://adsabs.harvard.edu/abs/2009MNRAS.398.1601F} {398, 1601}

\bibitem[\protect\citeauthoryear{{Feroz}, {Hobson}, {Cameron}  \&
  {Pettitt}}{{Feroz} et~al.}{2019}]{multinest3}
{Feroz} F.,  {Hobson} M.~P.,  {Cameron} E.,   {Pettitt} A.~N.,  2019, \mn@doi
  [The Open Journal of Astrophysics] {10.21105/astro.1306.2144}, \href
  {https://ui.adsabs.harvard.edu/abs/2019OJAp....2E..10F} {2, 10}

\bibitem[\protect\citeauthoryear{Finn}{Finn}{1992}]{finn92}
Finn L.~S.,  1992, \mn@doi [\prd] {10.1103/PhysRevD.46.5236}, 46, 5236

\bibitem[\protect\citeauthoryear{{Finn} \& {Chernoff}}{{Finn} \&
  {Chernoff}}{1993}]{finn1993}
{Finn} L.~S.,  {Chernoff} D.~F.,  1993, \mn@doi [\prd]
  {10.1103/PhysRevD.47.2198}, \href
  {https://ui.adsabs.harvard.edu/abs/1993PhRvD..47.2198F} {47, 2198}

\bibitem[\protect\citeauthoryear{{Fishbach} \& {Holz}}{{Fishbach} \&
  {Holz}}{2017}]{Fishbach17b}
{Fishbach} M.,  {Holz} D.~E.,  2017, \mn@doi [\apj] {10.3847/2041-8213/aa9bf6},
  \href {http://adsabs.harvard.edu/abs/2017ApJ...851L..25F} {851, L25}

\bibitem[\protect\citeauthoryear{{Flanagan} \& {Hinderer}}{{Flanagan} \&
  {Hinderer}}{2008}]{flanagan2008}
{Flanagan} {\'E}.~{\'E}.,  {Hinderer} T.,  2008, \mn@doi [\prd]
  {10.1103/PhysRevD.77.021502}, \href
  {https://ui.adsabs.harvard.edu/abs/2008PhRvD..77b1502F} {77, 021502}

\bibitem[\protect\citeauthoryear{{Fong} et~al.,}{{Fong} et~al.}{2019}]{fong19}
{Fong} W.,  et~al., 2019, \mn@doi [\apjl] {10.3847/2041-8213/ab3d9e}, \href
  {https://ui.adsabs.harvard.edu/abs/2019ApJ...883L...1F} {883, L1}

\bibitem[\protect\citeauthoryear{{Gabbard}, {Messenger}, {Heng}, {Tonolini}  \&
  {Murray-Smith}}{{Gabbard} et~al.}{2019}]{gabbard19}
{Gabbard} H.,  {Messenger} C.,  {Heng} I.~S.,  {Tonolini} F.,   {Murray-Smith}
  R.,  2019, arXiv e-prints, \href
  {https://ui.adsabs.harvard.edu/abs/2019arXiv190906296G} {p. arXiv:1909.06296}

\bibitem[\protect\citeauthoryear{{Galaudage}, {Talbot}  \&
  {Thrane}}{{Galaudage} et~al.}{2019}]{galaudage19}
{Galaudage} S.,  {Talbot} C.,   {Thrane} E.,  2019, arXiv e-prints, \href
  {https://ui.adsabs.harvard.edu/abs/2019arXiv191209708G} {p. arXiv:1912.09708}

\bibitem[\protect\citeauthoryear{{George} \& {Huerta}}{{George} \&
  {Huerta}}{2018}]{george18}
{George} D.,  {Huerta} E.~A.,  2018, \mn@doi [Physics Letters B]
  {10.1016/j.physletb.2017.12.053}, \href
  {https://ui.adsabs.harvard.edu/abs/2018PhLB..778...64G} {778, 64}

\bibitem[\protect\citeauthoryear{{Goncharov}, {Zhu}  \& {Thrane}}{{Goncharov}
  et~al.}{2019}]{goncharov19}
{Goncharov} B.,  {Zhu} X.-J.,   {Thrane} E.,  2019, arXiv e-prints, \href
  {https://ui.adsabs.harvard.edu/abs/2019arXiv191005961G} {p. arXiv:1910.05961}

\bibitem[\protect\citeauthoryear{{G{\'o}rski} \& {et al.}}{{G{\'o}rski} \& {et
  al.}}{1999}]{healpix}
{G{\'o}rski} K.~M.,  {et al.} 1999, in {Banday} A.~J.,  {Sheth} R.~K.,   {da
  Costa} L.~N.,  eds, Evolution of Large Scale Structure : From Recombination
  to Garching. p.~37 (\mn@eprint {arXiv} {astro-ph/9812350})

\bibitem[\protect\citeauthoryear{{G{\'o}rski}, {Hivon}, {Banday}, {Wandelt},
  {Hansen}, {Reinecke}  \& {Bartelmann}}{{G{\'o}rski} et~al.}{2005}]{healpy}
{G{\'o}rski} K.~M.,  {Hivon} E.,  {Banday} A.~J.,  {Wandelt} B.~D.,  {Hansen}
  F.~K.,  {Reinecke} M.,   {Bartelmann} M.,  2005, \mn@doi [\apj]
  {10.1086/427976}, 622, 759

\bibitem[\protect\citeauthoryear{Handley, Hobson  \& Lasenby}{Handley
  et~al.}{2015a}]{polychord1}
Handley W.~J.,  Hobson M.~P.,   Lasenby A.~N.,  2015a, \mn@doi [Monthly Notices
  of the Royal Astronomical Society: Letters] {10.1093/mnrasl/slv047}, 450, L61

\bibitem[\protect\citeauthoryear{Handley, Hobson  \& Lasenby}{Handley
  et~al.}{2015b}]{polychord2}
Handley W.~J.,  Hobson M.~P.,   Lasenby A.~N.,  2015b, \mn@doi [Monthly Notices
  of the Royal Astronomical Society] {10.1093/mnras/stv1911}, 453, 4384

\bibitem[\protect\citeauthoryear{{Hannam}, {Schmidt}, {Boh{\'e}}, {Haegel},
  {Husa}, {Ohme}, {Pratten}  \& {P{\"u}rrer}}{{Hannam}
  et~al.}{2014}]{hannam2014}
{Hannam} M.,  {Schmidt} P.,  {Boh{\'e}} A.,  {Haegel} L.,  {Husa} S.,  {Ohme}
  F.,  {Pratten} G.,   {P{\"u}rrer} M.,  2014, \mn@doi [\prl]
  {10.1103/PhysRevLett.113.151101}, \href
  {https://ui.adsabs.harvard.edu/abs/2014PhRvL.113o1101H} {113, 151101}

\bibitem[\protect\citeauthoryear{Hastings}{Hastings}{1970}]{hastings1970monte}
Hastings W.~K.,  1970, Biometrika, 57, 97

\bibitem[\protect\citeauthoryear{{Hernandez Vivanco}, {Smith}, {Thrane}  \&
  {Lasky}}{{Hernandez Vivanco} et~al.}{2019a}]{hernandezvivanco2019b}
{Hernandez Vivanco} F.,  {Smith} R.,  {Thrane} E.,   {Lasky} P.~D.,  2019a,
  \mn@doi [\prd] {10.1103/PhysRevD.100.043023}, \href
  {https://ui.adsabs.harvard.edu/abs/2019PhRvD.100d3023H} {100, 043023}

\bibitem[\protect\citeauthoryear{{Hernandez Vivanco}, {Smith}, {Thrane},
  {Lasky}, {Talbot}  \& {Raymond}}{{Hernandez Vivanco}
  et~al.}{2019b}]{hernandezvivanco2019}
{Hernandez Vivanco} F.,  {Smith} R.,  {Thrane} E.,  {Lasky} P.~D.,  {Talbot}
  C.,   {Raymond} V.,  2019b, \mn@doi [\prd] {10.1103/PhysRevD.100.103009},
  \href {https://ui.adsabs.harvard.edu/abs/2019PhRvD.100j3009H} {100, 103009}

\bibitem[\protect\citeauthoryear{Hogg \& Foreman-Mackey}{Hogg \&
  Foreman-Mackey}{2018}]{hogg18}
Hogg D.~W.,  Foreman-Mackey D.,  2018, \mn@doi [The Astrophysical Journal
  Supplement Series] {10.3847/1538-4365/aab76e}, 236, 11

\bibitem[\protect\citeauthoryear{{Hotokezaka}, {Nakar}, {Gottlieb}, {Nissanke},
  {Masuda}, {Hallinan}, {Mooley}  \& {Deller}}{{Hotokezaka}
  et~al.}{2019}]{hotokezaka19}
{Hotokezaka} K.,  {Nakar} E.,  {Gottlieb} O.,  {Nissanke} S.,  {Masuda} K.,
  {Hallinan} G.,  {Mooley} K.~P.,   {Deller} A.~T.,  2019, \mn@doi [\natast]
  {10.1038/s41550-019-0820-1}, \href
  {https://ui.adsabs.harvard.edu/abs/2019NatAs...3..940H} {3, 940}

\bibitem[\protect\citeauthoryear{{Hoy} \& {Raymond}}{{Hoy} \&
  {Raymond}}{2020}]{pesummary}
{Hoy} C.,  {Raymond} V.,  2020, arXiv e-prints, \href
  {https://ui.adsabs.harvard.edu/abs/2020arXiv200606639H} {p. arXiv:2006.06639}

\bibitem[\protect\citeauthoryear{{H{\"u}bner}, {Talbot}, {Lasky}  \&
  {Thrane}}{{H{\"u}bner} et~al.}{2020}]{huebner19}
{H{\"u}bner} M.,  {Talbot} C.,  {Lasky} P.~D.,   {Thrane} E.,  2020, \mn@doi
  [\prd] {10.1103/PhysRevD.101.023011}, \href
  {https://ui.adsabs.harvard.edu/abs/2020PhRvD.101b3011H} {101, 023011}

\bibitem[\protect\citeauthoryear{{Isi}, {Giesler}, {Farr}, {Scheel}  \&
  {Teukolsky}}{{Isi} et~al.}{2019}]{isi19}
{Isi} M.,  {Giesler} M.,  {Farr} W.~M.,  {Scheel} M.~A.,   {Teukolsky} S.~A.,
  2019, \mn@doi [\prl] {10.1103/PhysRevLett.123.111102}, \href
  {https://ui.adsabs.harvard.edu/abs/2019PhRvL.123k1102I} {123, 111102}

\bibitem[\protect\citeauthoryear{{Kasliwal} et~al.,}{{Kasliwal}
  et~al.}{2019}]{kasliwal19}
{Kasliwal} M.~M.,  et~al., 2019, \mn@doi [\mnras] {10.1093/mnrasl/slz007},
  \href {https://ui.adsabs.harvard.edu/abs/2019MNRAS.tmpL..14K} {p.~L14}

\bibitem[\protect\citeauthoryear{{Keitel}}{{Keitel}}{2019}]{keitel19}
{Keitel} D.,  2019, \mn@doi [Research Notes of the American Astronomical
  Society] {10.3847/2515-5172/ab0c0b}, \href
  {https://ui.adsabs.harvard.edu/abs/2019RNAAS...3...46K} {3, 46}

\bibitem[\protect\citeauthoryear{{Khan}, {Husa}, {Hannam}, {Ohme},
  {P{\"u}rrer}, {Forteza}  \& {Boh{\'e}}}{{Khan} et~al.}{2016}]{khan16}
{Khan} S.,  {Husa} S.,  {Hannam} M.,  {Ohme} F.,  {P{\"u}rrer} M.,  {Forteza}
  X.~J.,   {Boh{\'e}} A.,  2016, \mn@doi [\prd] {10.1103/PhysRevD.93.044007},
  \href {https://ui.adsabs.harvard.edu/abs/2016PhRvD..93d4007K} {93, 044007}

\bibitem[\protect\citeauthoryear{{Kimball}, {Talbot}, {Berry}, {Carney},
  {Zevin}, {Thrane}  \& {Kalogera}}{{Kimball} et~al.}{2020}]{Kimball20}
{Kimball} C.,  {Talbot} C.,  {Berry} C. P.~L.,  {Carney} M.,  {Zevin} M.,
  {Thrane} E.,   {Kalogera} V.,  2020, arXiv e-prints, \href
  {https://ui.adsabs.harvard.edu/abs/2020arXiv200500023K} {p. arXiv:2005.00023}

\bibitem[\protect\citeauthoryear{Kolmogorov}{Kolmogorov}{1933}]{kolmogorov1933}
Kolmogorov A.~N.,  1933, G. Ist. Ital. Attuari., 4, 83–91

\bibitem[\protect\citeauthoryear{{Krolak} \& {Schutz}}{{Krolak} \&
  {Schutz}}{1987}]{krolak1987}
{Krolak} A.,  {Schutz} B.~F.,  1987, \mn@doi [General Relativity and
  Gravitation] {10.1007/BF00759095}, \href
  {https://ui.adsabs.harvard.edu/abs/1987GReGr..19.1163K} {19, 1163}

\bibitem[\protect\citeauthoryear{Kullback \& Leibler}{Kullback \&
  Leibler}{1951}]{kullback1951}
Kullback S.,  Leibler R.~A.,  1951, \mn@doi [Ann. Math. Statist.]
  {10.1214/aoms/1177729694}, 22, 79

\bibitem[\protect\citeauthoryear{{LIGO Scientific Collaboration}}{{LIGO
  Scientific Collaboration}}{2018}]{LALSuite}
{LIGO Scientific Collaboration} 2018, {LIGO} {A}lgorithm {L}ibrary -
  {LALS}uite, free software (GPL), \mn@doi{10.7935/GT1W-FZ16}

\bibitem[\protect\citeauthoryear{{Lackey}, {P{\"u}rrer}, {Taracchini}  \&
  {Marsat}}{{Lackey} et~al.}{2019}]{Lackey19}
{Lackey} B.~D.,  {P{\"u}rrer} M.,  {Taracchini} A.,   {Marsat} S.,  2019,
  \mn@doi [\prd] {10.1103/PhysRevD.100.024002}, \href
  {https://ui.adsabs.harvard.edu/abs/2019PhRvD.100b4002L} {100, 024002}

\bibitem[\protect\citeauthoryear{Lange, O'Shaughnessy  \& Rizzo}{Lange
  et~al.}{2018}]{lange2018}
Lange J.,  O'Shaughnessy R.,   Rizzo M.,  2018, arXiv:1805.10457

\bibitem[\protect\citeauthoryear{{Lentati}, {Alexander}, {Hobson}, {Feroz},
  {van Haasteren}, {Lee}  \& {Shannon}}{{Lentati} et~al.}{2014}]{temponest}
{Lentati} L.,  {Alexander} P.,  {Hobson} M.~P.,  {Feroz} F.,  {van Haasteren}
  R.,  {Lee} K.~J.,   {Shannon} R.~M.,  2014, \mn@doi [\mnras]
  {10.1093/mnras/stt2122}, \href
  {https://ui.adsabs.harvard.edu/abs/2014MNRAS.437.3004L} {437, 3004}

\bibitem[\protect\citeauthoryear{{Lin}}{{Lin}}{1991}]{61115}
{Lin} J.,  1991, \mn@doi [IEEE Transactions on Information Theory]
  {10.1109/18.61115}, 37, 145

\bibitem[\protect\citeauthoryear{Littenberg \& Cornish}{Littenberg \&
  Cornish}{2015}]{Littenberg14}
Littenberg T.~B.,  Cornish N.~J.,  2015, \mn@doi [Phys. Rev. D]
  {10.1103/PhysRevD.91.084034}, 91, 084034

\bibitem[\protect\citeauthoryear{{Lower}, {Thrane}, {Lasky}  \&
  {Smith}}{{Lower} et~al.}{2018}]{lower18}
{Lower} M.~E.,  {Thrane} E.,  {Lasky} P.~D.,   {Smith} R.,  2018, \mn@doi
  [\prd] {10.1103/PhysRevD.98.083028}, 98, 083028

\bibitem[\protect\citeauthoryear{Macleod, Coughlin, Urban, Massinger
  et~al.}{Macleod et~al.}{2018}]{gwpy}
Macleod D.,  Coughlin S.,  Urban A.~L.,  Massinger T.,   et~al., 2018,
  gwpy/gwpy: 0.12.0, \mn@doi{10.5281/zenodo.1346349}

\bibitem[\protect\citeauthoryear{{Margutti} et~al.,}{{Margutti}
  et~al.}{2018}]{margutti18}
{Margutti} R.,  et~al., 2018, \mn@doi [\apjl] {10.3847/2041-8213/aab2ad}, \href
  {https://ui.adsabs.harvard.edu/abs/2018ApJ...856L..18M} {856, L18}

\bibitem[\protect\citeauthoryear{{Marsat}, {Baker}  \& {Dal Canton}}{{Marsat}
  et~al.}{2020}]{Marsat20}
{Marsat} S.,  {Baker} J.~G.,   {Dal Canton} T.,  2020, arXiv e-prints, \href
  {https://ui.adsabs.harvard.edu/abs/2020arXiv200300357M} {p. arXiv:2003.00357}

\bibitem[\protect\citeauthoryear{Metropolis, Rosenbluth, Rosenbluth, Teller  \&
  Teller}{Metropolis et~al.}{1953}]{metropolis1953equation}
Metropolis N.,  Rosenbluth A.~W.,  Rosenbluth M.~N.,  Teller A.~H.,   Teller
  E.,  1953, The journal of chemical physics, 21, 1087

\bibitem[\protect\citeauthoryear{{Mooley} et~al.,}{{Mooley}
  et~al.}{2018}]{mooley18}
{Mooley} K.~P.,  et~al., 2018, \mn@doi [\nat] {10.1038/s41586-018-0486-3},
  \href {https://ui.adsabs.harvard.edu/abs/2018Natur.561..355M} {561, 355}

\bibitem[\protect\citeauthoryear{{Most}, {Weih}, {Rezzolla}  \&
  {Schaffner-Bielich}}{{Most} et~al.}{2018}]{most18}
{Most} E.~R.,  {Weih} L.~R.,  {Rezzolla} L.,   {Schaffner-Bielich} J.,  2018,
  \mn@doi [\prl] {10.1103/PhysRevLett.120.261103}, \href
  {https://ui.adsabs.harvard.edu/abs/2018PhRvL.120z1103M} {120, 261103}

\bibitem[\protect\citeauthoryear{{Nagar} et~al.,}{{Nagar}
  et~al.}{2018}]{Nagar18}
{Nagar} A.,  et~al., 2018, \mn@doi [\prd] {10.1103/PhysRevD.98.104052}, \href
  {https://ui.adsabs.harvard.edu/abs/2018PhRvD..98j4052N} {98, 104052}

\bibitem[\protect\citeauthoryear{{Ossokine} et~al.,}{{Ossokine}
  et~al.}{2020}]{Ossokine20}
{Ossokine} S.,  et~al., 2020, arXiv e-prints, \href
  {https://ui.adsabs.harvard.edu/abs/2020arXiv200409442O} {p. arXiv:2004.09442}

\bibitem[\protect\citeauthoryear{{Pankow}, {Brady}, {Ochsner}  \&
  {O'Shaughnessy}}{{Pankow} et~al.}{2015}]{pankow15}
{Pankow} C.,  {Brady} P.,  {Ochsner} E.,   {O'Shaughnessy} R.,  2015, \mn@doi
  [\prd] {10.1103/PhysRevD.92.023002}, \href
  {https://ui.adsabs.harvard.edu/abs/2015PhRvD..92b3002P} {92, 023002}

\bibitem[\protect\citeauthoryear{{Pankow} et~al.,}{{Pankow}
  et~al.}{2018}]{pankow18}
{Pankow} C.,  et~al., 2018, \mn@doi [\prd] {10.1103/PhysRevD.98.084016}, \href
  {https://ui.adsabs.harvard.edu/abs/2018PhRvD..98h4016P} {98, 084016}

\bibitem[\protect\citeauthoryear{{Payne}, {Talbot}  \& {Thrane}}{{Payne}
  et~al.}{2019}]{payne2019}
{Payne} E.,  {Talbot} C.,   {Thrane} E.,  2019, \mn@doi [\prd]
  {10.1103/PhysRevD.100.123017}, \href
  {https://ui.adsabs.harvard.edu/abs/2019PhRvD.100l3017P} {100, 123017}

\bibitem[\protect\citeauthoryear{{Poisson} \& {Will}}{{Poisson} \&
  {Will}}{1995}]{poisson1995}
{Poisson} E.,  {Will} C.~M.,  1995, \mn@doi [\prd] {10.1103/PhysRevD.52.848},
  \href {https://ui.adsabs.harvard.edu/abs/1995PhRvD..52..848P} {52, 848}

\bibitem[\protect\citeauthoryear{{Powell}}{{Powell}}{2018}]{powell18_glitch}
{Powell} J.,  2018, \mn@doi [\cqg] {10.1088/1361-6382/aacf18}, \href
  {http://adsabs.harvard.edu/abs/2018CQGra..35o5017P} {35, 155017}

\bibitem[\protect\citeauthoryear{{Powell} \& {M{\"u}ller}}{{Powell} \&
  {M{\"u}ller}}{2019}]{powell19}
{Powell} J.,  {M{\"u}ller} B.,  2019, \mn@doi [\mnras] {10.1093/mnras/stz1304},
  \href {https://ui.adsabs.harvard.edu/abs/2019MNRAS.487.1178P} {487, 1178}

\bibitem[\protect\citeauthoryear{{Price-Whelan} et~al.}{{Price-Whelan}
  et~al.}{2018}]{astropy2}
{Price-Whelan} A.~M.,  et~al., 2018, \mn@doi [Astron. J.]
  {10.3847/1538-3881/aabc4f}, \href
  {https://ui.adsabs.harvard.edu/#abs/2018AJ....156..123T} {156, 123}

\bibitem[\protect\citeauthoryear{{Ramos-Buades}, {Husa}, {Pratten},
  {Estell{\'e}s}, {Garc{\'\i}a-Quir{\'o}s}, {Mateu-Lucena}, {Colleoni}  \&
  {Jaume}}{{Ramos-Buades} et~al.}{2020}]{ramos-buades2019}
{Ramos-Buades} A.,  {Husa} S.,  {Pratten} G.,  {Estell{\'e}s} H.,
  {Garc{\'\i}a-Quir{\'o}s} C.,  {Mateu-Lucena} M.,  {Colleoni} M.,   {Jaume}
  R.,  2020, \mn@doi [\prd] {10.1103/PhysRevD.101.083015}, \href
  {https://ui.adsabs.harvard.edu/abs/2020PhRvD.101h3015R} {101, 083015}

\bibitem[\protect\citeauthoryear{{Robitaille} et~al.}{{Robitaille}
  et~al.}{2013}]{astropy1}
{Robitaille} T.~P.,  et~al., 2013, \mn@doi [Astron. Astrophys.]
  {10.1051/0004-6361/201322068}, \href
  {http://adsabs.harvard.edu/abs/2013A%26A...558A..33A} {558, A33}

\bibitem[\protect\citeauthoryear{Romano \& Cornish}{Romano \&
  Cornish}{2017}]{romano17}
Romano J.~D.,  Cornish N.~J.,  2017, \mn@doi [Living Reviews in Relativity]
  {10.1007/s41114-017-0004-1}, 20, 2

\bibitem[\protect\citeauthoryear{{Romero-Shaw}, {Lasky}  \&
  {Thrane}}{{Romero-Shaw} et~al.}{2019}]{romero-shaw19}
{Romero-Shaw} I.~M.,  {Lasky} P.~D.,   {Thrane} E.,  2019, \mn@doi [\mnras]
  {10.1093/mnras/stz2996}, \href
  {https://ui.adsabs.harvard.edu/abs/2019MNRAS.tmp.2600R} {p.~2600}

\bibitem[\protect\citeauthoryear{Romero-Shaw et~al.}{Romero-Shaw
  et~al.}{2020b}]{Bilby-GWTC-1-Analysis-and-Verification}
Romero-Shaw I.,  et~al., 2020b, \url{https://doi.org/10.5281/zenodo.4017046},
  \mn@doi{10.5281/zenodo.4017046}

\bibitem[\protect\citeauthoryear{Romero-Shaw et~al.}{Romero-Shaw
  et~al.}{2020c}]{Bilby_samples}
Romero-Shaw I.,  et~al., 2020c, \url{https://dcc.ligo.org/LIGO-P2000193/public}

\bibitem[\protect\citeauthoryear{{Romero-Shaw}, {Farrow}, {Stevenson}, {Thrane}
   \& {Zhu}}{{Romero-Shaw} et~al.}{2020a}]{romero-shaw20}
{Romero-Shaw} I.~M.,  {Farrow} N.,  {Stevenson} S.,  {Thrane} E.,   {Zhu}
  X.-J.,  2020a, \mn@doi [\mnras] {10.1093/mnrasl/slaa084}, \href
  {https://ui.adsabs.harvard.edu/abs/2020MNRAS.tmpL..78R} {}

\bibitem[\protect\citeauthoryear{{R{\"o}ver}, {Meyer}  \&
  {Christensen}}{{R{\"o}ver} et~al.}{2006}]{rover06}
{R{\"o}ver} C.,  {Meyer} R.,   {Christensen} N.,  2006, \mn@doi [\cqg]
  {10.1088/0264-9381/23/15/009}, \href
  {https://ui.adsabs.harvard.edu/abs/2006CQGra..23.4895R} {23, 4895}

\bibitem[\protect\citeauthoryear{{R{\"o}ver}, {Meyer}  \&
  {Christensen}}{{R{\"o}ver} et~al.}{2007}]{rover07}
{R{\"o}ver} C.,  {Meyer} R.,   {Christensen} N.,  2007, \mn@doi [\prd]
  {10.1103/PhysRevD.75.062004}, \href
  {https://ui.adsabs.harvard.edu/abs/2007PhRvD..75f2004R} {75, 062004}

\bibitem[\protect\citeauthoryear{{Santamar{\'\i}a} et~al.,}{{Santamar{\'\i}a}
  et~al.}{2010}]{santamaria2010}
{Santamar{\'\i}a} L.,  et~al., 2010, \mn@doi [\prd]
  {10.1103/PhysRevD.82.064016}, \href
  {https://ui.adsabs.harvard.edu/abs/2010PhRvD..82f4016S} {82, 064016}

\bibitem[\protect\citeauthoryear{{Sarin}, {Lasky}  \& {Ashton}}{{Sarin}
  et~al.}{2020}]{sarin20}
{Sarin} N.,  {Lasky} P.~D.,   {Ashton} G.,  2020, \mn@doi [\prd]
  {10.1103/PhysRevD.101.063021}, \href
  {https://ui.adsabs.harvard.edu/abs/2020PhRvD.101f3021S} {101, 063021}

\bibitem[\protect\citeauthoryear{{Schmidt}, {Hannam}  \& {Husa}}{{Schmidt}
  et~al.}{2012}]{IMRPhenomP}
{Schmidt} P.,  {Hannam} M.,   {Husa} S.,  2012, \prd, 86, 104063

\bibitem[\protect\citeauthoryear{{Schmidt}, {Ohme}  \& {Hannam}}{{Schmidt}
  et~al.}{2015}]{schmidt2015}
{Schmidt} P.,  {Ohme} F.,   {Hannam} M.,  2015, \mn@doi [\prd]
  {10.1103/PhysRevD.91.024043}, \href
  {https://ui.adsabs.harvard.edu/abs/2015PhRvD..91b4043S} {91, 024043}

\bibitem[\protect\citeauthoryear{{Sidery} et~al.,}{{Sidery}
  et~al.}{2014}]{sidery14}
{Sidery} T.,  et~al., 2014, \mn@doi [\prd] {10.1103/PhysRevD.89.084060}, \href
  {https://ui.adsabs.harvard.edu/abs/2014PhRvD..89h4060S} {89, 084060}

\bibitem[\protect\citeauthoryear{{Singer} \& {Price}}{{Singer} \&
  {Price}}{2016}]{singer16a}
{Singer} L.~P.,  {Price} L.~R.,  2016, \mn@doi [\prd]
  {10.1103/PhysRevD.93.024013}, 93, 024013

\bibitem[\protect\citeauthoryear{Singer et~al.}{Singer
  et~al.}{2014}]{Singer:2014qca}
Singer L.~P.,  et~al., 2014, \mn@doi [Astrophys. J.]
  {10.1088/0004-637X/795/2/105}, 795, 105

\bibitem[\protect\citeauthoryear{{Singer} et~al.}{{Singer}
  et~al.}{2016}]{singer16b}
{Singer} L.~P.,  et~al., 2016, \mn@doi [\apj] {10.3847/2041-8205/829/1/L15},
  829, L15

\bibitem[\protect\citeauthoryear{Singer et~al.}{Singer et~al.}{2020}]{GWCelery}
Singer L.~P.,  et~al., 2020, {GWCelery},
  \url{https://git.ligo.org/emfollow/gwcelery}

\bibitem[\protect\citeauthoryear{Skilling}{Skilling}{2006}]{Skilling06}
Skilling J.,  2006, \mn@doi [Bayesian Analysis] {10.1214/06-BA127}, 1, 833

\bibitem[\protect\citeauthoryear{Smirnov}{Smirnov}{1948}]{smirnov1948}
Smirnov N.,  1948, \mn@doi [Ann. Math. Statist.] {10.1214/aoms/1177730256}, 19,
  279

\bibitem[\protect\citeauthoryear{Smith, Field, Blackburn, Haster, P{\"{u}}rrer,
  Raymond  \& Schmidt}{Smith et~al.}{2016}]{smith16}
Smith R.,  Field S.~E.,  Blackburn K.,  Haster C.-J.,  P{\"{u}}rrer M.,
  Raymond V.,   Schmidt P.,  2016, \mn@doi [\prd] {10.1103/PhysRevD.94.044031},
  94

\bibitem[\protect\citeauthoryear{{Smith}, {Ashton}, {Vajpeyi}  \&
  {Talbot}}{{Smith} et~al.}{2019}]{Smith:2019ucc}
{Smith} R.,  {Ashton} G.,  {Vajpeyi} A.,   {Talbot} C.,  2019, arXiv e-prints,
  \href {https://ui.adsabs.harvard.edu/abs/2019arXiv190911873S} {p.
  arXiv:1909.11873}

\bibitem[\protect\citeauthoryear{{Sokal}}{{Sokal}}{1994}]{sokal94}
{Sokal} A.~D.,  1994, arXiv e-prints, \href
  {https://ui.adsabs.harvard.edu/abs/1994hep.lat...5016S} {pp hep--lat/9405016}

\bibitem[\protect\citeauthoryear{{Speagle}}{{Speagle}}{2020}]{dynesty}
{Speagle} J.~S.,  2020, \mn@doi [\mnras] {10.1093/mnras/staa278}, \href
  {https://ui.adsabs.harvard.edu/abs/2020MNRAS.493.3132S} {493, 3132}

\bibitem[\protect\citeauthoryear{{Stevenson}, {Ohme}  \&
  {Fairhurst}}{{Stevenson} et~al.}{2015}]{stevenson15}
{Stevenson} S.,  {Ohme} F.,   {Fairhurst} S.,  2015, \mn@doi [\apj]
  {10.1088/0004-637X/810/1/58}, \href
  {https://ui.adsabs.harvard.edu/abs/2015ApJ...810...58S} {810, 58}

\bibitem[\protect\citeauthoryear{Suwa \& Todo}{Suwa \& Todo}{2010}]{Suwa2010}
Suwa H.,  Todo S.,  2010, \mn@doi [Phys. Rev. Lett.]
  {10.1103/PhysRevLett.105.120603}, 105, 120603

\bibitem[\protect\citeauthoryear{Talbot}{Talbot}{2020}]{talbot20}
Talbot C.,  2020, PhD thesis, Monash University,
  \mn@doi{10.26180/5e61a9fc39b73}, \url
  {https://bridges.monash.edu/articles/Astrophysics_of_Binary_Black_Holes_at_the_Dawn_of_Gravitational-Wave_Astronomy/11944914/1}

\bibitem[\protect\citeauthoryear{{Talbot}, {Smith}, {Thrane}  \&
  {Poole}}{{Talbot} et~al.}{2019}]{talbot19}
{Talbot} C.,  {Smith} R.,  {Thrane} E.,   {Poole} G.~B.,  2019, \mn@doi [\prd]
  {10.1103/PhysRevD.100.043030}, \href
  {https://ui.adsabs.harvard.edu/abs/2019PhRvD.100d3030T} {100, 043030}

\bibitem[\protect\citeauthoryear{{Talts}, {Betancourt}, {Simpson}, {Vehtari}
  \& {Gelman}}{{Talts} et~al.}{2018}]{talts18}
{Talts} S.,  {Betancourt} M.,  {Simpson} D.,  {Vehtari} A.,   {Gelman} A.,
  2018, arXiv e-prints, \href
  {https://ui.adsabs.harvard.edu/abs/2018arXiv180406788T} {p. arXiv:1804.06788}

\bibitem[\protect\citeauthoryear{{Tanvir} et~al.,}{{Tanvir}
  et~al.}{2017}]{tanvir17}
{Tanvir} N.~R.,  et~al., 2017, \mn@doi [\apjl] {10.3847/2041-8213/aa90b6},
  \href {https://ui.adsabs.harvard.edu/abs/2017ApJ...848L..27T} {848, L27}

\bibitem[\protect\citeauthoryear{{Thrane} \& {Talbot}}{{Thrane} \&
  {Talbot}}{2019}]{thrane18}
{Thrane} E.,  {Talbot} C.,  2019, \mn@doi [\pasa] {10.1017/pasa.2019.2}, \href
  {https://ui.adsabs.harvard.edu/abs/2019PASA...36...10T} {36, e010}

\bibitem[\protect\citeauthoryear{Veitch \& Del~Pozzo}{Veitch \&
  Del~Pozzo}{2013}]{veitch13}
Veitch J.,  Del~Pozzo W.,  2013, Technical Report LIGO-T1300326, {Analytic
  Marginalisation of Phase Parameter}, \url
  {https://dcc.ligo.org/LIGO-T1300326/public}.
\url {https://dcc.ligo.org/LIGO-T1300326/public}

\bibitem[\protect\citeauthoryear{{Veitch} \& {Vecchio}}{{Veitch} \&
  {Vecchio}}{2008}]{veitch08}
{Veitch} J.,  {Vecchio} A.,  2008, \mn@doi [\prd] {10.1103/PhysRevD.78.022001},
  \href {http://adsabs.harvard.edu/abs/2008PhRvD..78b2001V} {78, 022001}

\bibitem[\protect\citeauthoryear{{Veitch} \& {Vecchio}}{{Veitch} \&
  {Vecchio}}{2010}]{veitch10}
{Veitch} J.,  {Vecchio} A.,  2010, \mn@doi [\prd] {10.1103/PhysRevD.81.062003},
  \href {http://adsabs.harvard.edu/abs/2010PhRvD..81f2003V} {81, 062003}

\bibitem[\protect\citeauthoryear{{Veitch} et~al.}{{Veitch}
  et~al.}{2015}]{veitch15}
{Veitch} J.,  et~al., 2015, \mn@doi [\prd] {10.1103/PhysRevD.91.042003}, 91,
  042003

\bibitem[\protect\citeauthoryear{Veitch, Pozzo, Cody, Pitkin  \&
  ed1d1a8d}{Veitch et~al.}{2017}]{Veitch17-cpnest}
Veitch J.,  Pozzo W.~D.,  Cody Pitkin M.,   ed1d1a8d 2017, johnveitch/cpnest:
  Minor optimisation, \mn@doi{10.5281/zenodo.835874}, \url
  {https://doi.org/10.5281/zenodo.835874}

\bibitem[\protect\citeauthoryear{{Viets} et~al.,}{{Viets}
  et~al.}{2018}]{viets18}
{Viets} A.~D.,  et~al., 2018, \mn@doi [\cqg] {10.1088/1361-6382/aab658}, \href
  {https://ui.adsabs.harvard.edu/abs/2018CQGra..35i5015V} {35, 095015}

\bibitem[\protect\citeauthoryear{{Vigeland} \& {Vallisneri}}{{Vigeland} \&
  {Vallisneri}}{2014}]{VigelandVallisneri}
{Vigeland} S.~J.,  {Vallisneri} M.,  2014, \mn@doi [\mnras]
  {10.1093/mnras/stu312}, \href
  {https://ui.adsabs.harvard.edu/abs/2014MNRAS.440.1446V} {440, 1446}

\bibitem[\protect\citeauthoryear{Vitale, Del~Pozzo, Li, Van Den~Broeck, Mandel,
  Aylott  \& Veitch}{Vitale et~al.}{2012}]{vitale12}
Vitale S.,  Del~Pozzo W.,  Li T. G.~F.,  Van Den~Broeck C.,  Mandel I.,  Aylott
  B.,   Veitch J.,  2012, \mn@doi [Phys. Rev. D] {10.1103/PhysRevD.85.064034},
  85, 064034

\bibitem[\protect\citeauthoryear{{Wade}, {Creighton}, {Ochsner}, {Lackey},
  {Farr}, {Littenberg}  \& {Raymond}}{{Wade} et~al.}{2014}]{wade2014}
{Wade} L.,  {Creighton} J. D.~E.,  {Ochsner} E.,  {Lackey} B.~D.,  {Farr}
  B.~F.,  {Littenberg} T.~B.,   {Raymond} V.,  2014, \mn@doi [\prd]
  {10.1103/PhysRevD.89.103012}, \href
  {https://ui.adsabs.harvard.edu/abs/2014PhRvD..89j3012W} {89, 103012}

\bibitem[\protect\citeauthoryear{{Wang} \& {Zhao}}{{Wang} \&
  {Zhao}}{2020}]{Wang2020}
{Wang} S.,  {Zhao} Z.-C.,  2020, arXiv e-prints, \href
  {https://ui.adsabs.harvard.edu/abs/2020arXiv200200396W} {p. arXiv:2002.00396}

\bibitem[\protect\citeauthoryear{{Watson} et~al.,}{{Watson}
  et~al.}{2019}]{watson19}
{Watson} D.,  et~al., 2019, \mn@doi [\nat] {10.1038/s41586-019-1676-3}, \href
  {https://ui.adsabs.harvard.edu/abs/2019Natur.574..497W} {574, 497}

\bibitem[\protect\citeauthoryear{{Yunes} \& {Siemens}}{{Yunes} \&
  {Siemens}}{2013}]{yunes13}
{Yunes} N.,  {Siemens} X.,  2013, \mn@doi [Living Reviews in Relativity]
  {10.12942/lrr-2013-9}, \href
  {https://ui.adsabs.harvard.edu/abs/2013LRR....16....9Y} {16, 9}

\bibitem[\protect\citeauthoryear{{Yunes}, {Yagi}  \& {Pretorius}}{{Yunes}
  et~al.}{2016}]{yunes16}
{Yunes} N.,  {Yagi} K.,   {Pretorius} F.,  2016, \mn@doi [\prd]
  {10.1103/PhysRevD.94.084002}, \href
  {https://ui.adsabs.harvard.edu/abs/2016PhRvD..94h4002Y} {94, 084002}

\bibitem[\protect\citeauthoryear{{Zackay}, {Dai}  \& {Venumadhav}}{{Zackay}
  et~al.}{2018}]{zackay18}
{Zackay} B.,  {Dai} L.,   {Venumadhav} T.,  2018, arXiv e-prints, \href
  {https://ui.adsabs.harvard.edu/abs/2018arXiv180608792Z} {p. arXiv:1806.08792}

\bibitem[\protect\citeauthoryear{{Zevin}, {Pankow}, {Rodriguez}, {Sampson},
  {Chase}, {Kalogera}  \& {Rasio}}{{Zevin} et~al.}{2017}]{Zevin2017}
{Zevin} M.,  {Pankow} C.,  {Rodriguez} C.~L.,  {Sampson} L.,  {Chase} E.,
  {Kalogera} V.,   {Rasio} F.~A.,  2017, \mn@doi [Astrophys. J.]
  {10.3847/1538-4357/aa8408}, \href
  {http://adsabs.harvard.edu/abs/2017ApJ...846...82Z} {846, 82}

\bibitem[\protect\citeauthoryear{{Zevin}, {Berry}, {Coughlin}, {Chatziioannou}
  \& {Vitale}}{{Zevin} et~al.}{2020}]{zevin2020}
{Zevin} M.,  {Berry} C. P.~L.,  {Coughlin} S.,  {Chatziioannou} K.,   {Vitale}
  S.,  2020, \mn@doi [\apjl] {10.3847/2041-8213/aba8ef}, \href
  {https://ui.adsabs.harvard.edu/abs/2020ApJ...899L..17Z} {899, L17}

\bibitem[\protect\citeauthoryear{{Zhao}, {Lin}  \& {Chang}}{{Zhao}
  et~al.}{2019}]{zhao19}
{Zhao} Z.-C.,  {Lin} H.-N.,   {Chang} Z.,  2019, \mn@doi [Chinese Physics C]
  {10.1088/1674-1137/43/7/075102}, \href
  {https://ui.adsabs.harvard.edu/abs/2019ChPhC..43g5102Z} {43, 075102}

\bibitem[\protect\citeauthoryear{{van der Sluys}, {Raymond}, {Mandel},
  {R{\"o}ver}, {Christensen}, {Kalogera}, {Meyer}  \& {Vecchio}}{{van der
  Sluys} et~al.}{2008a}]{vandersluys08a}
{van der Sluys} M.,  {Raymond} V.,  {Mandel} I.,  {R{\"o}ver} C.,
  {Christensen} N.,  {Kalogera} V.,  {Meyer} R.,   {Vecchio} A.,  2008a,
  \mn@doi [Classical and Quantum Gravity] {10.1088/0264-9381/25/18/184011},
  \href {https://ui.adsabs.harvard.edu/abs/2008CQGra..25r4011V} {25, 184011}

\bibitem[\protect\citeauthoryear{{van der Sluys} et~al.,}{{van der Sluys}
  et~al.}{2008b}]{vandersluys08b}
{van der Sluys} M.~V.,  et~al., 2008b, \mn@doi [\apjl] {10.1086/595279}, \href
  {http://adsabs.harvard.edu/abs/2008ApJ...688L..61V} {688, L61}

\makeatother
\end{thebibliography}

\bsp	
\label{lastpage}
\end{document}